\documentclass[longauth]{aa}
\usepackage[english]{babel}
\usepackage{graphicx}
\usepackage{newlfont}
\usepackage{amssymb}
\usepackage{subfig}
\usepackage{commath}

\usepackage{amsmath}
\usepackage{latexsym}
\usepackage{amsthm}
\usepackage{natbib}
\usepackage{microtype}
\bibpunct{(}{)}{;}{a}{}{,} 
\usepackage{longtable}
\usepackage{lscape}
\usepackage{rotating}
\usepackage{supertabular}
\usepackage{placeins}
\usepackage[usenames,dvipsnames]{xcolor}
\usepackage[bookmarks, colorlinks, breaklinks]{hyperref}  
\hypersetup{linkcolor=blue,citecolor=MidnightBlue,filecolor=black,urlcolor=MidnightBlue} 

\usepackage[varg]{txfonts}

\newcommand{\figiac}[1]{Fig. \ref{#1}}

\newfont{\gwpfont}{cmssq8 scaled 1000}
\newcommand{\rexcess}{{\gwpfont REXCESS}}

\def\YSZ {Y_{\textrm SZ}}

\def\Mv {M_{500}}
\def\Mv {M_{500}}
\def\Rv {R_{500}}

\def\Rvvyx {R_{200}^\mathrm{Y_{X}}}
\def\Rvysz {R_{500}^\mathrm{Y_{\textrm SZ}}}
\def\Rvvysz {R_{200}^\mathrm{Y_{\textrm SZ}}}

\def\Mvysz {M_{500}^\mathrm{Y_{\textrm SZ}}}

\def\Mvher500{M^\mathrm{HE}}

\def\MYSZ {$\Mv$--$\YSZ$}

\def\xmm{XMM-{\it Newton}}
\def\planck{{\it Planck}}

\def\inter{{\it total}}
\def\intra{{\it projection}}
\usepackage{color}

\def\chxmt{CHEX--MATE}
\def\threehun{{\sc The Three Hundred}}
\def\EMS{$\widetilde{\mathrm{EM}}$}
\def\EMSS{${\mathrm{EM_S}}$}

\begin{document}

\title{CHEX-MATE: Constraining the origin of the scatter in galaxy cluster radial X-ray surface brightness profiles}
\author{
I. Bartalucci\inst{\ref{milano_iasf}},
S. Molendi\inst{\ref{milano_iasf}}, 
E. Rasia\inst{\ref{trieste_inaf},\ref{trieste_ifpu}},  
G.W. Pratt\inst{\ref{saclay}}, 
M. Arnaud\inst{\ref{saclay}}, 
M. Rossetti\inst{\ref{milano_iasf}}, 
F. Gastaldello\inst{\ref{milano_iasf}},
D. Eckert\inst{\ref{ginevra}}, 
M. Balboni\inst{\ref{milano_iasf},\ref{insubria}},
S. Borgani\inst{\ref{trieste_inaf},\ref{trieste_ifpu},\ref{trieste_astronomyunit}},
H. Bourdin\inst{\ref{torvergata}},
M.G. Campitiello\inst{\ref{bologna_inaf},\ref{bologna_uni}},
S. De Grandi\inst{\ref{merate}}, 
M. De Petris\inst{\ref{sapienza}},
R.T. Duffy\inst{\ref{saclay}}, 
S. Ettori\inst{\ref{bologna_inaf},\ref{bologna_infn}}, 
A. Ferragamo\inst{\ref{sapienza}},
M. Gaspari\inst{\ref{princeton}},
R. Gavazzi\inst{\ref{iap_paris},\ref{cambridge}}, 
S. Ghizzardi\inst{\ref{milano_iasf}}, 
A. Iqbal\inst{\ref{saclay}},
S.T. Kay\inst{\ref{manchester}},
L. Lovisari\inst{\ref{bologna_inaf},\ref{cfa}}, 
P. Mazzotta\inst{\ref{torvergata}}, 
B.J. Maughan\inst{\ref{bristol}}, 
E. Pointecouteau\inst{\ref{toulouse}}, 
G. Riva\inst{\ref{milano_iasf},\ref{unimi}},
M. Sereno\inst{\ref{bologna_inaf},\ref{bologna_infn}} 
}
\authorrunning{I. Bartalucci}
\institute{
INAF, IASF-Milano, via A. Corti 12, I-20133 Milano, Italy
\label{milano_iasf} 
\and
INAF – Osservatorio Astronomico di Trieste, via Tiepolo 11, I-34131 Trieste, Italy
\label{trieste_inaf} 
\and
IFPU – Institute for Fundamental Physics of the Universe, via Beirut 2, 34151, Trieste, Italy
\label{trieste_ifpu} 
\and
Université Paris-Saclay, Université Paris Cité, CEA, CNRS, AIM, 91191, Gif-sur-Yvette, France
\label{saclay} 
\and
Department of Astronomy, University of Geneva, ch. d’\'Ecogia 16, CH-1290 Versoix Switzerland
\label{ginevra} 
\and
DiSAT, Università degli Studi dell’Insubria, via Valleggio 11, I-22100 Como, Italy 
\label{insubria} 
\and
Astronomy Unit, Department of Physics, University of Trieste, via
Tiepolo 11, 34131 Trieste, Italy
\label{trieste_astronomyunit}
\and
Università degli studi di Roma ‘Tor Vergata’; Via della ricerca scientifica, 1; 00133 Roma, Italy
\label{torvergata}
\and
INAF, Osservatorio di Astrofisica e Scienza dello Spazio, via Piero Gobetti 93/3, 40129 Bologna, Italy 
\label{bologna_inaf}
\and
Dipartimento di Fisica e Astronomia, Universitá di Bologna, via Gobetti 92/3, 40121 Bologna, Italy
\label{bologna_uni}
\and
INAF,  Osservatorio Astronomico di Brera, via E. Bianchi 46, 23807 Merate, Italy
\label{merate} 
\and
Dipartimento di Fisica, Sapienza Universitá di Roma, Piazzale Aldo Moro 5, 00185 Roma, Italy
\label{sapienza}
\and
INFN, Sezione di Bologna, viale Berti Pichat 6/2, I-40127 Bologna, Italy
\label{bologna_infn} 
\and
Department of Astrophysical Sciences, Princeton University, Princeton, NJ 08544, USA
\label{princeton}
\and
Laboratoire d'Astrophysique de Marseille, Aix-Marseille Université, CNRS, CNES, Marseille, France
\label{iap_paris}
\and
Institut d’Astrophysique de Paris, CNRS, Sorbonne Université, Paris, France
\label{cambridge}
\and
Jodrell Bank Centre for Astrophysics, Department of Physics and Astronomy, School of Natural Sciences, The University of Manchester, Manchester M13 9PL, UK
\label{manchester} 
\and
Center for Astrophysics $|$ Harvard $\&$ Smithsonian, 60 Garden Street, Cambridge, MA 02138, USA
\label{cfa} 
\and
HH Wills Physics Laboratory, University of Bristol, Tyndall Ave, Bristol, BS8 1TL, UK
\label{bristol} 
\and
IRAP, Universit\'e de Toulouse, CNRS, CNES, UPS, 9 av du colonel Roche, BP44346, 31028 Toulouse cedex 4, France
\label{toulouse} 
\and
Dipartimento di Fisica, Universitá degli Studi di Milano, Via G. Celoria 16, 20133 Milano, Italy
\label{unimi}
}
\date{received -- accepted}
\abstract{We investigate the statistical properties and the origin of the scatter within the spatially resolved surface brightness profiles of the \chxmt\ sample, formed by 118 galaxy clusters selected via the SZ effect. These objects have been drawn from the \textit{Planck} SZ catalogue and cover a wide range of masses, M$_{500}=[2-15] \times 10^{14} $M$_{\odot}$, and redshift, z=[0.05,0.6].
We derived the surface brightness and emission measure profiles and determined the statistical properties of the full sample and sub-samples according to their morphology, mass, and redshift. We found that there is a critical scale, R$\sim 0.4 \Rv$, within which morphologically relaxed and disturbed object profiles diverge. The median of each sub-sample differs by a factor of $\sim 10$ at $0.05\,R_{500}$. There are no significant differences between mass- and redshift-selected sub-samples once proper scaling is applied.

We compare \chxmt\ with a sample of 115 clusters drawn from the \threehun\ suite of cosmological simulations. We found that simulated emission measure profiles are systematically steeper than those of observations. For the first time, the simulations were used to break down the components causing the scatter between the profiles. 
We investigated the behaviour of the scatter due to object-by-object variation. We found that the high scatter, approximately 110\%, at $R<0.4\Rvysz$ is due to a genuine difference between the distribution of the gas in the core of the clusters. The intermediate scale, $\Rvysz =[0.4-0.8]$, is characterised by the minimum value of the scatter on the order of  $0.56,$ indicating a region where cluster profiles are the closest to the self-similar regime. Larger scales are characterised by increasing scatter due to the complex spatial distribution of the gas. Also for the first time, we verify that the scatter due to projection effects is smaller than the scatter due to genuine object-by-object variation in all the considered scales.}
\keywords{intracluster medium -- X-rays: galaxies: clusters}
\maketitle
\section{Introduction}\label{sec:introduction}
Galaxy clusters represent the ultimate manifestation of large-scale structure formation. Dark matter comprises $80\%$ of the total mass in a cluster and is the main actor of the gravitation assembly prcoess \citep{voit2005,allen2011,borgani2011}. This  influences the prevalent baryonic component represented by a hot and rarefied plasma that fills the cluster volume, that is, the intracluster medium (ICM). 
This plasma's properties are affected by the individual assembly history and ongoing merging activities. The study of its observational properties is thus fundamental to study how galaxy clusters form and evolve. The ideal tool for investigating this component is X-ray observations, as the ICM emits in this band via thermal Bremsstrahlung.

The radial profiles of the X-ray surface brightness (SX) of a galaxy cluster and the derived emission measure (EM) are  direct probes of the plasma properties. These two quantities can be easily measured in the X-ray band and have played a crucial role in the characterisation of the ICM distribution since the advent of high spatial resolution X-ray observations (e.g. \citealt{vikhlinin99}). \citealt{neumann99} and \citealt{neumann01} compared SX profiles with expectations from theory to test the self-similar evolution scenario and investigate the relation between the cluster luminosity and its mass and temperature. \cite{arnaud2001} tested the self-similarity of the EM profiles of 25 clusters in the [0.3-0.8] redshift range, finding that clusters evolve in a self-similar scenario, which deviates from the simplest models because of the individual formation history. The SX and EM profiles have been used to investigate the properties of the outer regions of galaxy clusters, both in observations (e.g. \citealt{vikhlinin99,neumann05,ettori09}) and in a suite of cosmological simulations (see e.g. \citealt{roncarelli06}). These regions are of particular interest because of the plethora of signatures from the accretion phenomena, but they are hard to observe because of their faint signal. More recent works based on large catalogues (see e.g. \citealt{rossetti2017} and \citealt{santos2017}) have determined the effects of the X-ray versus the Sunyaev Zel'Dovich (SZ; \citealt{sunyaev1980}) selection by studying the concentration of the surface brightness profiles in the central regions of galaxy clusters.
Finally, the SX radial profile represents the baseline for any study envisaging to derive the thermodynamical  properties of the ICM, such as the 3D spatial distribution of the gas \citep{sereno2012,sereno2017,sereno2018}. This information can be combined with the radial profile of the temperature, and together, they can be used to derive quantities such as the entropy (see, e.g. \citealt{voit2005entropy}), pressure, and mass of the galaxy cluster under the assumption of hydrostatic equilibrium \citep{ettori2013,pratt22}. 

In this paper, we used the exceptional data quality of the 118 galaxy clusters from the Cluster HEritage project with XMM-Newton - Mass Assembly and Thermodynamics at the Endpoint of structure formation (CHEX-MATE\footnote{\href{http://xmm-heritage.oas.inaf.it/}{xmm-heritage.oas.inaf.it}}, PI; S. Ettori and G.W. Pratt). Specifically, we investigate for the first time the statistical properties of the X-ray surface brightness and emission measure  radial profiles of a sample of galaxy clusters observed with unprecedented and homogeneous deep \xmm\ observations. The sample, being based on the Planck catalogue, is SZ selected and thus predicted to be tightly linked to the mass of the cluster (e.g. \citealt{planelles17} and \citealt{leb18}), and hence it should yield a minimally biased sample of the underlying cluster population.

Our analysis is strengthened by the implementation of the results from a mass-redshift equivalent sample from cosmological and hydrodynamical simulations of the \threehun\ collaboration \citep{cui2016}. We used a new approach to understand the different components of the scatter, considering the population (i.e. cluster-to-cluster) scatter and the single object scatter inherent to projection effects.

In Sect. \ref{sec:data_sample}, we present the \chxmt\ sample.
In Sects. \ref{sec:analysis} and \ref{sec:cosmo_analysis}, we describe the methodology used to prepare the data and the derivation of the radial profiles of the \chxmt\ and numerical datasets, respectively. In Sect. \ref{sec:profile_shape}, we discuss the shape of the profiles. In Sect. \ref{sec:profile_scatter}, we present the scatter within the \chxmt\ sample. In Sect. \ref{sec:scatter_origin}, we investigate the origin of the scatter of the EM profiles, and finally in Sect. \ref{sec:conclusions}, we discuss
our results and present our conclusions. 

We adopted a flat $\Lambda$-cold dark matter cosmology with $\Omega_{\mathrm{M}}(0) = 0.3$, $\Omega_\Lambda = 0.7$, H$_{0} = 70$ km Mpc s$^{-1}$, E(z)$ = (\Omega_{\mathrm{M}} (1+z)^3 + \Omega_\Lambda)^{1/2}$, and $\Omega_{\mathrm{M}}(z) = \Omega_{\mathrm{M}}(0)(1+z)^3/$E(z)$^2$ throughout. The same cosmology was used for the numerical simulations, except for $h=0.6777$.
Uncertainties are given at the 68\% confidence level (i.e. $1\sigma$).
All the fits were performed via $\chi^2$ minimisation. 
We characterised the statistical properties of a sample by computing the median and the $68\%$ dispersion around it. This dispersion was computed by ordering the profiles according to their $\chi^2$ with respect to the median and by considering the profile at $\pm 34\%$ around it. We use natural logarithm throughout the work except for where we state otherwise.


\section{The \chxmt\ sample}\label{sec:data_sample}
\subsection{Definition}
\begin{figure}[!ht]
\begin{center}
\resizebox{1\columnwidth}{!}{
\includegraphics[]{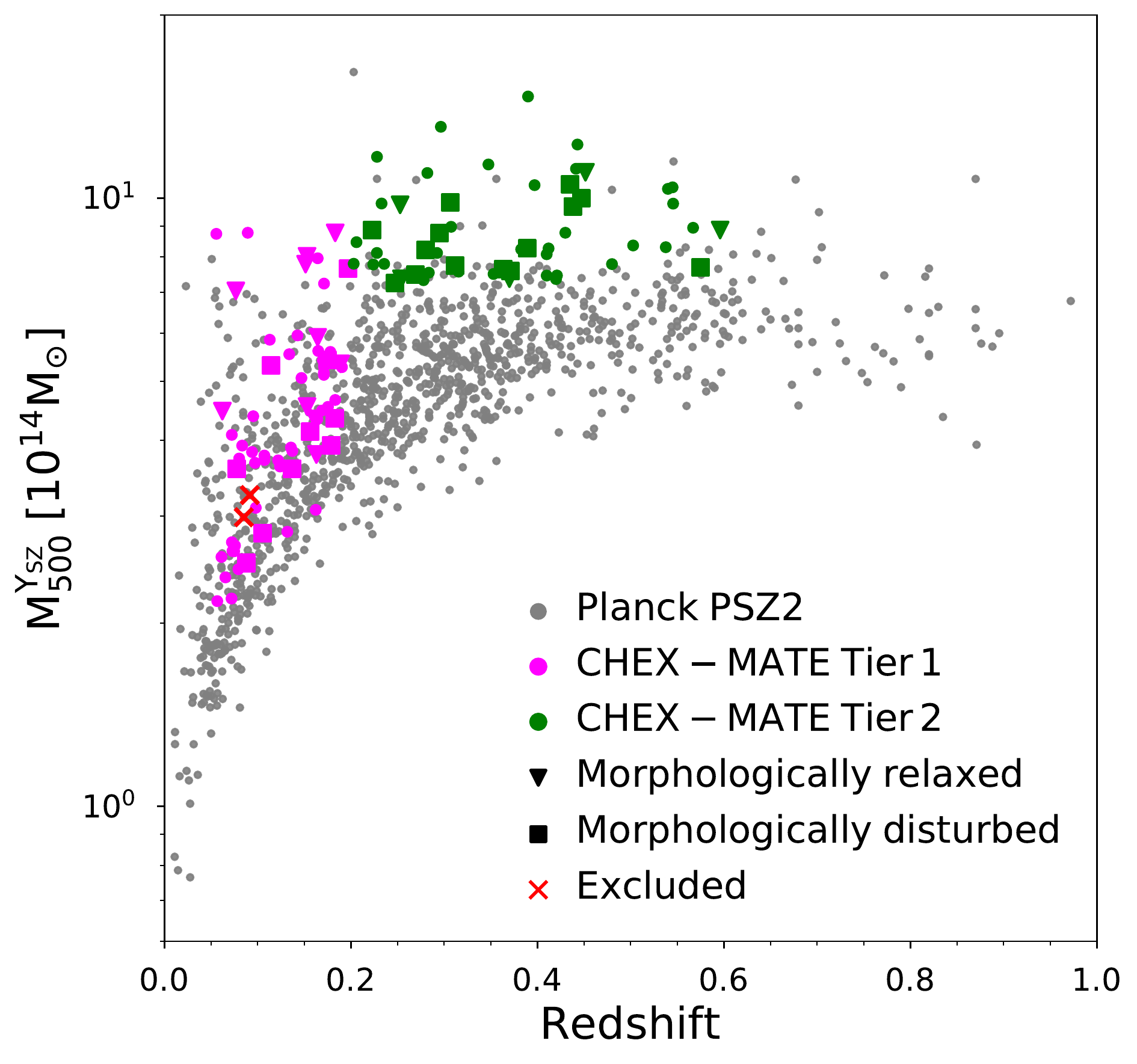} 
}
\end{center}
\caption{\footnotesize{Distribution of the clusters published in the PSZ2 \planck\ catalogue \citep{PSZ2} in the mass-redshift plane. The masses in the \planck\ catalogue were derived iteratively from the \MYSZ\ relation calibrated using hydrostatic masses from \xmm; they were not corrected for the hydrostatic equilibrium bias. The magenta and green points represent the Tier 1 and Tier 2 clusters of the \chxmt\ sample, respectively \citep{chexmate1}. The triangles and squares identify the morphologically relaxed and disturbed clusters, respectively, which were identified  according to the classification scheme in \citet{campitiello2022}. The two red crosses identify the clusters excluded from the analysis of this work.}}
\label{fig:planck_mz_plane}
\end{figure}
     This work builds on the sample defined for the \xmm\ heritage programme accepted in AO-17. We briefly report the sample definition and selection criteria here that are detailed in \citet{chexmate1}. 
    The scientific objective of this programme is to investigate the ultimate manifestation of structure formation in mass and time by observing and characterising the radial thermodynamical and dynamic properties of a large, minimally biased and S/N-limited sample of galaxy clusters. This objective is achieved by selecting 118 objects from the \planck\ PSZ2 catalogue \citep{PSZ2}, applying an SNR threshold of 6.5 in the SZ identification, and folding the \xmm\ visibility criteria. 
    
    The key quantity $\Mvysz$,  defined as the mass enclosed within the radius $\Rvysz$ of the cluster where its average total matter density is 500 times the critical density of the Universe, is measured by the \planck\ collaboration using the MMF3 SZ detection algorithm detailed in \citet{planck_psz2}. This algorithm measures the $\YSZ$ flux associated to each detected cluster, and it is used to derive the $\Mvysz$ using the \MYSZ\ relation calibrated in \citet{arnaud2010}, assuming self-similar evolution.  We note that while the clusters' precise mass determination is one of the milestones of the multi-wavelength coverage of the CHEX-MATE programme, in this paper we consider the radii and mass values directly from the \planck\ catalogue. The impact of this choice will be discussed in Sect. \ref{sec:mediansim_vs_chexsim}.
    
    The \chxmt\ sample is split in two sub-samples according to the cluster redshift.\\
   \textbf{Tier 1} provides a local sample of  61 objects in the [0.05-0.2] redshift range in the northern sky (i.e. DEC > 0), and their $\Mvysz$ span the $[2-9] \times 10^{14} M_{\odot}$ mass range. These objects represent a local anchor for any evolution study.\\
  \textbf{Tier 2} offers a sample of the massive clusters, $\Mvysz > 7.25 \times 10^{14} M_{\odot}$ in the [0.2-0.6] redshift range. These objects represent the culmination of cluster evolution in the Universe. 
   
 The distribution in the mass and redshift plane of the \chxmt\ sample and its sub-samples are shown in  \figiac{fig:planck_mz_plane}. The exposure times of these observations were optimised to allow the determination of spatially resolved temperature profiles at least up to $\Rv$ with a precision of $15\%$. 
    
The clusters PSZ2 G028.63+50.15 and PSZ2 G283.91+73.87 were excluded from the analysis presented in this work since their radial analysis could introduce large systematic errors without increasing the statistical quality of the sample. Indeed, the former system presents a complex morphology (see \citealt{schellenberger22} for a detailed analysis), and it has a background cluster at $z=0.38$ within its extended emission. The latter is only $\sim 30$ arcmin from M87, and thus its emission is heavily affected by the extended emission of Virgo.
The basic properties of the final sample of 116 objects are listed in Table \ref{tab:500_prop}. 

\subsection{Sub-samples}
\label{sec:subsamples}
We defined \chxmt\ sub-samples based on key quantities: mass, redshift, and morphological status.
The analysis of the morphology of the \chxmt\ clusters  sample is described in detail in \citet{campitiello2022}. The authors use a combination of morphological parameters (see \citealt{rasia2013} for the definition of these parameters) to classify the clusters as morphologically relaxed, disturbed, or mixed. Following the criteria described in Sect. 8.2 of \citet{campitiello2022}, the authors identified the 15 most relaxed and 25 most disturbed clusters. We adopted their classification in this paper and refer to the former group as morphologically relaxed clusters and the latter group as disturbed clusters. 

We defined the sub-samples of nearby and distant clusters considering the 85 and 31 clusters at $z \leq 0.33$ and $z > 0.33$, respectively, the value 0.33 being the mean redshift of the sample. Similarly, we built the sub-samples of high- and low-massive clusters considering the 40 and 76 clusters with $\Mvysz \leq 5 \times 10^{14} M_{\odot}$ and  $\Mvysz > 5 \times 10^{14} M_{\odot}$, respectively.

\section{Data analysis}\label{sec:analysis}
\subsection{Data preparation}
\subsubsection{\xmm\ data}
    The clusters used in this work were observed using the European Photon Imaging Camera (EPIC; \citealt{turner2001} and \citealt{struder2001}). The instrument comprises three CCD arrays, namely, MOS1, MOS2, and \textit{pn}, that simultaneously observe the target. 
    Datasets were reprocessed using the Extended-Science Analysis System (ESAS\footnote{\href{http://cosmos.esa.int/web/xmm-newton}{cosmos.esa.int/web/xmm-newton}}; \citealt{snowden08}) embedded in SAS version 16.1. The emchain and epchain tools were used to apply the latest calibration files made available January $2021$ and produce pn out-of-time datasets.
    Events in which the keyword PATTERN is greater than four for the MOS$1$ and MOS$2$ cameras and greater than 12 for the \textit{pn} camera were filtered out from the analysis. The CCDs showing an anomalous count rate in the MOS1 and MOS2 cameras were also removed from the analysis.
    Time intervals affected by flares were removed using the tools mos-filter and pn-filter by extracting the light curves in the [2.5-8.5] keV band and removing the time intervals where the count rate exceeded $3\sigma$ times the mean count rate from the analysis. 
     Point sources were filtered from the analysis following the scheme detailed in Section 2.2.3 of \citet{ghirardini19}, which we summarise as follows. Point sources were identified by running the SAS wavelet detection tool \textit{ewavdetect} on $[0.3-2]$ keV and $[2-7]$ keV images obtained from the combination of the three EPIC cameras and using  wavelet scales in the range of 1–32 pixels and an S/N threshold of five, with each bin width being $\sim 2$ arcsec. The PSF and sensitivity of \xmm\ depends on the off-axis angle. For this reason, the fraction of unresolved point sources forming the Cosmic X-ray Background (CXB; \citealt{giacconi2001}) is spatially dependent.  We used a threshold in the LogN-LogS distribution of detected sources, below which we deliberately left the point source in the images to ensure a constant CXB flux across the detector. Catalogues produced from the two energy band images were then merged. 
     At the end of the procedure, we inspected the identified point sources by eye to check for false detections in CCD gaps.
     We also identified extended bright sources other than the cluster itself by eye and removed them from the analysis. We identified 13 clusters affected by at least one sub-structure within $\Rvvysz$ that were masked by applying circular masks of $\sim 3$ arcmin radius on average. 

\subsubsection{Image preparation}

     We undertook the following procedures to generate the images from which we derived the profiles. 
     Firstly, we extracted the photon count images in the $[0.7-1.2]$ keV band for each camera, this energy band maximises the source to background ratio \citep{ett10}. An exposure map for each camera folding the vignetting effect was produced using the ESAS tool eexpmap.
     
     The background affecting the X-ray observations was due to a sky and instrumental component. The former was from the local Galactic emission and the CXB \citep{kuntzsnowden2000}, and its extraction is described in detail in Sect. \ref{sec:radial_analysis}. The latter was due to the interaction of high energy particles with the detector. We followed the strategy described in \cite{ghirardini19} to remove this component by producing background images that accounted for the particle background and the residual soft protons.

     The images, exposure, and background maps of the three cameras were merged to maximise the statistic. The \textit{pn} exposure map was multiplied by a factor to account for the ratio of the effective area MOS to \textit{pn} in the $[0.7-1.2]$ keV band when merging the exposure maps. This factor was computed using XSPEC by assuming a mean temperature and using the hydrogen column absorption value, ${\mathrm{N_{H}}}$, reported in Table \ref{tab:500_prop}. Henceforth, we refer to the combined images of the three cameras  and the background maps simply as the observation images and the particle background datasets, respectively.
     
\begin{figure*}[!ht]
\begin{center}
\resizebox{1\textwidth}{!}{
\includegraphics[]{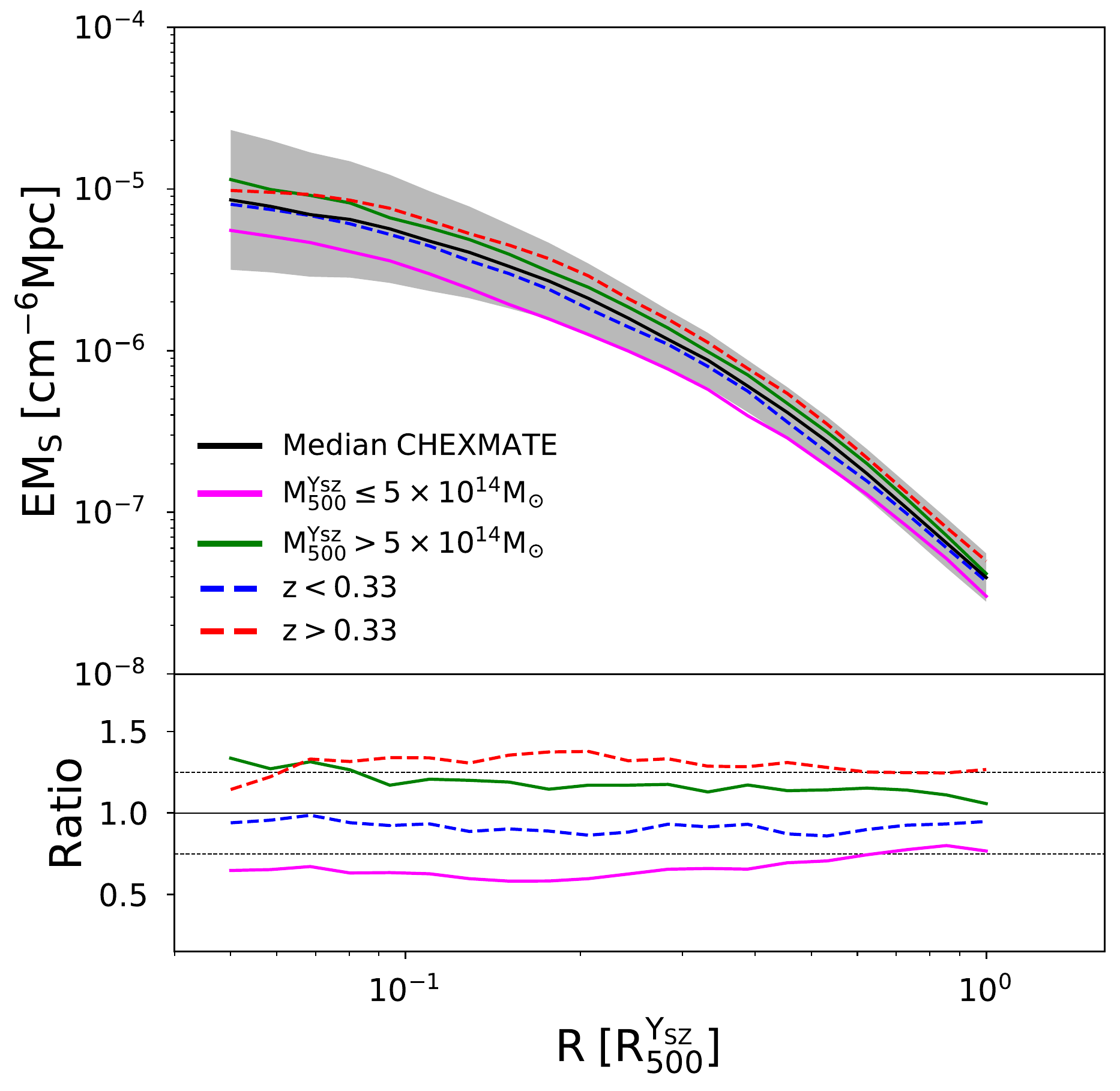}
\includegraphics[]{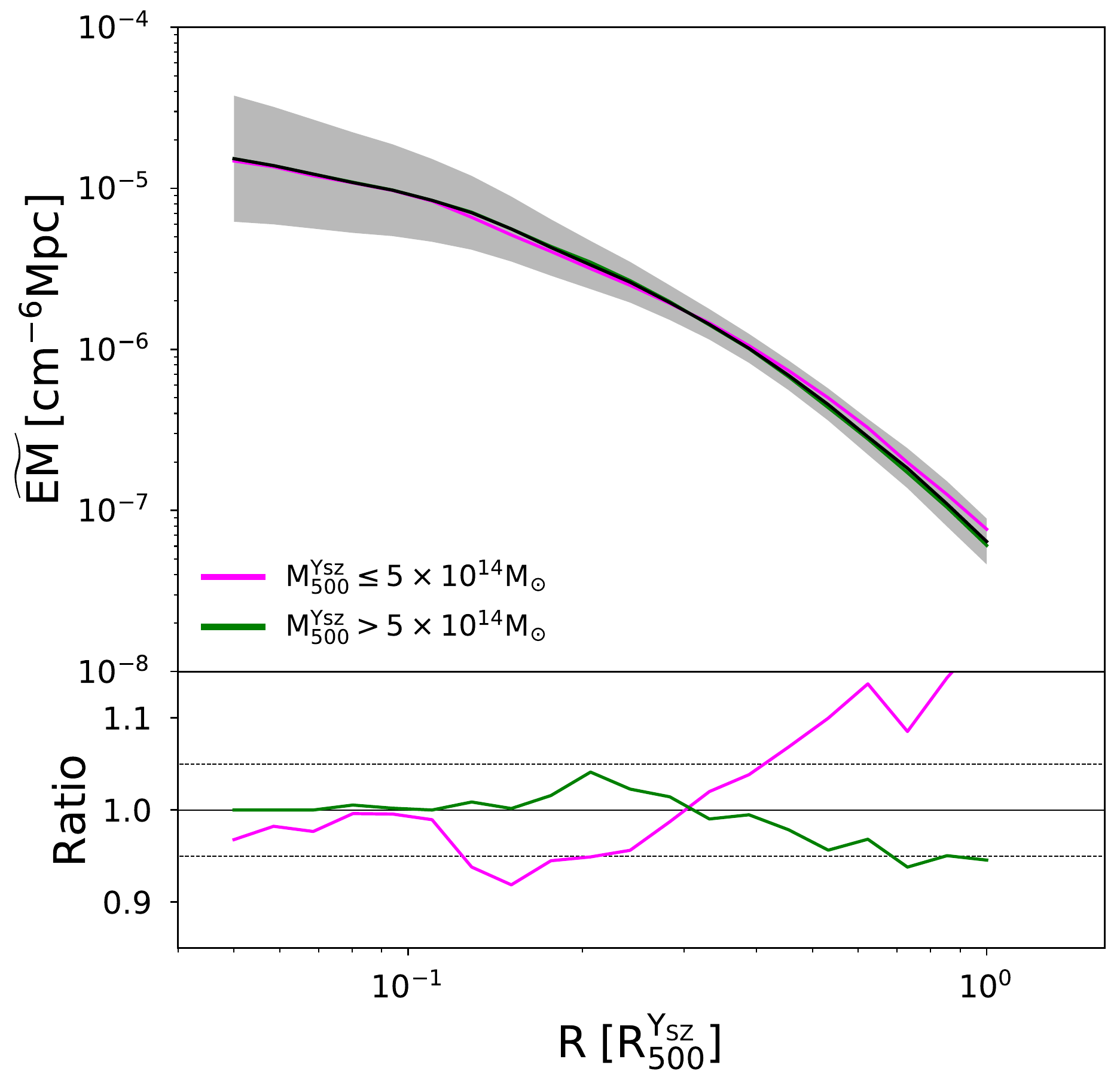}
\includegraphics[]{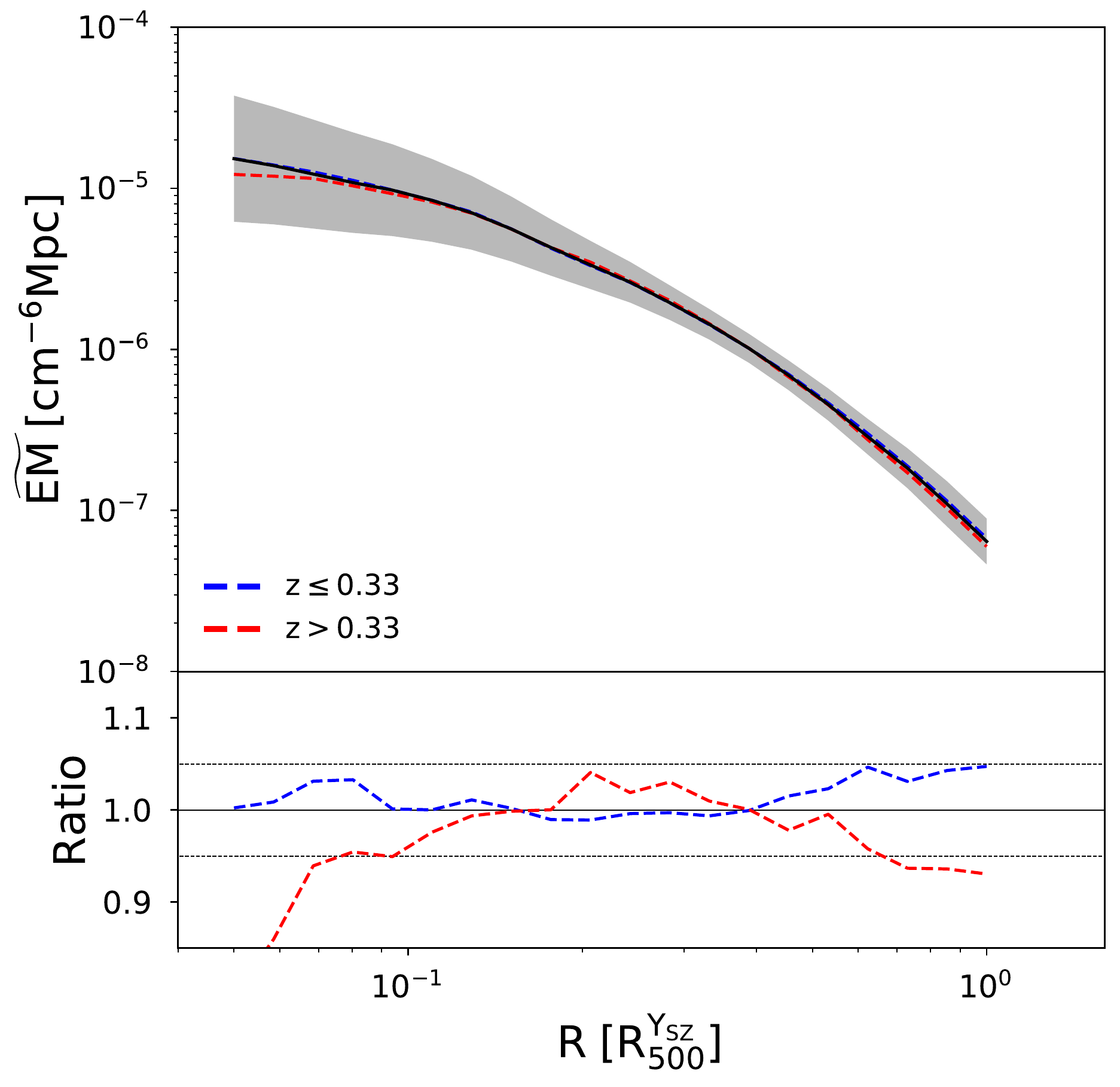}
}
\end{center}
\caption{\footnotesize{Comparison of the statistical properties of the \chxmt\ EM profiles divided in redshift and mass selected sub-samples. Top-left panel: Median of the \EMSS\ profiles centred on the X-ray peak scaled for self-similar evolution using Eq. \ref{eq:self_similar_scaling}. The dispersion is shown with the black solid line and the grey envelope. The magenta and green solid lines represent the median of the low-mass, $\Mvysz \leq 5 \times 10^{14} $M$_{\odot}$, and high-mass, $\Mvysz > 5 \times 10^{14} $M$_{\odot}$, sub-samples, respectively. The red and blue dotted lines represent the medians of the low-, $z \leq 0.33$, and high-redshift, $z > 0.33$, sub-samples, respectively. Bottom-left panel: Ratio of the sub-samples medians with respect to the full \chxmt\ sample median. The dotted horizontal lines represent the 25\% plus and minus levels.
Middle panels: Comparison of the statistical properties of the \EMS\ profiles scaled to account also for mass and redshift evolution divided in mass selected sub-samples. On the top panel we show the same as in the top left panel except the medians and the dispersion were computed using profiles scaled with Eq. \ref{eq:scale_formula}. On the bottom we show the ratio between the medians of the sub-samples as respect to the median of the full sample. The dotted horizontal lines in the lower-middle panel represent the plus and minus 5\% levels.
Right panels: Comparison of the statistical properties of the scaled \EMS\ profiles divided in redshift selected sub-samples. On the top panel we show the same as in the top left panel except that the medians and the dispersion were computed using the profiles scaled with Eq. \ref{eq:scale_formula}. On the bottom we show the ratio between the medians of the sub-samples as respect to the median of the full \chxmt\ sample.
}}
\label{fig:em_medians_mass_red_dep}
\end{figure*} 
\subsection{Global quantities}
\subsubsection{Average temperature}\label{sec:global_quantities}
We estimated the average temperature, T$_{\mathrm{avg}}$, of each cluster by applying the definition of the temperature of a singular isothermal sphere with mass $M_{500}$ as described in Appendix A of \citet{arnaud2010}:
\begin{equation}\label{eq:t200}
    \mathrm{T}_{\mathrm{avg}} = 0.8 \times \mathrm{T_{500}} = 0.8 \times \frac{\mu m_p G \Mvysz}{2 \Rvysz},
\end{equation} 
where $\mu = 0.59$ is the mean molecular weight, $m_p$ is the proton mass, $G$ is the gravitational constant, and the 0.8 factor represents the average value of the universal temperature profile derived by \citet{ghirardini19} with respect to T$_{500}$. 
These temperatures are reported in Table \ref{tab:500_prop}.

\subsubsection{Cluster coordinates}\label{sec:cluster:coordinates}
We produced point source free emission images by filling the holes from the masking procedure with the local mean emission estimated in a ring around each excluded region by using the tool dmfilth. 
We then performed the vignetting correction by dividing them for the exposure map. We used these images to determine the peak by identifying the maximum of the emission after the convolution of the map with a Gaussian filter with $\sim 10$ arcsec width. The centroid of the cluster was determined by performing a weighted-mean of the pixel positions using the counts as weight within a circular region centred on the peak and with its radius as $R_{500}^{Y_{SZ}}$. This has been done to avoid artefact contamination near the detector edges. The coordinates obtained are reported in Table \ref{tab:500_prop}.

\subsection{Radial profiles}\label{sec:radial_analysis}
\subsubsection{Surface brightness profiles}\label{sec:sx_profiles}
\textbf{Azimuthal mean profiles.} The surface brightness radial profiles, S$(\Theta)$, were extracted using the following technique. We defined concentric annuli centred on the X-ray peak and the centroid. The minimum width was set to $4^{\prime\prime}$ and was increased using a logarithmic factor. In each annulus, we computed the sum of the photons from the observation image as well as from the particle background  datasets. The particle background-subtracted profile was divided by the exposure folding the vignetting in the same annulus region. We estimated the sky background component as the average count rate between $R_{200}=1.49 \Rvysz$ and $13.5$ arcmin and subtracted it from the profile. If $R_{200}$ was outside the field of view, we estimated the sky background component using the \xmm-ROSAT background relation described in Appendix \ref{appendix:rosat}. The sky background-subtracted profiles were re-binned to have at least nine counts per bin after background subtraction. 
We corrected the profiles for the PSF using the model developed by \citet{ghizzardi2001}. We refer hereafter to these profiles as the mean SX profiles.

\noindent
\textbf{Azimuthal median profiles.} We also computed the surface brightness radial profiles considering the median in each annulus following the procedure detailed in Section 3 of \citet{eckert15}. This procedure has been introduced to limit the bias caused by the emission of sub-clumps and sub-structures too faint to be identified and masked (e.g. \citealt{roncarelli13,zuv13}).
Briefly, we applied the same binning scheme and point source mask to the particle background dataset to perform the background subtraction. Employing the procedure of \citet{cappellari03} and \citet{diehl06}, we first produced Voronoi-binned maps to ensure 20 counts per bin on average to apply the Gaussian approximation. We then extracted the surface brightness median profile with the same annular binning of the mean profile, considering in each radial bin the median count rate of the Voronoi cells, whose centre lies within the annulus. The sky background was estimated with the same approach used for the mean profile except that we estimated the median count rate. Finally, the sky background-subtracted profiles were re-binned using the same 3$\sigma$ binning of the mean profiles.  The four resulting types of surface brightness profiles are shown in \ref{fig:sx_all_profiles}.
We were able to measure the profiles beyond $\Rvysz$ for 107 of the 116   (i.e. $\sim92\%$) \chxmt\ objects.

We report in Table \ref{tab:relative_errors} the median relative errors at fixed radii to illustrate the excellent data quality.
From now on, we refer to these profiles as the median SX profiles, and throughout the paper, we use these profiles centred on the X-ray peak unless stated otherwise.
\begin{table}
\caption{ {\footnotesize Average of the relative errors of the \chxmt\ EM profiles.}}\label{tab:relative_errors}
\begin{center}
\resizebox{0.9\columnwidth}{!} {
\begin{tabular}{lcc}
\hline        
\hline
Radius $[\Rvysz]$     & Average relative error [\%] & Number of profiles used \\
0.2                   &  1.7                       & 116 \\
0.5                   &  2.1                       & 116  \\
0.7                   &  3.0                       & 116  \\
1.0                   &  6.0                       & 107    \\
\hline
\end{tabular}
}
\end{center}
\footnotesize{\textbf{Notes}: We used the EM median profiles centred on the X-ray peak. We also report the number of profiles that have been used to compute the relative error in the third column.}
\end{table}

\subsubsection{Emission measure profiles}\label{sec:emission_profiles}
We computed the $\mathrm{EM}$ radial profiles using Equation 1 of \citet{arnaud02}:
\begin{equation}\label{eq:monique_em_formula}
     \mathrm{EM(r)} =  \mathrm{S}(\Theta )\frac{4 \pi (1 + z)^4}{\epsilon {(T,z)}},
\end{equation}
where $\Theta =  r / \mathrm{d_A} (z),$ with $\mathrm{d_A} (z)$ as the angular diameter distance, and $\epsilon$ is the emissivity integrated in the E$_1=0.7$ keV and E$_2=1.2$ keV band and is defined as 
\begin{equation}
 \epsilon(T,z) = \int_{E_1}^{E_2} \Sigma(E) e^{-\sigma(E) N_H} f_T((1+z)E)(1+z)^2dE,
\end{equation}
where  $\Sigma$(E) is the detector effective area at energy $E$, $\sigma (E)$ is the absorption cross section, $N_\mathrm{H}$ is the hydrogen column density along the line of sight, and $f_T((1+z)E)$ is the emissivity at energy $(1+z)E$ for a plasma at temperature $T$.
The $\epsilon$ factor was computed using an absorbed Astrophysical Plasma Emission Code (APEC) within the XSPEC environment. The absorption was calculated using the phabs model folding the Hydrogen absorption column reported in Table \ref{tab:500_prop}.
The dependency of $\epsilon$ on temperature and abundance in the band we used to extract the profile is weak (e.g. \citealt{lovisari21}). Therefore, for APEC we used the average temperature, $k\mathrm{T_{avg}}$, of the cluster within $\Rvysz$ and the abundance fixed to 0.25 \citep{ghizzardi21} with respect to the solar abundance table of \cite{anders89}. Finally, we used the redshift values reported in Table \ref{tab:500_prop}.
 We obtained EM azimuthal mean and azimuthal median profiles centred on the X-ray peak and centroid, converting the respective surface brightness profiles. 
The EM profiles were first scaled considering only the self-similar evolution scenario, $\mathrm{EM_S}$, as in \citet{arnaud02}:
\begin{equation}\label{eq:self_similar_scaling}
  \mathrm{  EM_S}(r,T,z) = E(z)^{-3} \times \left( \frac{kT_{\mathrm{avg}}}{10} \right) ^{-1/2} \times \mathrm{EM}(r),  
\end{equation}
where $x$=r/$\Rvysz$ and T$_{avg}$ is the average temperature of the cluster, as in Equation \ref{eq:t200}.
The left panel of \figiac{fig:em_medians_mass_red_dep} shows the median of the $\mathrm{EM_S}$ profiles centred on the X-ray peak as well as its $68\%$ dispersion. In the same plot, we also show the medians of the sub-samples introduced in Sect~\ref{sec:subsamples}. 

\begin{figure}[!ht]
\begin{center}
\resizebox{1\columnwidth}{!}{
\includegraphics[]{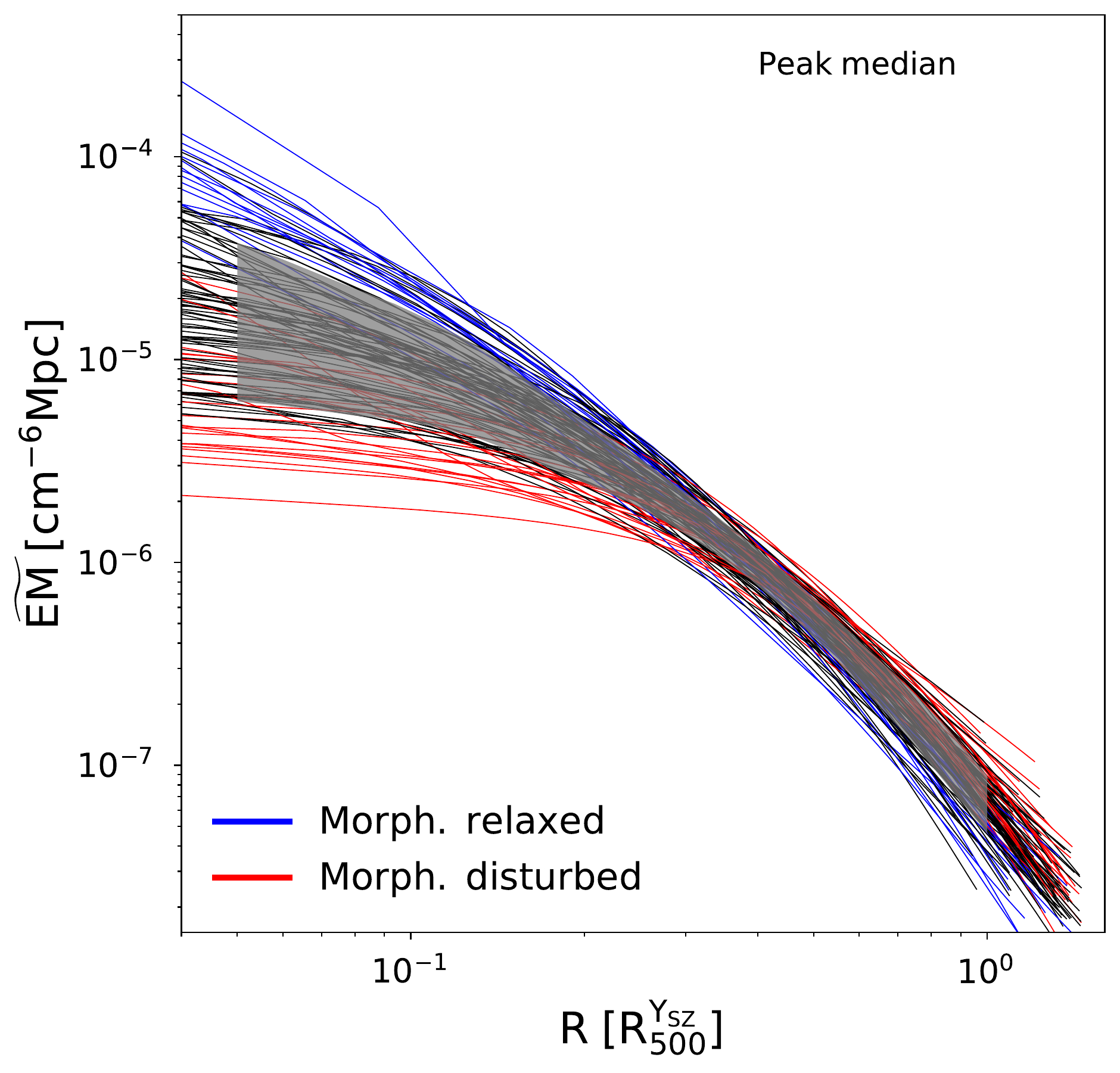} 
}
\end{center}
\caption{\footnotesize{Scaled emission measure median profiles centred on the X-ray peak. The blue and red solid lines indicate morphologically relaxed and disturbed clusters, respectively. The profiles extracted from clusters with mixed morphology are shown with black solid lines.  The selection criteria was based on the classification made by \citet{campitiello2022} in their Section 8.2. The grey-shaded envelope represents the dispersion at the $68\%$ level.}}
\label{fig:profili_mediani_chexmate}
\end{figure}

Their ratio with respect to the CHEX-MATE median shown in the bottom panel demonstrates that the employed re-scaling is not optimal since for all sub-samples there are variations with respect to the median that range between 10\% and 50\% at all scales.  We therefore tested another re-scaling following \citet{pratt22} and \citet{ettori22}, who point out how the mass dependency is not properly represented by the self-similar scenario and had a small correction also with respect to the redshift evolution. 
%
The final scaling that we considered is given by the following:
\begin{equation}\label{eq:scale_formula}
    \mathrm{\widetilde{EM}}(r,T,z) = E(z)^{-3.17} \left( \frac{kT_{\mathrm{\mathrm{avg}}}}{10 \; \mathrm {keV} } \right)^{-1.38} \times  \mathrm{EM}(r).
\end{equation}
The effect of this scaling on the mass and redshift residual dependency is shown in the middle and right panel of \figiac{fig:em_medians_mass_red_dep}, respectively. The medians of the sub-samples show little variations in relation to the whole sample within the order of a few percentage points on average. 
We show the individual scaled median radial profiles centred on the X-ray peak in  \figiac{fig:profili_mediani_chexmate} together with the $68\%$ dispersion. The discussion of the difference between the relaxed and disturbed sub-samples is detailed in Section \ref{sec:the_median_chex_profiles}.

\section{Cosmological simulations data}\label{sec:cosmo_analysis}
The main scientific goal of this paper is to investigate the origin of the diversity of the EM profiles. The main source of the scatter between the profiles is expected to be due to a genuine different spatial distribution of the ICM related to the individual formation history of the cluster. The other sources that impact the observed scatter are related to how we observe clusters. There are systematic errors associated with X-ray analysis and observing clusters in projection. This latter point is of crucial importance when computing the scatter within a cluster sample. For instance, a system formed by two merging halos of similar mass will appear as a merging system if the projection is perpendicular to the merging axis but will otherwise appear regular if the projection is parallel. In this work, we employed cosmological simulations from the \threehun\ collaboration \citep{cui18} to evaluate this effect. 
\begin{figure}[!ht]
\begin{center}
\resizebox{1\columnwidth}{!}{
\includegraphics[]{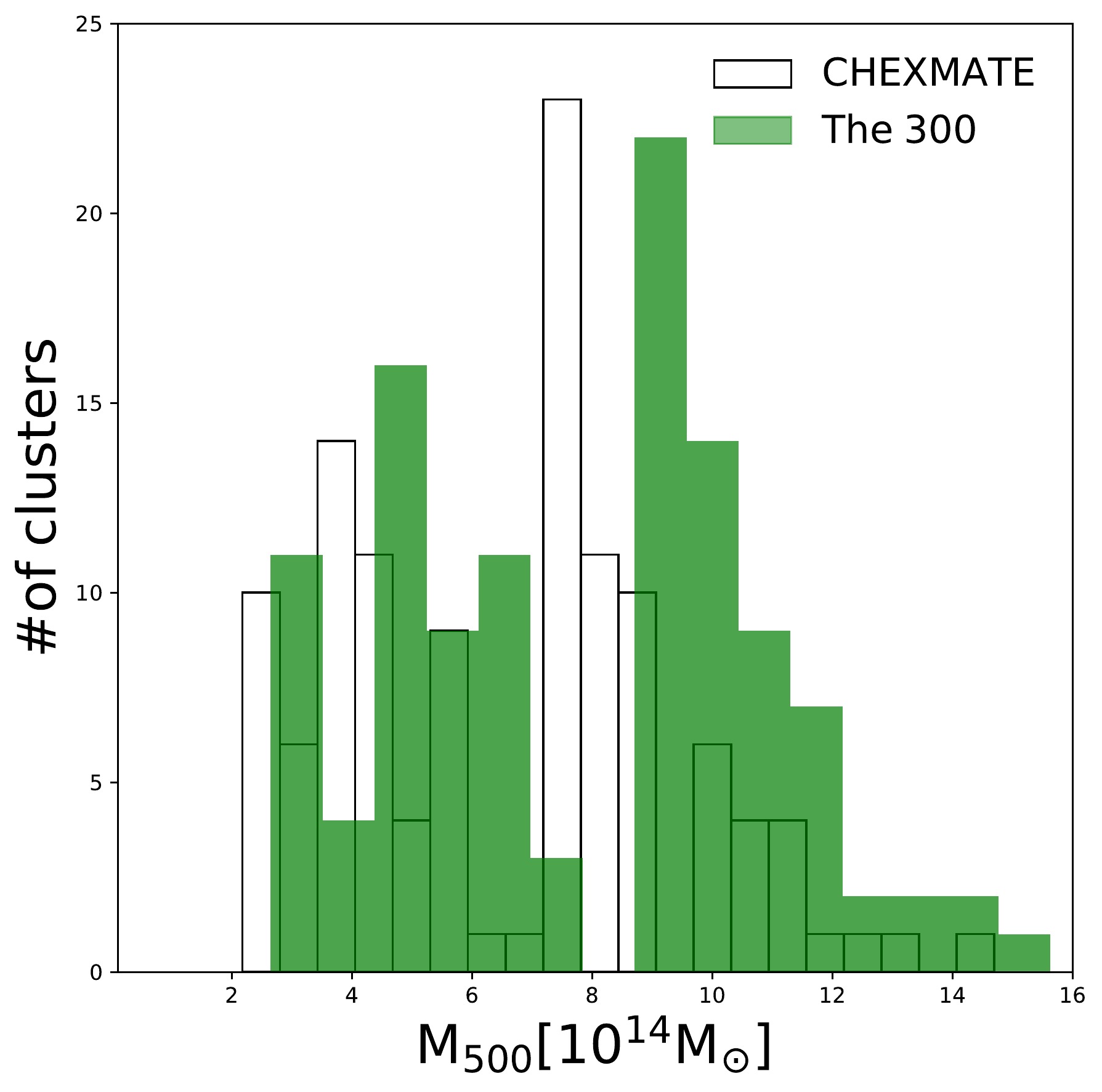} 
}
\end{center}
\caption{\footnotesize{Mass distribution of the \chxmt\ and \threehun\ samples. These are shown with black empty and green polygons, respectively. The gap between $6-7 \times 10^{14} \Mvysz$ is an artefact from the \chxmt\ sample being divided into two tiers, shown in \figiac{fig:planck_mz_plane}. The shift between the two distributions is due to the fact that the \chxmt\ masses are assumed to be 0.8 lower than the true mass due to the hydrostatic bias. For more details, refer to Section \ref{sec:cosmo_analysis}.}}
\label{fig:simulations_prop}
\end{figure}

Specifically, we study the GADGET-X version of \threehun\ suite. This is composed of re-simulations of the 324 most massive clusters identified at $z=0$ within the dark matter-only MULTIDARK simulation \citet{klypin2016}, and thus it constitutes an ideal sample of massive clusters from which to extract a CHEX-MATE simulated counterpart. The cosmology assumed in the MULTIDARK simulation is that of the Planck collaboration \citet{planck2016} and is similar to what is assumed in this paper. The adopted baryon physics include metal-dependent radiative gas cooling, star formation, stellar feedback, supermassive black hole growth, and active galactic nuclei (AGN) feedback \citep{rasia2015}.
To cover the observational redshift range, the simulated sample  was extracted from six different snapshots corresponding to $z=0.067,0.141,0.222,0.333,0.456$, and $0.592$. For each observed object, in addition to the redshift, we matched the cluster mass $M_{500}$ imposed to be close to $M_{500}^{\mathrm Y_{SZ}}/0.8$. With this condition, we followed the indication of the Planck collaboration (see \citealt{planck2014}) that assumed a baseline mass bias of 20\% ($1-b=0.8$). We also checked whether the selected simulated clusters have a strikingly inconsistent   morphological appearance, such as a double cluster associated to a relaxed system. In such cases, we considered the second closest mass object. In the final sample, the standard deviation of the $M_{500,\mathrm{sim}}/(\Mvysz / 0.8)$ is equal to $0.037$. Due to the distribution of the CHEX-MATE sample in the mass-redshift place, we allowed a few Tier 2 clusters to be matched to the same simulated clusters taken from different cosmic times. Even with this stratagem, which will not impact the results of this investigation, we observed that a very massive cluster at $z=0.4$ remained unmatched. The final simulated sample thus includes 115 objects.

The simulation sample mass distribution is shown in \figiac{fig:simulations_prop}. For each simulated cluster, we generated 40 EM maps centred on the cluster {\it total} density peak and integrating the emission along different lines of sight for a distance equal to $6\Rv$ using the \textit{Smac} code (\citealt{dolag2005,ansarifard20}). Henceforth, we refer to these maps as "sim EM" and they are in units of [Mpc cm$^{-6}$].
\begin{figure*}[!ht]
\begin{center}
\resizebox{0.8\textwidth}{!}{
\includegraphics[]{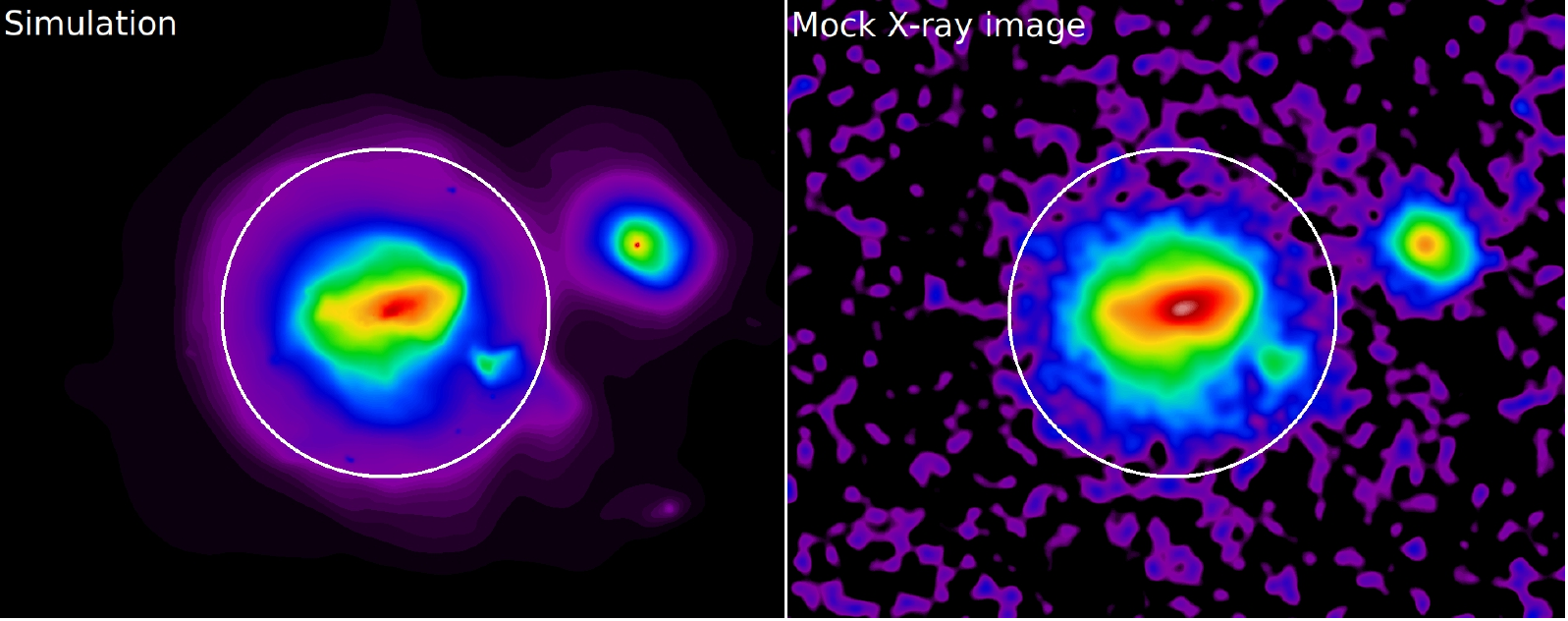}
}
\end{center}
\caption{\footnotesize{Example of the creation of the X-ray mock images. Left panel: EM map of a simulated cluster of our sample. The white circle encompasses R$_{500}$. Right panel: Mock X-ray background-subtracted image in the [0.5-2] keV band of the same object shown in the left panel after we applied the procedures simulating typical X-ray observation effects. These are  described in detail in Section \ref{sec:xlike_images}.  }}
\label{fig:mock_image_example}
\end{figure*}

\subsection{X-ray mock images of simulated clusters}\label{sec:xlike_images}
 We produced mock X-ray observations by applying observational effects to the \threehun\ maps. Firstly, we transformed the EM in surface brightness maps by inverting Equation \ref{eq:monique_em_formula}. The emissivity factor $\epsilon$(T,z) was computed using the same procedure as in Section \ref{sec:emission_profiles}. The absorption was fixed to the average value of the \chxmt\ sample, N$_\mathrm{H} = 2 \times 10^{20}$ cm$^{-2}$, and the average temperature was computed by using the $M_{500}$ of the cluster and applying Equation \ref{eq:t200}. The instrumental effects were accounted for by folding in the \textit{pn} instrumental response files computed at the aimpoint.
 We produced the count rate maps by multiplying the surface brightness maps by the median exposure time of the \chxmt\ programme, $4 \times 10^4$ s, and by the size of the pixel in arcminutes$^2$.
 We added to these maps a spatially non-uniform sky background whose count rate is $ \langle cr_{sky} \rangle = 5.165 \times 10^{-3}$ [ct/s/arcmin$^2$], as measured by \textit{pn} in the $[0.5-2]$ keV band. 
 We then included the \xmm\ vignetting as derived from the calibration files, and we simulated the PSF effect by convolving the map with a Gaussian function with a width of ten arcsec. Finally,  we drew a Poisson realisation of the expected counts in each pixel and produced a mock X-ray observation.  %
We divided the field of view into square tiles with sides of 2.6 arcmin within which we introduced $3\%$ variations to the mean sky background count rate to mimic the mean variations of the sky on the field of view of \xmm. We multiplied these maps by 1.07 and 0.93 to create over- and underestimated background maps, respectively, which account for the systematic error related to the background estimation. We randomly chose the over- or underestimated map and subtracted it from the mock X-ray observation. 
After the subtraction, we corrected for the vignetting by using a function obtained through the fit of the calibration values to those we randomly added a $1 \pm 0.05$ factor to mimic our imprecision in the calibration of the response as a function of the off-axis angle.

The typical effects introduced by the procedures described above are shown in \figiac{fig:mock_image_example}. The EM map produced using the simulation data is shown in the left panel where there is a large sub-structure in the west sector and a small one in the south-west sector within R$_{500}$. The right panel shows the mock X-ray image where the degradation effects are evident. The spatial features within the central regions were lost due to the PSF. Despite the resolution loss, the ellipsoidal spatial distribution of the ICM is clearly visible, and the presence of features such as the small sub-structure in the south-west are still visible. The emission outside R$_{500}$ is dominated by the background, and the small filament emission in the south-west was too faint to remain visible. The large sub-structure is still evident, but the bridge connecting it to the main halo has become muddled into the background. 

\subsection{Simulation emission measure profiles}\label{sec:em_sim_profile_production}

\begin{figure}[!ht]
\resizebox{1\columnwidth}{!}{
\includegraphics[]{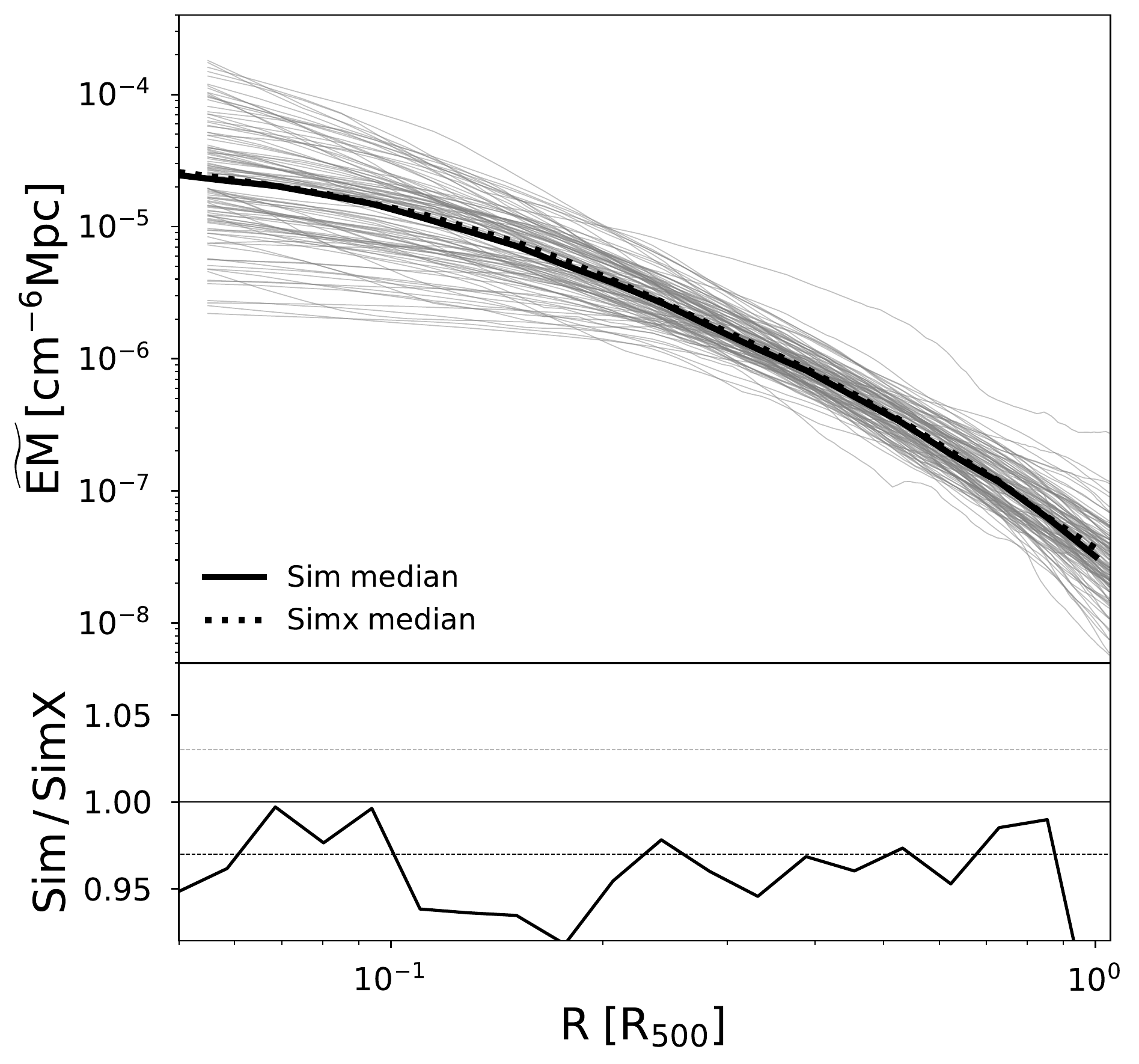} 
}
\caption{\footnotesize{Comparison between the Sim and Simx profiles.
Top panel: Comparison between the medians of the Sim and Simx \EMS\ profiles extracted from random projections. These are shown with black solid and dotted lines, respectively. The grey solid lines represent the Sim profiles. Bottom panel: Ratio between the Sim over the Simx median. The black solid and dotted lines correspond to the identity and the $\pm$ 2\% levels, respectively.}}
\label{fig:sim_vs_simx}
\end{figure}
We extracted the EM profiles from the \threehun\ maps by computing the median EM of all the pixels within concentric annuli, the bin width being $2$ arcsec. These annuli are centred on the  map centre (i.e. the peak of the halo total density).
We obtained the \EMS\ profiles by applying this process to our sample of 115 simulated clusters and for each of the 40 projections, and we scaled them according to Equation \ref{eq:scale_formula}.
From hereon, we refer to these profiles as the "Sim" profiles. Similarly, we extracted the X-ray mock profiles, henceforth the "Simx" profiles, from the synthetic X-ray maps. These are shown along one randomly selected projection with a grey solid line in \figiac{fig:sim_vs_simx}.
We show the emission measure profile projected along only one line of sight because the results along the other projections are similar. 

The comparison of the sample medians of the Sim and Simx profiles is shown in \figiac{fig:sim_vs_simx}. The two sample medians are in excellent agreement, up to $\sim 0.7\Rv$. Beyond that radius, the Simx median is flatter. This is an effect of the PSF, which redistributes on larger scales the contribution of sub-halos and local inhomogeneities. However, the fact that the medians are similar after the application of the X-ray effects is likely due to the combination of the good statistics of the \chxmt\ programme and the procedure used to derive each cluster EM profile, which considers the medians of all pixels. 
%
%
The former ensures that the extraction of profiles is not affected by large statistical scatter, at least up to $\Rv$, and the latter tends to hamper effects related to the presence of sub-structures. 

There are key differences between the analysis of the Simx and \chxmt\ profiles despite our underlying strategy of applying the same procedures. 
For example, the centre used in the simulations introduces a third option with respect to the X-ray centroid and peak. Furthermore, in simulated clusters, we computed the azimuthal median on pixels instead of on the Voronoi cells. We expected that the centre offsets would affect the profiles at small scales,  $R< 0.1 $R$_{500}$, as shown in the left panel of \figiac{fig:cen_vs_peak}. Finally, the X-ray analysis masks the emission associated to sub-halos, while this is not possible in simulations, as the development of an automated procedure to detect extended sources in the large number of images of our simulations, $4600=115 \times 40$, was beyond the scope of this paper. The impact of this difference on the scatter is discussed in Appendix \ref{sec:appendix_subhalo}.
\section{The profile shape}\label{sec:profile_shape}

In this section, we study the shape of the emission measure profiles by checking the impact of the centre definition (as in Sect. \ref{sec:sx_profiles}) and of the radial profile procedure (as described in Sect. \ref{sec:radial_analysis}). Subsequently, we compare the sample median profiles of the relaxed and disturbed sub-samples and compare the CHEX-MATE median profile with the literature and the \threehun\ simulations. 

\subsection{The impact of the profile centre}\label{sec:cen_vs_peak}
\begin{figure*}[!ht]
\begin{center}

\resizebox{0.8\textwidth}{!}{
\includegraphics[]{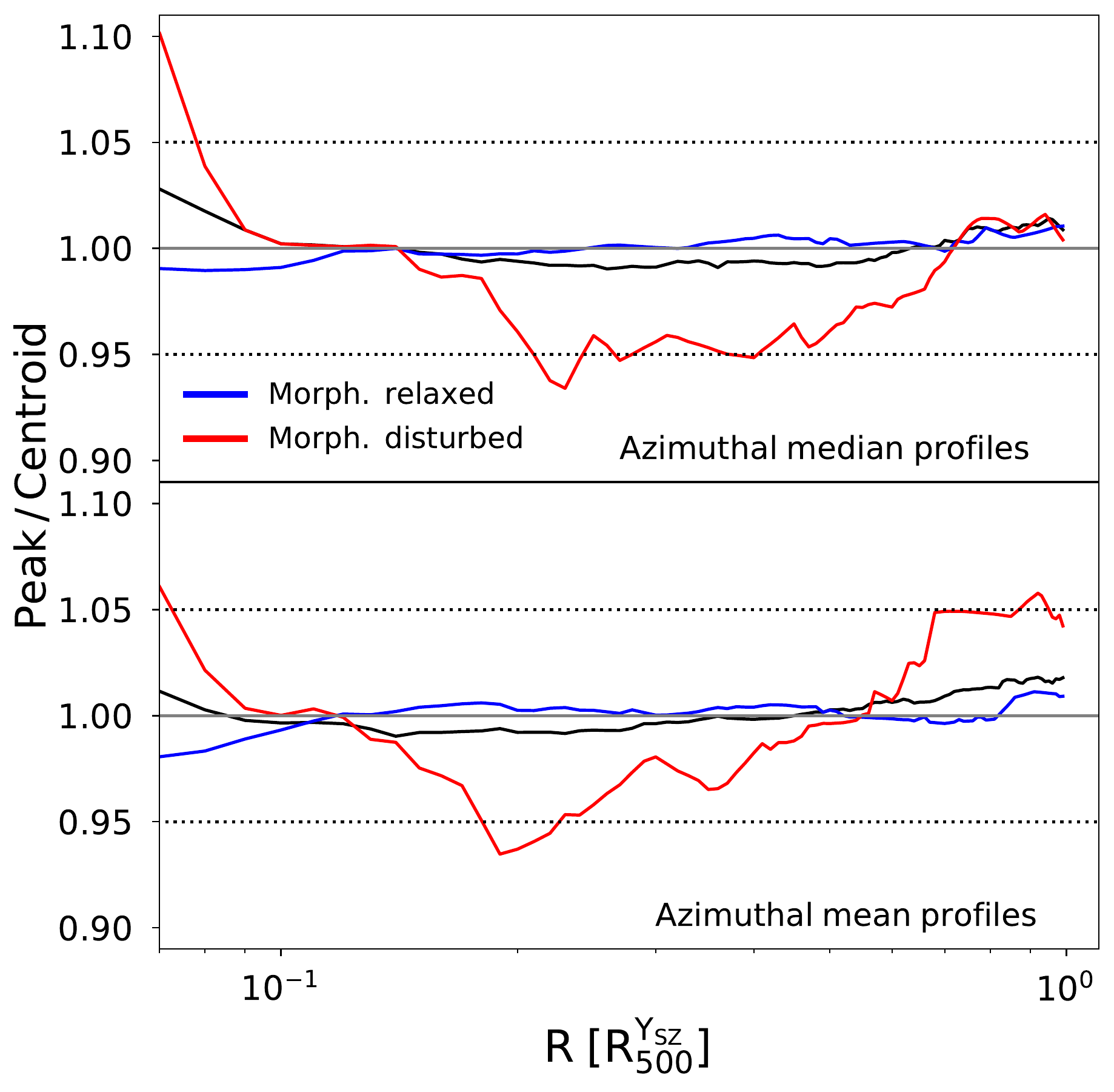}
\includegraphics[]{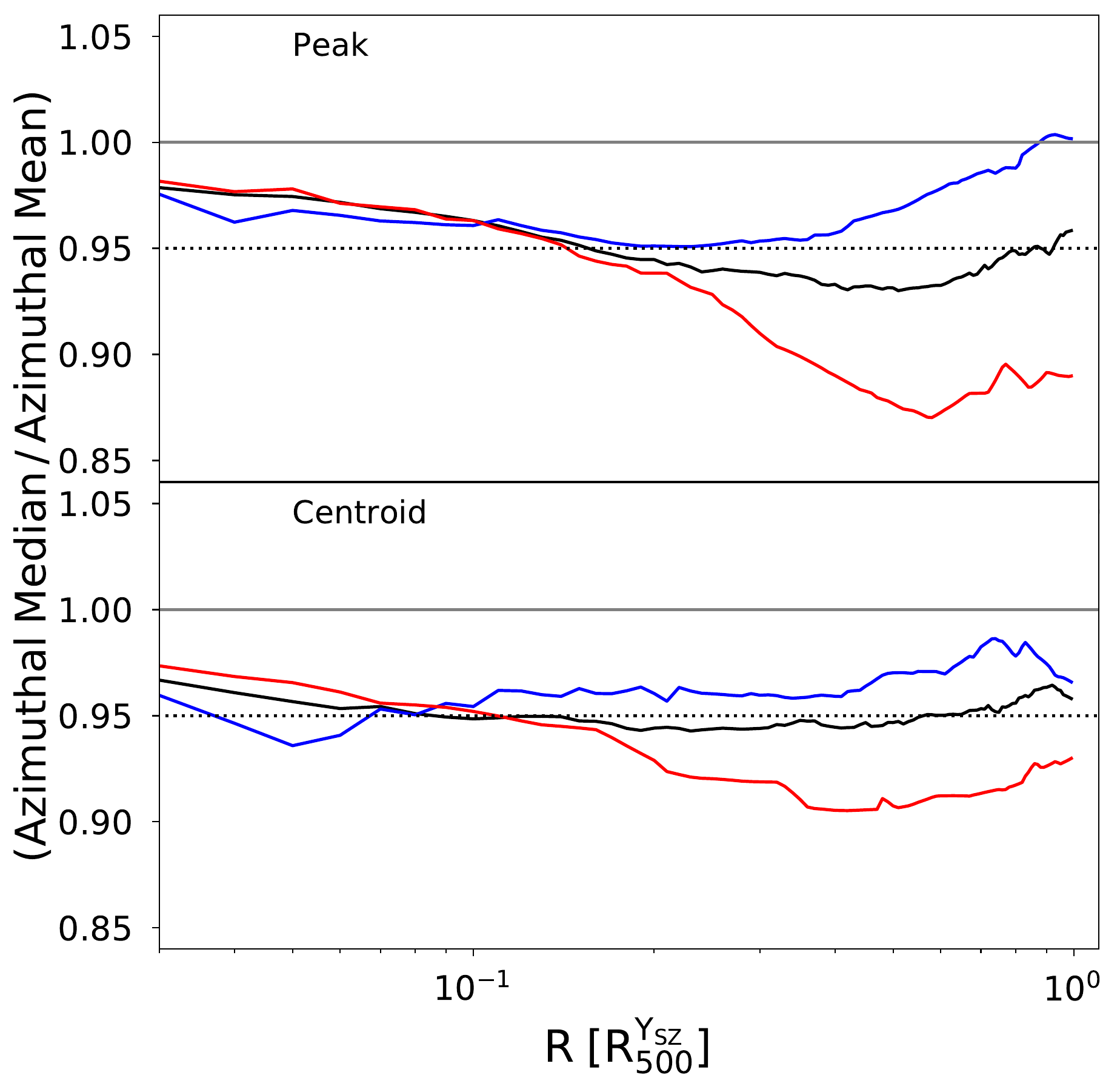}
}
\end{center}
\caption{\footnotesize{ Ratio between the medians of the EM profiles obtained using the X-ray peak or the centroid as centre and using the azimuthal average or median. Left panels: Ratio between the medians of the profiles centred on the peak and centroid. The top and bottom panels show the ratio computed using the azimuthal median and azimuthal mean \EMS\ profiles, respectively. The black solid lines represent the median of the ratio considering the whole sample. The blue and red solid lines show the ratio considering only the morphologically relaxed and disturbed clusters, respectively. The grey solid lines indicate the identity line, and the dotted lines represent the plus and minus 5\% levels. Right panels: Same as the left panels except we show the ratio between the medians of the azimuthal median and mean \EMS\ profiles. The top and bottom panels show the ratio computed using the profiles extracted with the X-ray peak and the centroid  as centre, respectively.
The legend of the solid and dotted lines is the same as in the left panels except for the fact that we show only the minus 5\% level.}}
\label{fig:cen_vs_peak}
\end{figure*}
The impact of the choice of the centre for the profile extraction is crucial for any study that builds on the shape of profiles, such as the determination of the hydrostatic mass profile (see \citealt{pratt19} for a recent review). The heterogeneity of morphology and the exquisite data quality of the \chxmt\ sample offer a unique opportunity to assess how the choice of the centre affects the overall shape of the profile.

We show in the top part of the left panel in \figiac{fig:cen_vs_peak} the ratio between the medians of the \EMS\ azimuthal {\it median} profiles centred on the peak and those centred on the centroid. The colours of the lines respectively refer to the entire sample and the morphologically relaxed and disturbed sub-samples. 
 The bottom panel shows a similar ratio where the azimuthal {\it mean} profiles are considered. From the figure, we noticed that the results obtained using mean or median profiles are similar, with the exception of the outskirts of the disturbed systems, which will be discussed below. 
On average, the relaxed sub-sample shows little deviation from one at all radial scales, as would be expected since for these systems the X-ray peak likely coincides with the centroid. 
The variations of the disturbed objects are up to $10\%$ in the centre, where the profiles centred on the X-ray peak are steeper, and about $5\%$ in the [0.15-0.5]$\Rvysz$ region, where the centroid profiles have greater emission. 
These variations are not reflected in the entire \chxmt\ sample, despite the fact that it includes approximately $87\%$ of the disturbed and morphologically mixed systems. Indeed, in this case, all deviations are within 2\%, implying that the choice of referring to the X-ray peak does not influence the shape of the sample median profile.

\subsection{Mean versus median}
We proceeded by testing the radial profile procedure (Sect. \ref{sec:sx_profiles}) next, comparing
the azimuthal median and the azimuthal mean profiles (\figiac{fig:cen_vs_peak}) and centring both on either the X-ray peak (top panel) or, for completeness, on the centroid (bottom panel). As expected from the previous results, there are little differences between the two panels. Overall, we noticed that the azimuthal mean profiles are greater than the azimuthal medians, implying that greater density fluctuations are present at all scales and that they play a larger role in the outskirts where a larger number of undetected clumps might be present. The differences between the two profiles are always within 5\% for the relaxed systems. The deviations are more important for the disturbed objects, especially when centred on the X-ray peak. This last remark implies that the regions outside $\sim 0.4 \Rvysz$ of the \chxmt\ disturbed objects not only have greater density fluctuations but are also spherically asymmetric in their gas distribution; otherwise, the same mean-median deviations would be detected when considering the centroid as centre. The global effect on the \chxmt\ sample is that the median profiles are about 7\% lower than the mean profiles at $R > 0.3-0.4 \Rvysz$. 
We noticed that similar results were obtained by \citet{eckert15} (cfr. Fig.6).
This test confirmed that our choice of using the azimuthal medians for each cluster profile is more robust for our goal of describing the overall \chxmt\ radial profiles.




\subsection{The median \chxmt\ profiles and the comparison between relaxed and disturbed systems}\label{sec:the_median_chex_profiles}
\begin{figure*}[!ht]
\begin{center}
\resizebox{0.8\textwidth}{!}{
\includegraphics[]{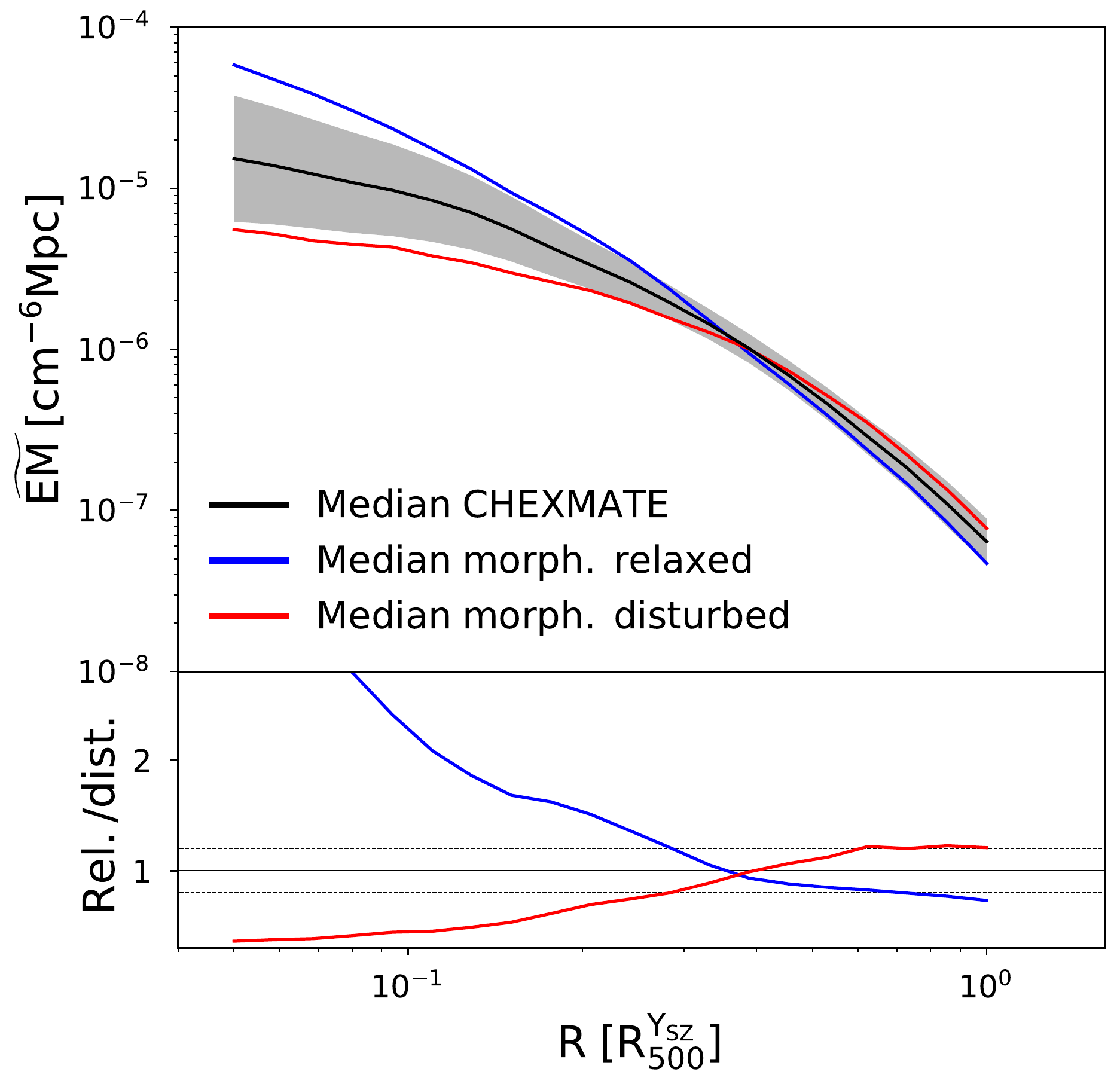}
\includegraphics[]{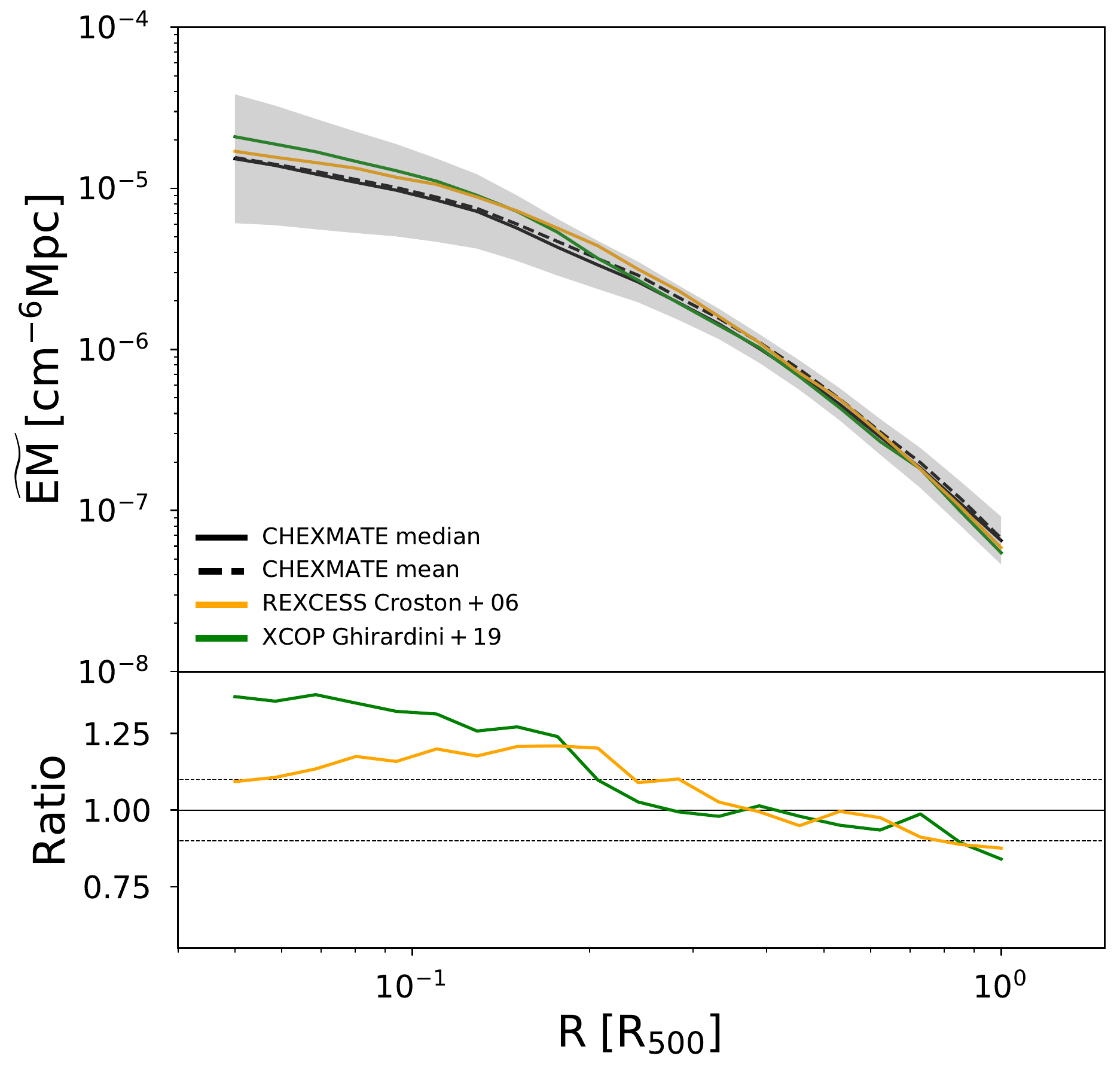}
}
\end{center}
\caption{\footnotesize{ Comparison of the statistical properties of the \chxmt\ \EMS\ profiles with morphologically selected sub-samples and the X-COP and \rexcess\ samples.
Top-left panel: Median of the \EMS\ peak median profiles of the \chxmt\ sample. Its dispersion is shown with the black solid line and the grey envelope. The blue and red solid lines represent the median of the profiles derived from the morphologically relaxed and disturbed objects, respectively. Bottom-left panel: Ratio of the median of the morphologically relaxed and disturbed \EMS\ profiles over the median of the full \chxmt\ sample. The same colour coding as above is used. The dotted lines represent 0.8 and 1.2 values and are shown for reference.
Top-right panel: Same as the left panel except the  median of the azimuthal mean profiles is shown. The median is indicated with the dotted black line. The median of the X-COP \citep{ghirardini19} and REXCESS \citep{croston2008} \EMS\ profiles are shown with green and orange solid lines, respectively. Bottom-left panel: Ratio between the median of X-COP and REXCESS samples to the \chxmt\ median. The REXCESS profiles were extracted performing an azimuthal average. For this reason, we show the ratio between the median of the REXCESS profiles and the median of the \chxmt\ azimuthal mean profiles.}}
\label{fig:em_medians_and_xcop}
\end{figure*} 

In \figiac{fig:em_medians_mass_red_dep}, we show the behaviour of the \chxmt\ \EMS\ median as well as the medians of the mass and redshift sub-samples. In the left panel of  \figiac{fig:em_medians_and_xcop}, we compare the medians of the relaxed and disturbed sub-samples whose individual profiles are shown in \figiac{fig:profili_mediani_chexmate}.
The former is approximately two times greater than the median of the whole sample at $R\sim0.1 \Rvysz$ 
and is not within the dispersion. 
The morphologically disturbed clusters are on average within the dispersion, being $70\%$ smaller than the whole sample median at $R<0.2 \Rvysz$.
The morphologically relaxed profiles become steeper than the disturbed profiles at $R>0.4\Rvysz$. A similar behaviour has been observed in several works,  such as \citet{arnaud2010}, \citet{pra10}, \citet{maughan2012}, and \citet{eckert12} (cfr. Figure 4), when comparing cool core systems with non-cool core systems.

Combining these results with those of \figiac{fig:em_medians_mass_red_dep}, we concluded that \chxmt\ Eq. \ref{eq:scale_formula} provides reasonable mass normalisation and captures the evolution of the cluster population well. The large sample dispersion seen in the cluster cores is linked to the variety of morphologies present in the sample. The medians of the relaxed and disturbed sub-samples indeed differ by more than a factor of ten at R$<0.1\Rvysz$. At around $\Rvysz$, we also noticed some different behaviours in our sub-sample: The most massive objects are approximately 25\% larger than the least massive ones, and the morphologically disturbed clusters are 50\% larger than the relaxed ones (see also \citealt{sayers2022}).

\subsection{Comparison with other samples}\label{sec:comparison_with_other_samples}

In this section, the statistical properties of the \chxmt\ profiles are compared to SZ and X-ray selected samples at $z\lesssim 0.3$ with similar mass ranges in order to investigate the impact of different selection effects. The SZ-selected sample is the \xmm\ Cluster Outskirts Project (X-COP; \citealt{eckert17}) sample that contains 12 SZ-selected clusters in the [0.05-0.1] redshift range and has  a total mass range similar to \chxmt \ but with a greater median mass ($\sim 6\times 10^{14}$M$_{\odot}$). The individual  \EMS\ profiles for X-COP were computed using the same procedure as described in  this work. The profiles were scaled by applying Equation \ref{eq:scale_formula}, with $T_{\mathrm{avg}}$ given by Equation \ref{eq:t200}, using the masses presented in Table~1 of \citet{ettori2019}.

We also compare the \chxmt\ profile properties to 
the X-ray selected Representative \xmm\ Cluster Structure Survey (REXCESS; \citealt{bohringer2007}) sample, which is composed of 31 X-ray selected clusters in the [0.05-0.3] redshift range, with a mass range spanning the [1-8]$\times 10^{14} $M$_{\odot}$ and the median mass being $2.7 \times 10^{14}\, $M$_{\odot}$. The \rexcess\ \EMS\ profiles were obtained from the surface brightness presented in Appendix A of \citep{croston2008}. These profiles were  computed using the azimuthal average in each annulus. For this reason, we compare the \rexcess\ profiles with the mean \chxmt\ profiles. The REXCESS profiles were scaled using Equation \ref{eq:scale_formula} with $T_{\mathrm{avg}}$ from \citet{pratt2009}.


The median and its dispersion for each of these samples were computed using the procedure described above, and their comparison with \chxmt\ is shown in the right panel of \figiac{fig:em_medians_and_xcop}. Both sample medians present an overall good agreement that is within $10\%$ at $R > 0.2 \Rvysz$. The X-COP median is $25\%$ more peaked in the central regions at $R<0.2 \Rvysz$ with respect to both \chxmt\ and REXCESS. 
Nevertheless, the X-COP median is well within the dispersion of the \chxmt\ sample and variations of such order are expected in the core where the \EMS\ values are comprised in the wide range $\sim[6 ,30]  \times 10^{-5} $ cm $^{-6}$ Mpc.
\begin{figure}[!ht]
\begin{center}
\resizebox{1\columnwidth}{!}{
\includegraphics[]{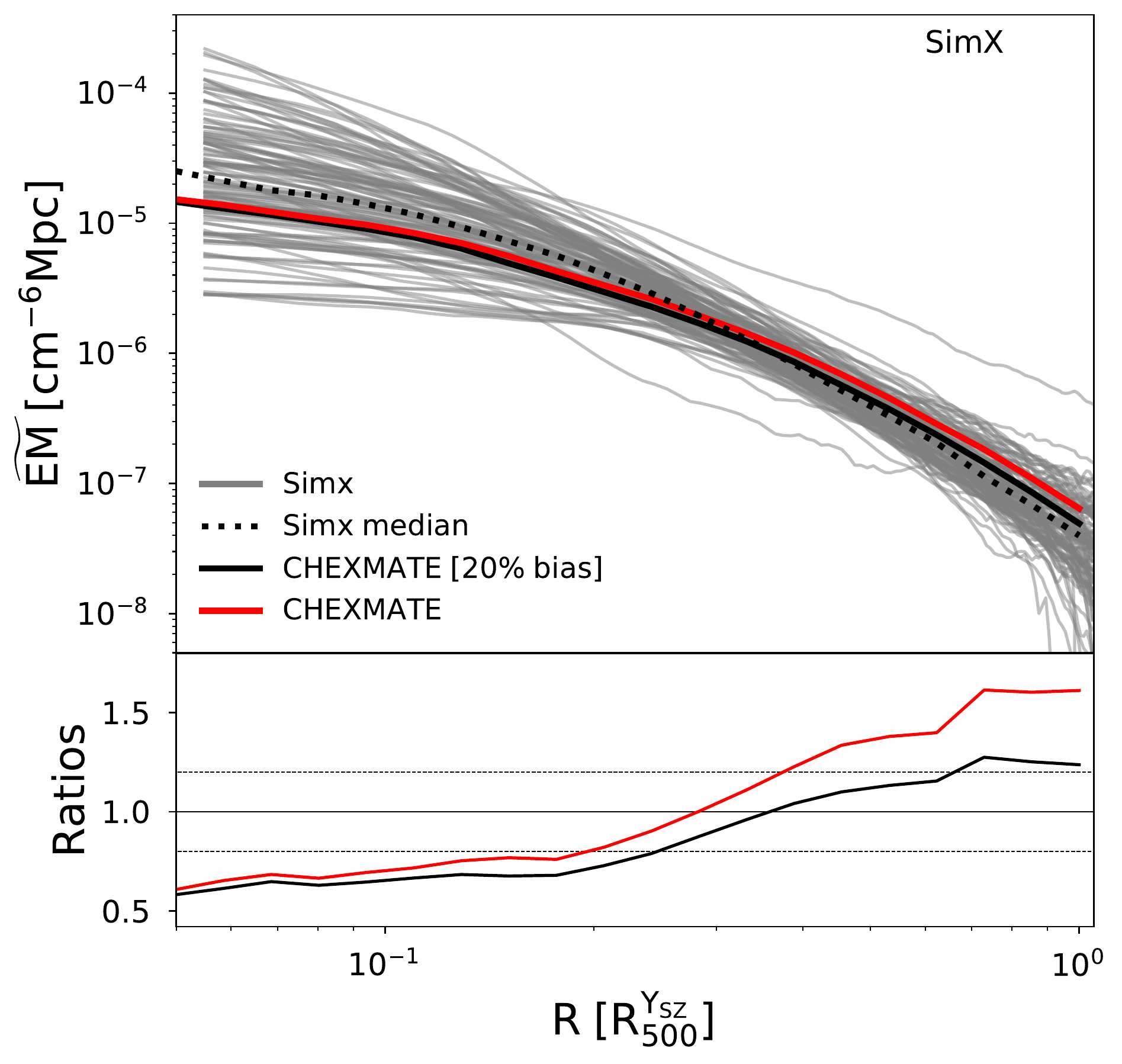} 
}
\end{center}
\caption{\footnotesize{Top panel: Simx \EMS\ profiles and their median. The profiles are shown with grey solid lines and their median with a black dashed line. The Simx profiles were extracted from a randomly chosen line of sight for each cluster. The red solid line identifies the median of the \chxmt\ sample. The black solid line is the median of the \chxmt\ sample assuming a 20\% bias on the mass (i.e. each profile has been scaled  for $\Rvysz/0.8^{1/3}$; see Sect. \ref{sec:cosmo_analysis} for details).   Bottom panel: Ratio of the median of the \chxmt\ with and without correction for hydrostatic bias and the median of the Simx simulations. The CHEX-MATE median with the correction is shown with the red line and the median without it is shown with the black line.}}
\label{fig:simulation_vs_chexprofiles}
\end{figure}
\begin{figure*}[!ht]
\begin{center}
\resizebox{0.9\textwidth}{!}{
\includegraphics[]{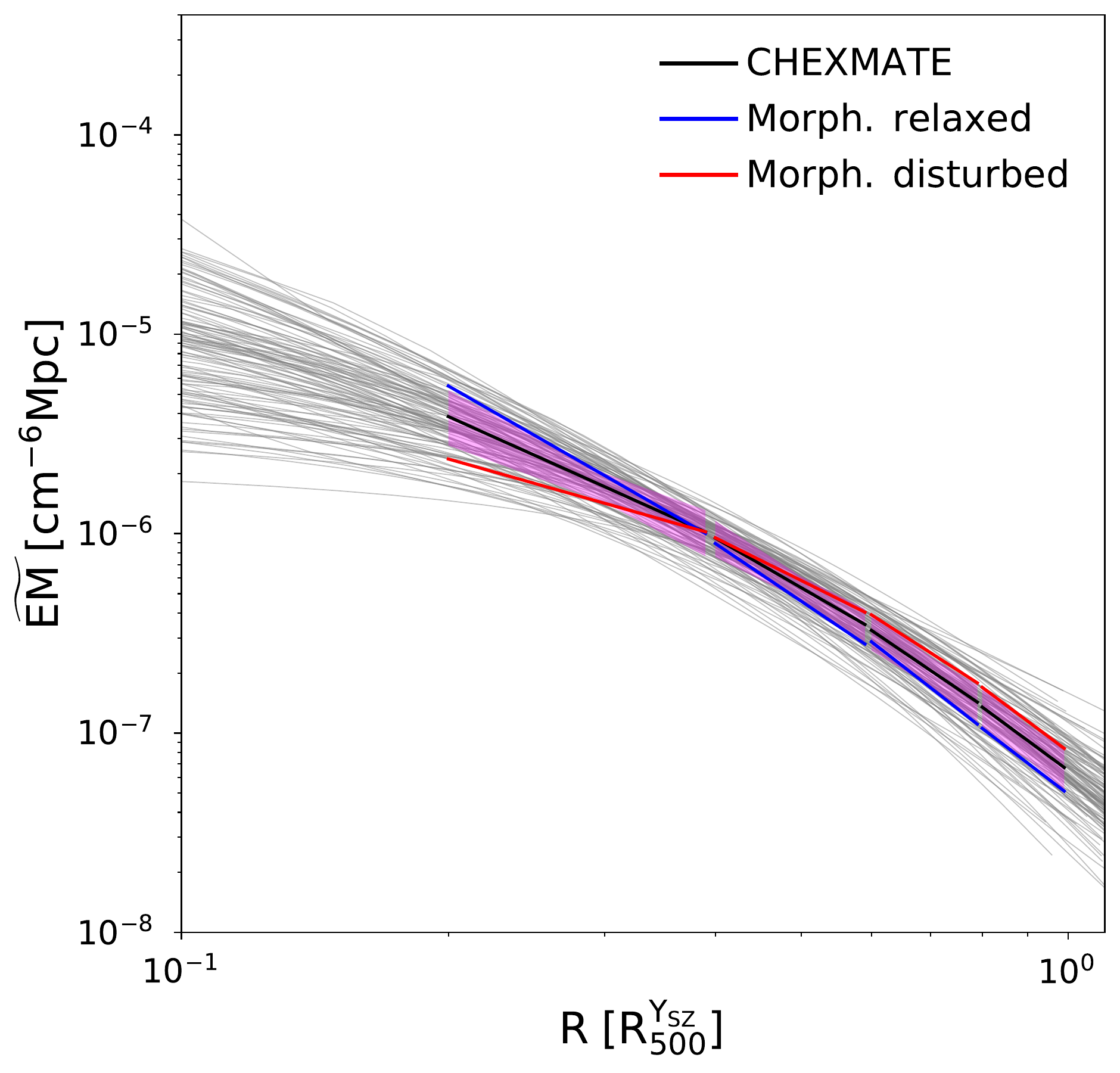} 
\includegraphics[]{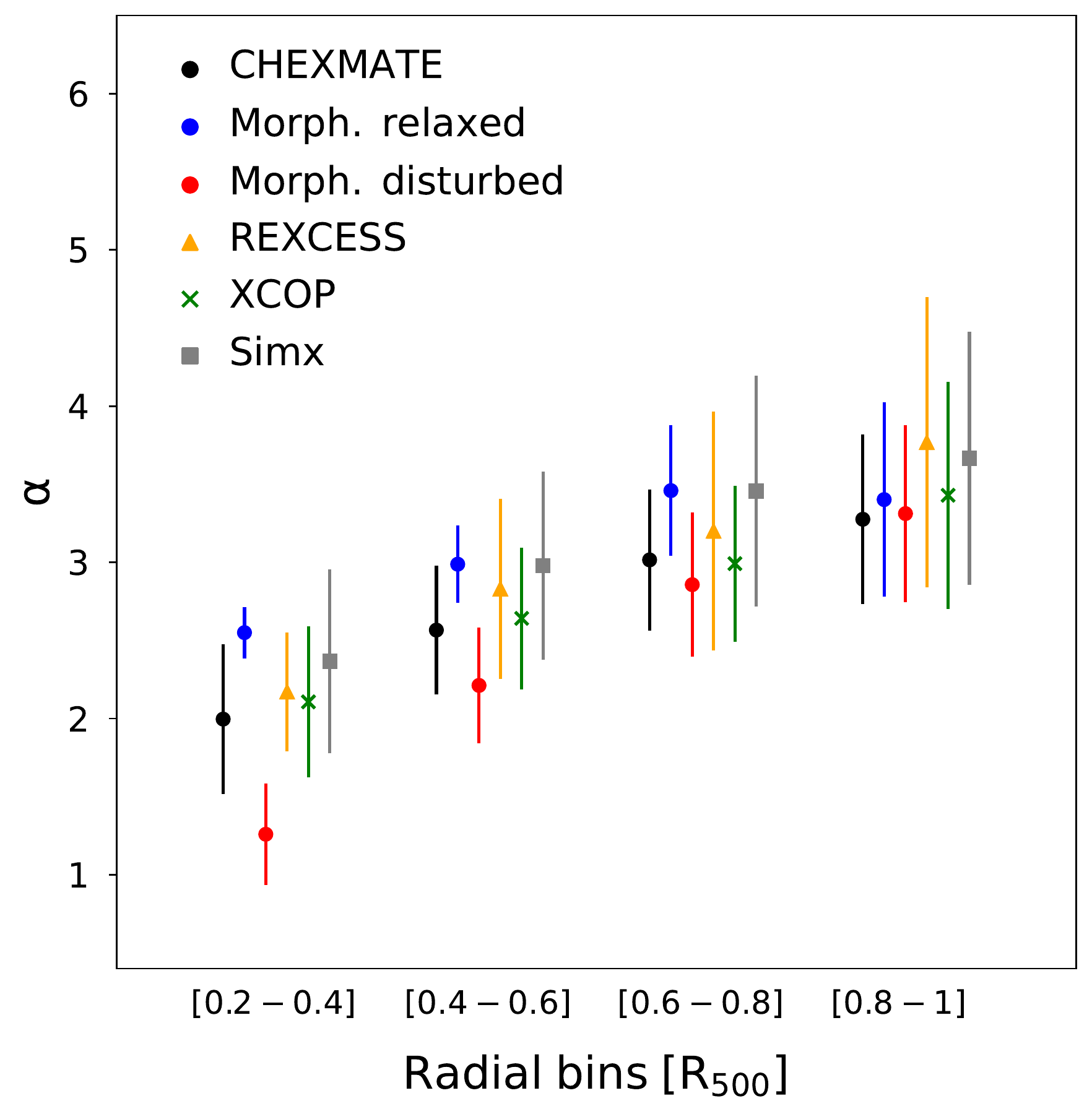} 
}
\end{center}
\caption{\footnotesize{Results of the fit of the \chxmt\ \EMS\ profiles using broken power laws. Left panel: Fit of the median \chxmt\ \EMS\ profiles shown with grey lines with a power law in four radial bins. The bins are $[0.2-0.4], [0.4-0.6], [0.6-0.8]$, and $[0.8-1]$ and are in units of $\Rvysz$. For each radial bin, the black solid line represents the best fit of the power law shown in Eq. \ref{eq:pow}. The magenta envelope was obtained considering the dispersion of the fitted parameters (A and $\alpha$ in Eq. \ref{eq:pow}). The blue and red solid lines represent the best fit of the profiles of the morphologically relaxed and disturbed clusters, respectively. Right panel: Median values of the power law indexes, $\alpha$, obtained from the fit of the \chxmt, X-COP, REXCESS, and Simx samples in the four radial bins shown in the left panel. For each value we report its dispersion.}}
\label{fig:profile_slopes}
\end{figure*}
 
\subsection{Comparison with simulations}\label{sec:mediansim_vs_chexsim}
The 115 Simx \EMS\ profiles extracted from random projections for each cluster are shown together with their median value in  \figiac{fig:simulation_vs_chexprofiles}. The median of the \chxmt\ sample is also shown. Overall, the CHEX-MATE median is flatter than the medians of Simx in the  [0.06-1]$\Rvysz$ radial range, and specifically it is $\sim 50\%$ smaller in the centre and $\sim 50\%$ larger in the outskirts. 
Part of the difference in the external regions might be caused by the re-scaling of the observational sample. Indeed, each \chxmt\ profile has been scaled using the $\Rvysz$ derived from $\Mvysz$, which is expected to be biased low by 20\%. 
Factoring in this aspect, a more proper re-scaling should be done with respect to 
$\Rvysz/(0.8^{1/3})$. The agreement between the \chxmt\ data and the simulations increases at $R>0.5 \Rvysz$, with relative variations of about 40\% in the [0.2-1] $\Rvysz$ radial range. 
These considerations do not have any repercussion on the central regions, which remain larger in the simulated profiles, confirming the results found in \cite{campitiello2022} and \cite{darragh2023}. 

Providing precise measures of the mass of the observed clusters is one of the goals of the \chxmt\ collaboration. For this paper, it is sufficient to prove that \threehun\ clusters have a profile in reasonable agreement with the observed sample in order to employ them for the study of the \EMS\ scatter. 

\subsection{Measuring the slopes}\label{sec:slopes}
We measured the slopes of the \chxmt\ \EMS\ profiles adopting the technique described in Section 3.1 of \citet{ghirardini19}. Briefly, we considered four radial bins in the [0.2-1]$\times \Rvysz$ radial range and with widths equal to $0.2\Rvysz$. We excluded the innermost bin [0.-0.2]$\Rvysz$ because of the very high dispersion of the profiles within this region. 
We measured the slope, $\alpha$, and normalisation, $A$, of each profile by performing the fit within each radial bin using the following expression:
\begin{equation}\label{eq:pow}
     Q(x) = A x^{\alpha} e^{\pm \sigma_{int}},
\end{equation}
where $x=R/\Rvysz$ and $\sigma_{int}$ is the intrinsic scatter. 
The error on each parameter was estimated via a Monte Carlo procedure, producing 100 realisations of each profile. The left panel of \figiac{fig:profile_slopes} shows the power law  computed using the median of the $\alpha$ and A within each radial bin. 

The fit of the [0.2-0.4]$\Rvysz$ bin revealed that there is a striking difference between the morphologically relaxed and disturbed objects. This result is notable because the considered region is far from the cooling region at $\sim 0.1-0.15 \Rvysz$.
The median power law index of the morphologically relaxed object profiles is $\alpha_{rel} = 2.57 \pm 0.15$ and is not consistent with the morphologically disturbed one, which is $\alpha_{dis} = 1.37 \pm 0.2$ at more than the $3 \sigma$ level. That is, the shape of the most disturbed and relaxed objects differ at least up to $0.4\Rvysz$. However, the fitted power law is within the dispersion of the full sample, whose median index is $\alpha = 2.02 \pm 0.36$.
The median values in the $[0.4-0.6] \Rvysz$ region are $\alpha_{rel}=3 \pm 0.22$ and $\alpha_{dis} = 2.3 \pm 0.2$ for the morphologically relaxed and disturbed objects, respectively. The indexes are consistent at the $2\sigma$ level, implying that the profiles are still affected by the morphology in the centre. 
The overall scenario changes at $R > 0.6 \Rvysz$. The power law index of the morphologically relaxed and disturbed objects are consistent with the median obtained from fitting the whole sample.
\citet{ettori09} found that the average slope of a sample of 11 clusters at $0.4R_{200} (\sim 0.6 \Rvysz$) and $0.7R_{200} (\sim \Rvysz$)  is $3.15 \pm 0.46$ and  $3.86 \pm 0.7$, respectively. These values are consistent within $1\sigma$ with our measurements.

We show in the right panel of \figiac{fig:profile_slopes}  the comparison between the median $\alpha$ computed from the \chxmt\ sample in each radial bin with the same quantity obtained using REXCESS and X-COP. There is an excellent agreement in all the considered radial bins. Interestingly, there is also a good agreement in the shape between a sample selected in X-ray (REXCESS) and SZ. One could expect to see more differences in the central parts, as X-ray selection should favour peaked clusters. 

The comparison with the median $\alpha$ obtained using the Simx profiles is also shown in the right panel of \figiac{fig:profile_slopes}. The Simx median $\alpha$ is systematically greater than the median of the observed sample. As discussed in Sect. \ref{sec:mediansim_vs_chexsim}, the bias introduced by using $\Mvysz$ 
might play a role when comparing  \chxmt\ to Simx and partly contributes to this systematic difference. However, we stress the fact that the slopes are consistent within $1 \sigma$ in the four radial bins.

 \citet{campitiello2022} find similar results when comparing the concentration of surface brightness profiles within fixed apertures of simulated and \chxmt\ clusters. This quantity measures how concentrated the cluster core is with respect to the outer regions (i.e. more concentrated clusters show a steeper profile). The concentration of the simulations is systematically higher by approximately $20-30\% $ (cfr. Table 1 of \citealt{campitiello2022}). 

\section{The EM radial profile scatter}\label{sec:profile_scatter}
\subsection{Computation of the scatter}\label{sec:scatter_computation}
Departures from self-similarity are linked to individual formation history as well as non-gravitational processes such as AGN feedback (outflows, jets, cavities, shocks) and feeding (cooling, multi-phase condensation; e.g. \citealt{gaspari20}). The additional terms used to obtain the \EMS\ profiles can partly account for these effects. The scatter of these profiles offers the opportunity to quantify such departures, and the \chxmt\ sample is ideal to achieve this goal since the selection function is simple and well understood. 

We computed the intrinsic scatter of the \chxmt\ radial profiles by applying the following procedure. First, we interpolated each scaled profile on a common grid formed by ten logarithmically spaced radial bins in the $[0.05-1.1] \Rvysz$ radial range. We used the model for which the observed distribution of the points, $S_{obs}$, in each radial bin is the realisation of an underlying normal distribution, $S_{true}$, with log-normal intrinsic scatter, $\sigma_{int}$:
\begin{align}
 & \ln \mathrm{S_{true}} \sim \mathrm{Normal}(\ln\mu, \sigma_{\mathrm{int}}),
\end{align}
 with $\mu$ as the mean value of the distribution. We set broad priors on the parameters we are interested in:
\begin{align}
    &\ln \mu \sim \mathrm{Normal}(\ln \langle \mathrm{\widetilde{EM}(r)} \rangle, \sigma=10), \\
    &\sigma_{int} \sim \mathrm{Half-Cauchy}(\beta=1.0),
\end{align}
where $\langle \mathrm{\widetilde{EM}(r)} \rangle$ is the mean value of the interpolated EM profiles at the radius r. 
We assumed a Half-Cauchy distribution for the scatter, as this quantity is defined as positive. Since $\sigma(\ln X) = \sigma(X)/X$, the intrinsic scatter in linear scale becomes:
\begin{equation}
\sigma_{lin} = \sigma_{int} * \mu, 
\end{equation}
and the total scatter, $\sigma_{tot}$, is the quadratic sum of $\sigma_{lin}$ and the statistical scatter $\sigma_{stat}$:
\begin{equation}
   \sigma_{tot} = \sqrt{\sigma_{lin}^2 + \sigma_{stat}^2}.
 \end{equation}
The observed data were then assumed to be drawn from a normal realisation of the mean value and total scatter:
\begin{equation}\label{eq:likehood}
 \mathrm{S_{obs}} \sim \mathrm{Normal}(\mu, \sigma_{\mathrm{tot}}).
\end{equation}
We determined the intrinsic scatter $\sigma_{int}$ and its $1\sigma$ error by applying the No U-Turn Sampler (NUTS) as implemented in the Python package PyMC3 \citep{pymc3} and  using 1,000 output samples.
\begin{figure*}[!ht]
\begin{center}
\resizebox{0.7\textwidth}{!}{
\includegraphics[]{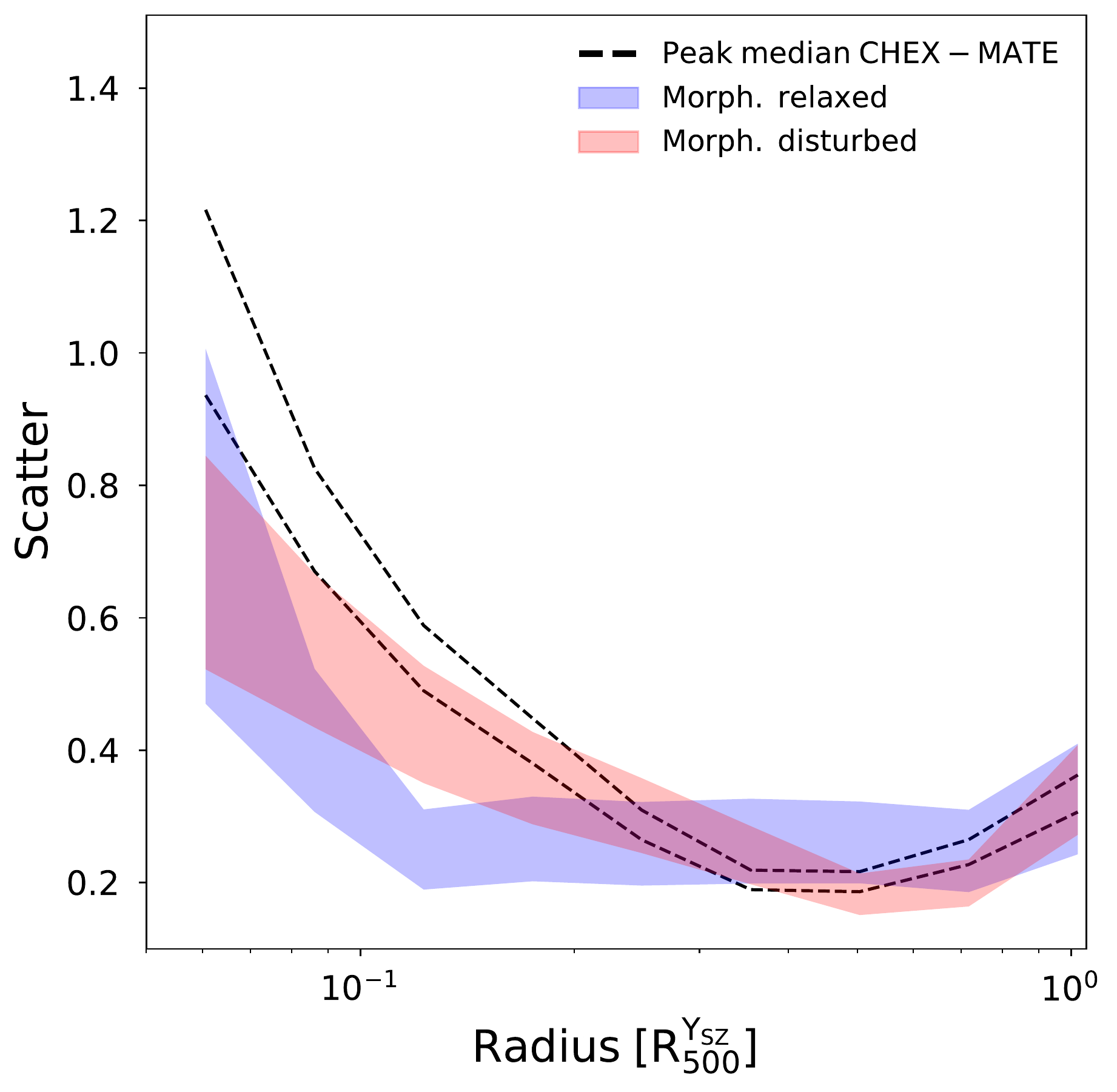}
\includegraphics[]{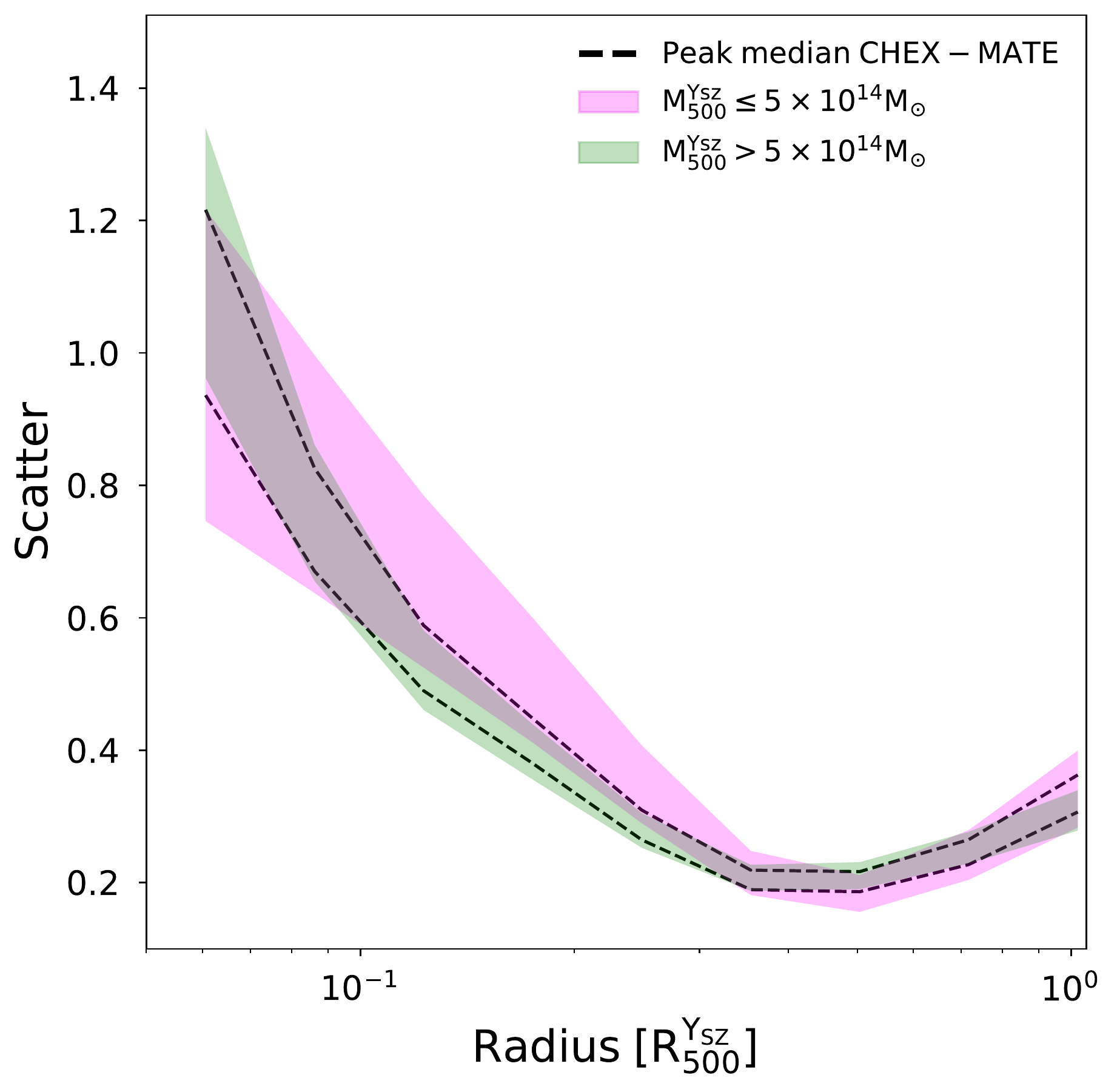} 
}
\resizebox{0.7\textwidth}{!}{
\includegraphics[]{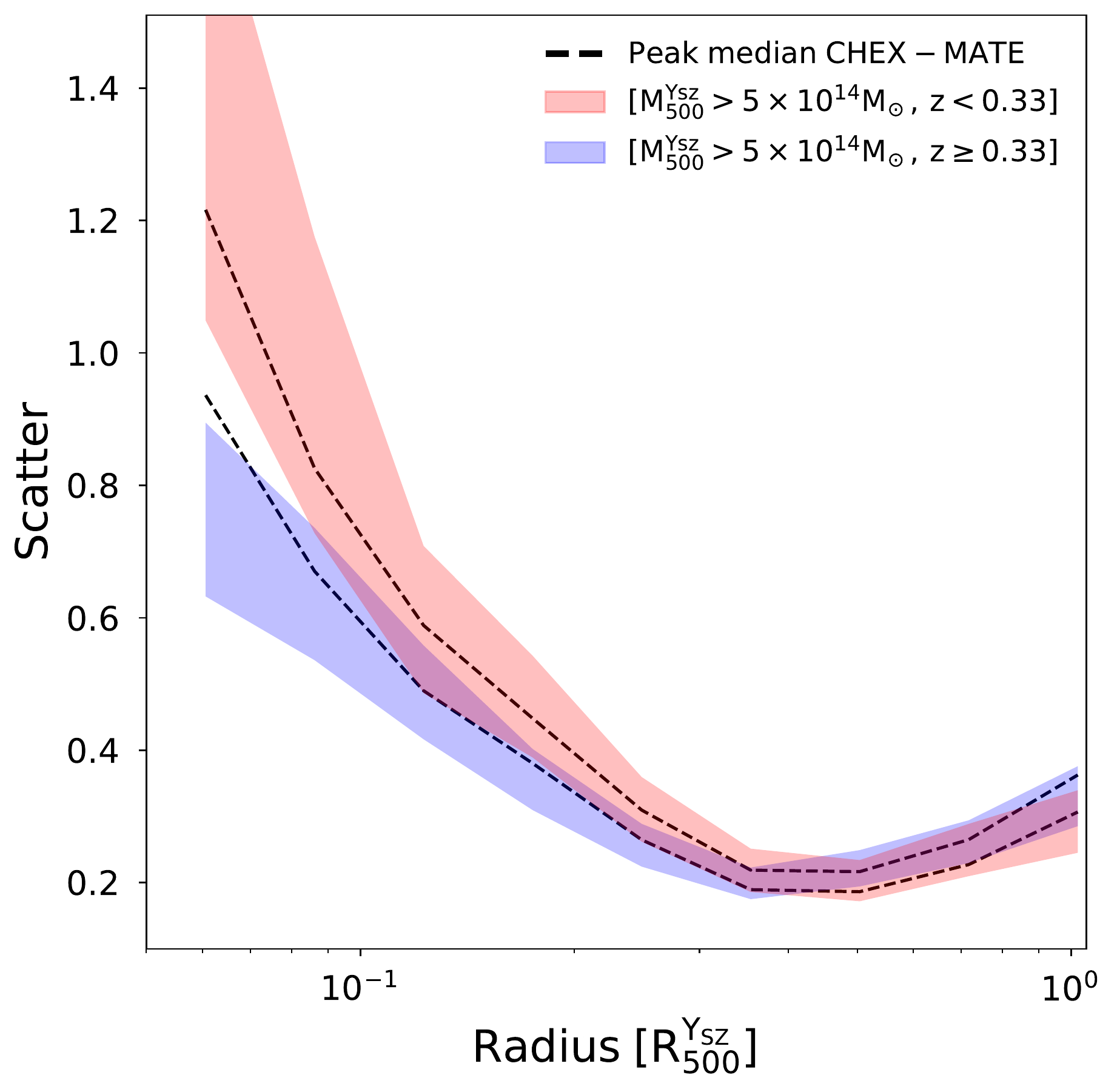} 
\includegraphics[]{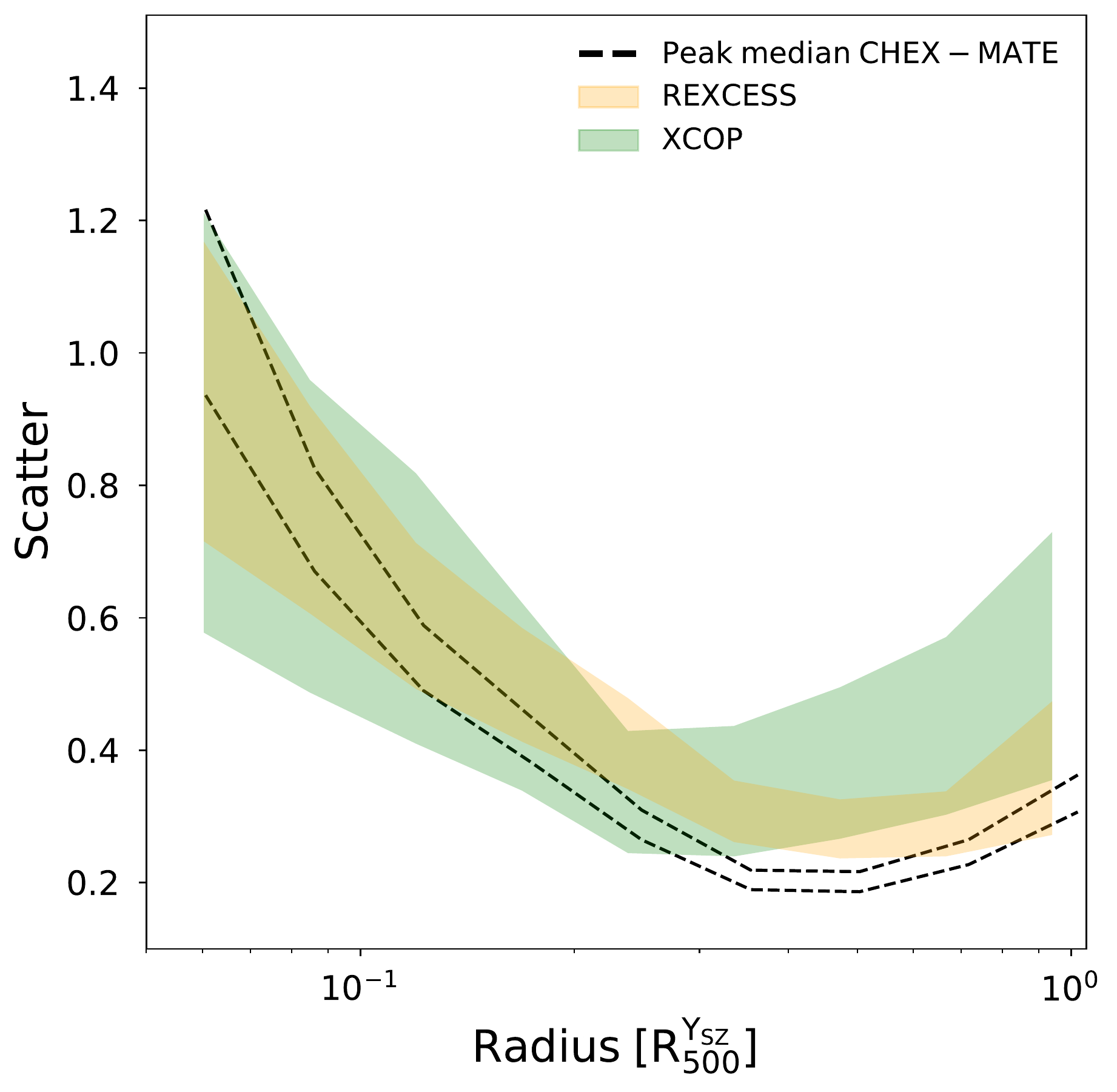}
}
\end{center}
\caption{\footnotesize{ Scatter of the \EMS\ \chxmt, X-COP, and \rexcess\  profiles. Top-left panel: Comparison between the scatter of the \chxmt\ sample and the morphologically selected sub-samples. Black dotted lines identify the $\pm 1\sigma$ scatter between the \EMS\ profiles of the \chxmt\ sample. The scatter between the profiles of morphologically relaxed and disturbed clusters are shown with blue and red envelopes, respectively. The width of the envelopes corresponds to the $1\sigma$ uncertainty. 
Top-right panel: Comparison between the scatter of the \chxmt\ sample and the mass selected sub-samples.  Green and magenta envelopes represent the scatter between the \EMS\ profiles of the low- and high-mass sub-samples, respectively. 
Bottom-left panel: Investigation of evolution of the scatter. Blue and red envelopes represent the scatter between the profiles of the hi-mass clusters, $\Mvysz > 5 \times 10^{14} M_{\odot}$, of the low- and high- redshift samples, respectively. 
Bottom-right panel: Comparison between the scatter of the \chxmt\ with X-COP \citep{ghirardini19} and \rexcess\ \citep{croston2008} samples. The X-COP and \rexcess\ scatters are shown with green and orange envelopes, respectively. We recall that the scaling of the X-COP and REXCESS profiles was performed using temperatures obtained differently than those for \chxmt, see Section \ref{sec:comparison_with_other_samples} for details. 
}
} 
\label{fig:scatter_obs}
\end{figure*}

Our sample contains nine objects for which we were not able to measure the profile above $\Rvysz$. Six of these objects are less massive than $\Mvysz \lesssim 4\times 10^{14}$M$_{\odot}$ and are classified as "mixed morphology" objects. We investigated the impact of excluding these profiles from the computation of the scatter by comparing the scatter computed within $0.9 \Rvysz$ using the full sample with the scatter computed excluding the nine objects. We noticed that this exclusion reduces the scatter by a factor of approximately$15\%$ at $\Rvysz$ starting from $\sim 0.4 \Rvysz$. We argue that the reduction of the scatter is linked to the fact that the nine clusters contribute positively to the total scatter being morphologically mixed. 
For this reason, we corrected for this effect, defining a correction factor, $cf$, that quantifies the difference in the scatter due to the exclusion of these profiles. We computed the ratio between the scatter including and excluding the nine profiles in the [$0.06-0.9 \Rvysz$] radial range, where we extracted the profiles for the whole \chxmt\ sample. We fitted this ratio via the mpcurvefit routine using a two degree polynomial function of the form $cf(r) = ar^2 + br + c$ and obtained the coefficients $[a,b,c] = [0.410,-0.117,0.993]$. 
We multiplied the scatter of the whole sample by $cf$ in the  [$0.06-1] \Rvysz$ radial range. From hereon, we refer to this scatter as the "corrected intrinsic scatter".

\subsection{The \chxmt\ scatter}
The corrected intrinsic scatter of the \EMS\ profiles is reported in the top-left panel of \figiac{fig:scatter_obs}. The scatter computed using the profiles centred either on the centroid or on the peak gave consistent results.

The intrinsic scatter of the scaled \EMS\ profiles substantially depends on the scale considered. In the central regions, the large observed scatter of $\sim 0.8$, at $R \sim 0.1 \Rvysz$ reflects the complexity of the cluster cores in the presence of non-gravitational phenomena, such as cooling and AGN feedback. On top of that, merging events are known to redistribute gas properties between the core and the outskirts, which flattens the gas density profiles in cluster cores. The scatter reaches a minimum value of $\sim 0.2$ in the [0.3-0.7]$\Rvysz$ radial range, where the scatter remains almost constant. This result confirms the behaviour observed in the left panel of \figiac{fig:em_medians_and_xcop}, where the dispersion of the profiles shown is minimal in this radial range and the scaled profiles converge to very similar values. The scatter increases at $R > 0.7 \Rvysz$ from 0.2 to 0.35.

The scatter of the morphologically disturbed and relaxed clusters considered separately are shown in the top-left panel of \figiac{fig:scatter_obs}. The scatter of the morphologically disturbed clusters is higher but consistent at $R<0.3\Rvysz$ with the relaxed one. This is expected, as the scatter originates from the combination of non-gravitational processes in the core and merging phenomena. This reinforces the scenario in which the differences between the \EMS\ profiles of relaxed and disturbed objects disappear in cluster outskirts, as already shown with the study of the shapes in Sect. \ref{sec:profile_shape}. The dependency of the scatter on cluster mass was identified by comparing the scatter between high- and low-mass objects, as shown in the top-right panel of \figiac{fig:scatter_obs}. No significant differences could be seen.
We investigated the evolution of the scatter by comparing the most massive clusters, $\Mvysz > 5 \times 10^{14} M_{\odot}$, in the low- and high-redshift samples. This is shown in the bottom-left panel of the figure, and as for the mass sub-samples, we found no significant differences except in the very inner core at $R<0.1\Rvysz$, where the local objects indicate larger variation.
\begin{figure*}[!ht]
\begin{center}
\resizebox{1\textwidth}{!}{\includegraphics[]{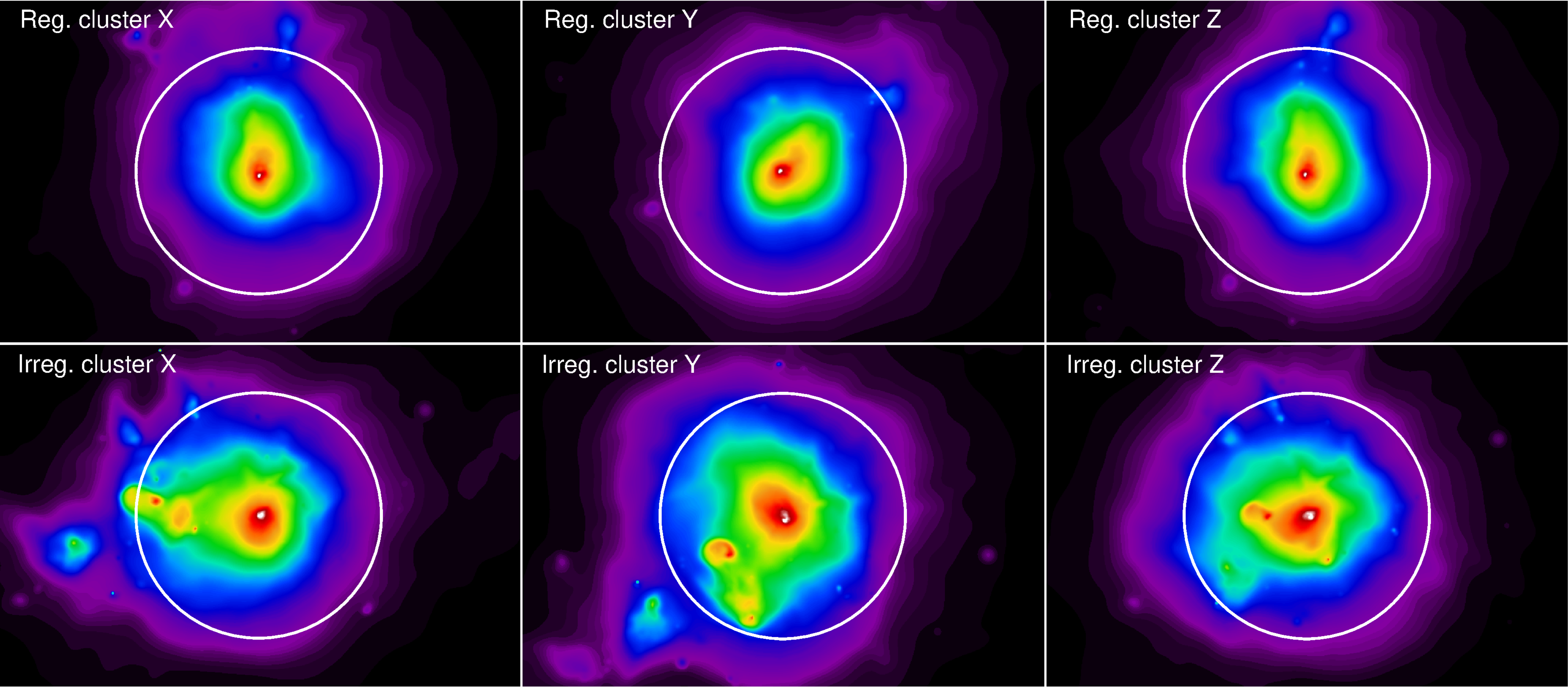}}
\end{center}
\caption{\footnotesize{EM maps of two of the simulated clusters used in this work projected along the lines of sight X, Y, and Z. The top row shows a cluster whose morphology appears roundish in the three projections considered. On the bottom we show, on the contrary, a cluster whose morphology is particularly complex and appears different in each of the three projections. We refer to the cluster in the top row as regular and the latter as irregular. The white circle indicates $\Rv$.}}
\label{fig:scatter_projection_example}
\end{figure*}
\subsection{Comparison with other samples}
We computed the scatters of the profiles of the REXCESS and X-COP sample following the same procedure we used for the \chxmt\ sample. These are shown in the bottom-right panel of \figiac{fig:scatter_obs}. The width of the envelope corresponds to 1$\sigma$ uncertainty. 
Overall, the \chxmt, \rexcess,\ and X-COP scatters are consistent at the [0.07-0.6] $\Rvysz$ radial range. 
This excellent agreement is due to the fact that the samples are representative of the wide plethora of \EMS\ profile shapes in the core of clusters.
There is slight disagreement at a larger scale between the samples, with the \chxmt\ scatter being lower at more than $1\sigma$, which could be do to the re-scaling. This is one important issue that will be investigated in forthcoming papers recurring also to multi-wavelength data. 

\section{Investigating the origin of the scatter}\label{sec:scatter_origin}
\subsection{Simulation scatters}

In this section, we turn our attention to the \threehun\ dataset. The cosmological simulations allowed us to break down the sample scatter, or \inter\ scatter, into two components: the genuine cluster-to-cluster scatter, which would be the sample scatter measured between the true 3D profiles of the objects, and the \intra\ scatter. The latter measures the differences that various observers across the Universe would detect when looking at the same object from distinct points of view.

In this work, we scaled the \chxmt\ EM profiles using the results of \citet{pratt22} and \citet{ettori22}, which were derived using empirical ad-hoc adaptation of the self-similar scaling predictions. However, the same scaling is less suitable for the simulations, which agree better with the self-similar evolution of Eq. \ref{eq:self_similar_scaling} since this expression minimises their scatter. For this reason, all the scatters presented from this point on were derived from EM profiles scaled assuming only self-similar evolution both for \threehun\ and \chxmt\ samples.

\subsection{The \intra\ scatter term}\label{sec:intra_scatter_term}
The evaluation of this term requires the knowledge of the 3D spatial distribution of the ICM. 
A perfectly spherical symmetric object would appear identical from all perspectives, and the \intra\ scatter would be equal to zero. 
On the other hand, an object whose ICM spatial distribution presents a complex morphology will produce a large \intra\ scatter. 
This can be visualised by looking at the three EM maps obtained for three orthogonal lines of sight for two objects of the \threehun\ collaboration in \figiac{fig:scatter_projection_example}.
In detail, the cluster shown in the top rows is roundish and does not show evident traces of merging activity within a radius of R=R$_{500}$ (white circle). The cluster in the bottom row, however, exhibits a complex morphology due to ongoing merging activities and the presence of sub-structures, which cause it to appear different in the three projections.

\begin{figure}[!t]
\begin{center}
\resizebox{1\columnwidth}{!}{\includegraphics[]{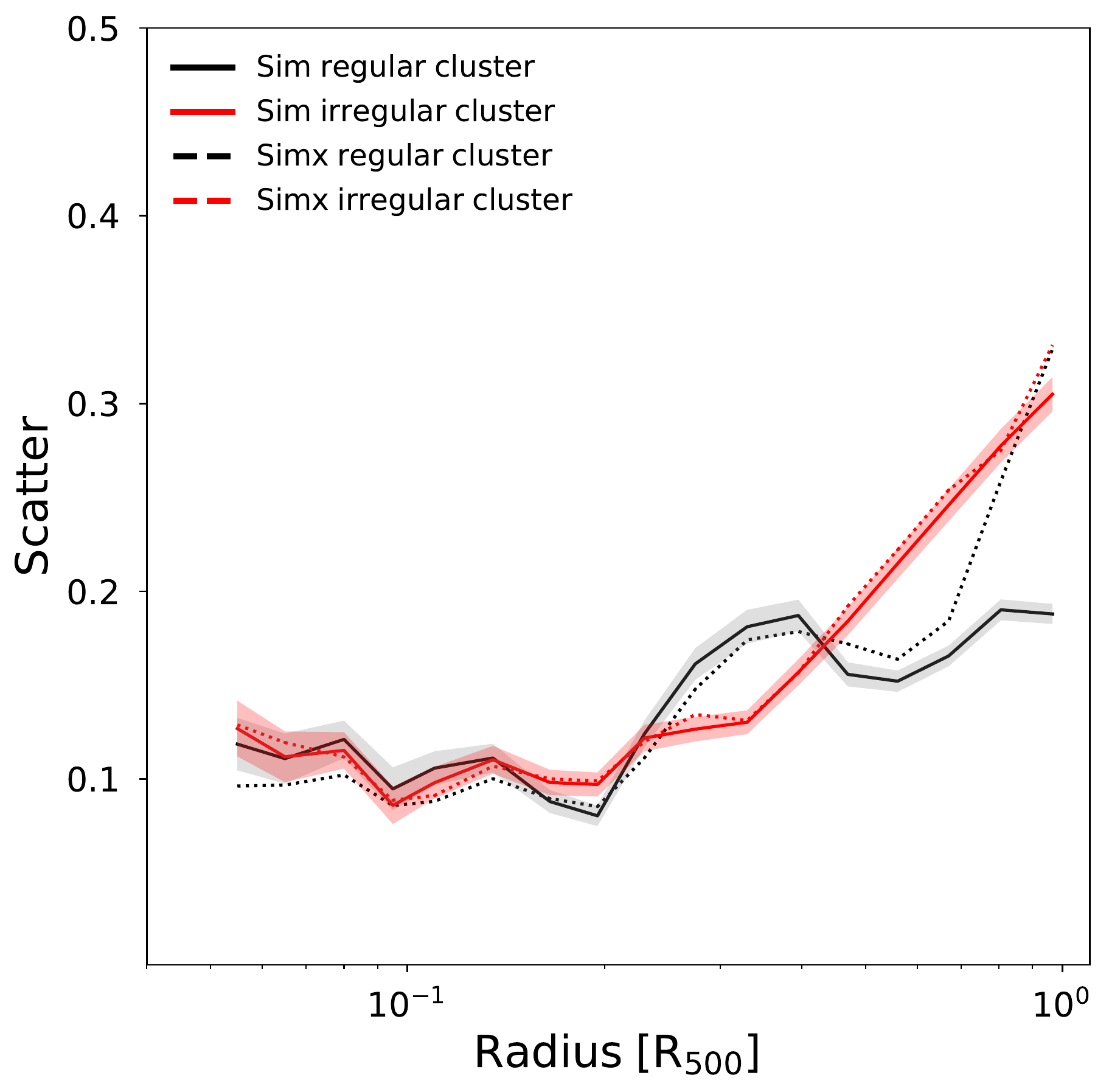}}
\end{center}
\caption{\footnotesize{Projection scatter of the regular and irregular clusters in \figiac{fig:scatter_projection_example}. The scatters are shown with black and red solid lines, respectively. The black and red envelops represent the dispersion. The black and red dotted lines refer to the \intra\ scatter computed using the Simx profiles.}}
\label{fig:example_scatter}
\end{figure}

This complexity is reflected in the \intra\ scatters shown in \figiac{fig:example_scatter}, which was computed considering the 40 lines of projection for the two objects and not only the three shown with the images. The scatters are similar within $R<0.2 \Rv$. At R>0.4R$_{500}$, the irregular cluster scatter diverges, while the one of the regular object remains almost constant. In particular, in the case of the irregular object, $\sim 0.4\Rv$ corresponds to the position of the big sub-structure visible in the bottom row of the left panel of \figiac{fig:scatter_projection_example}.
Interestingly, the Simx \intra\ scatter increases rapidly at R$\sim 0.8\Rv$ also for the regular cluster, while the Sim one remains mostly constant. This difference in the behaviour is due to the deliberate 7\% over- and underestimated background correction explained in Section \ref{sec:em_sim_profile_production}. The over-and underestimation of the background yields profiles that are steeper or flatter than the correct profiles, respectively, and hence they increase the scatter between the profiles. This effect is particularly important at R$\sim \Rv$ because the cluster signal reaches the background level. 

We calculated the \intra\ scatter between the 40 projections for each of the 115 objects from  \threehun\ sample, and these profiles are shown in \figiac{fig:inter_vs_intra}. 
On average, the scatter starts from a value of $0.15$ at $R \sim 0.1 \Rv$ and then reaches the value of $0.3$ at $\Rv$, with a rapid increase from 0.2 to 0.3 at $R \sim 0.9 \Rv$. This rapid increase is due to the complex spatial ICM distribution at large radii.
There are approximately five outliers that exhibit a larger scatter from the envelope and a complex behaviour. 
These clusters are characterised by the presence of sub-structures that happen to be behind or in front of the main halo of the cluster along some lines of projection. For this reason, the sub-structure emission is not visible as it blends with the emission from the core of the cluster. On the contrary, if the sub-halo is on a random position as respect to the main halo it will appear as a sub-structure in different position depending on the projection. In this case, the cluster morphology is complex. For these reasons, the resulting profile for these clusters can show remarkable differences depending on the line of sight.

\subsection{The \inter\ scatter term}
 \begin{figure}[!t]
\begin{center}
\resizebox{1\columnwidth}{!}{\includegraphics[]{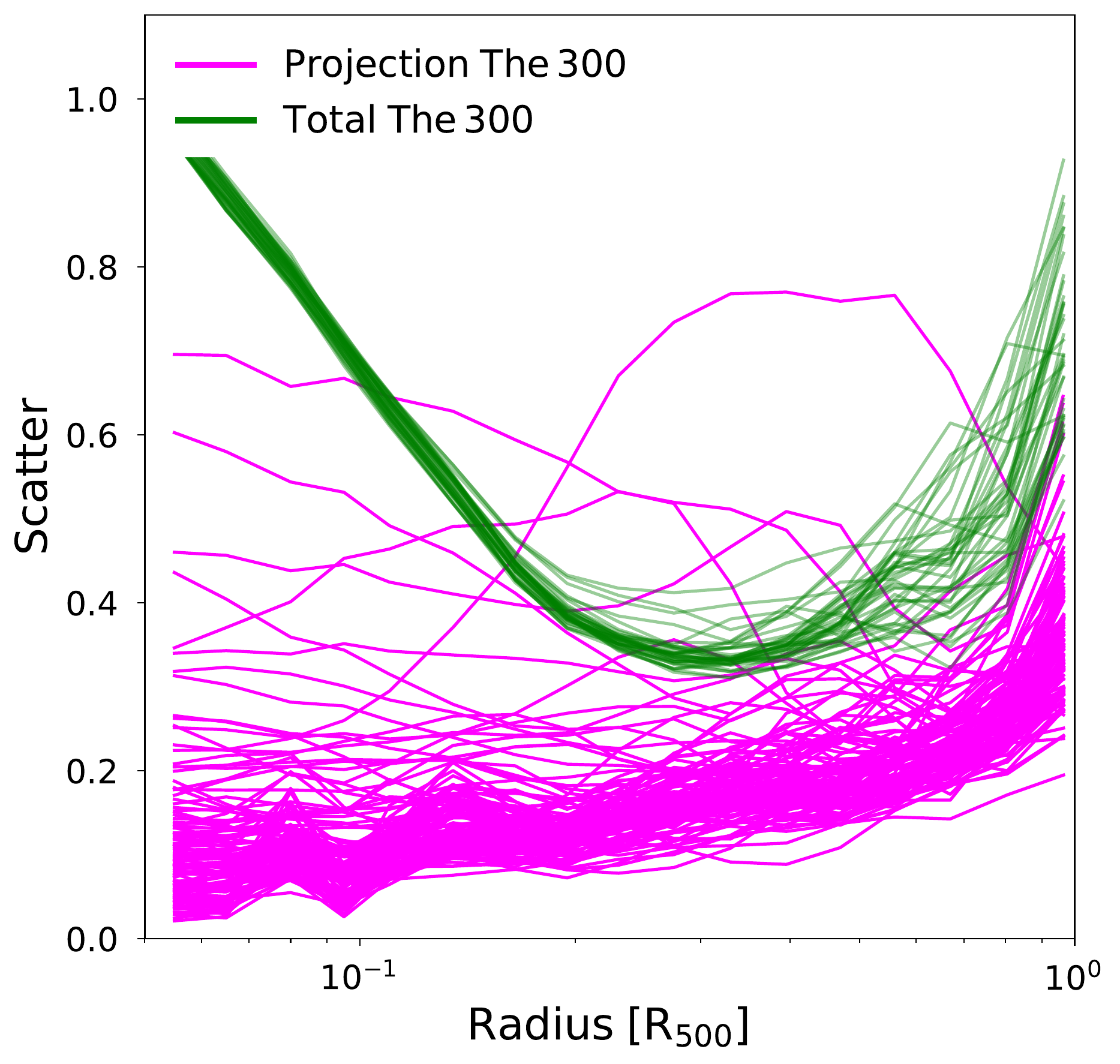}}
\end{center}
\caption{\footnotesize{Comparison between the \inter\ and \intra\ scatters of the Simx profiles. The scatters are shown with solid green and magenta lines, respectively.}}
\label{fig:inter_vs_intra}
\end{figure}
The total scatter term measures the differences between the cluster EM profiles within a sample. We recall that each of our simulated objects is seen along 40 lines of sight. With this possibility in hand, we created 40 realisations of the same sample of 115 objects and computed the scatter for each realisation. The 40 \inter\ scatters of the Simx profiles are shown in \figiac{fig:inter_vs_intra} . 

The average high value of $0.9$ at  $R < 0.3 \Rv$ of the \inter\ scatter captures the wide range of the profile shapes within the inner core. The scatter reaches its minimum value of $\sim 0.4$ at $R \sim 0.5 \Rvysz$ and then rapidly increases afterwards, due to the presence of sub-structures in the outskirts and the phenomena related to merging activities as well as the background subtraction discussed in Section \ref{sec:intra_scatter_term}.
 
\subsection{Comparison between \inter\ and \intra}\label{sec:inter_intra_explanation}
Direct comparison of the \intra\ and \inter\ scatter terms in the simulated sample allowed us to investigate the origin of the scatter as predicted by numerical models. The two scatters are shown in \figiac{fig:inter_vs_intra}.
The \inter\ scatter is almost eight times greater than the \intra\ at $R \sim 0.1 \Rv$ and rapidly decreases to be only two times greater at $R \sim 0.3 \Rv,$ as shown in \figiac{fig:inter_vs_intra}. This indicates that differences between clusters dominate with respect to the variations from the projection along different lines of sight at such scale. 

The \inter\ scatter is only $20\%$ greater than the \intra\ term in the $[0.4-0.8]\Rvysz$ radial range. This scale is where the cluster differences are smaller. At $R> 0.8 \Rvysz$, both scatters increase, implying that merging phenomena and sub-structures are impacting the distribution of the gas. Furthermore, we argue that the deliberate background over- and under-subtraction discussed in Section \ref{sec:intra_scatter_term} contributes to increasing both scatter terms by enlarging the distribution of the profiles where the signal of the cluster reaches the background level. The \inter\ scatters obtained using the Sim are similar to the ones obtained using Simx up to $0.9\Rv$ but remain below $\sim 0.45$ at $\Rv$.
\subsection{Simulation versus observations}
\begin{figure}[!t]
\begin{center}
\resizebox{\columnwidth}{!}{\includegraphics[]{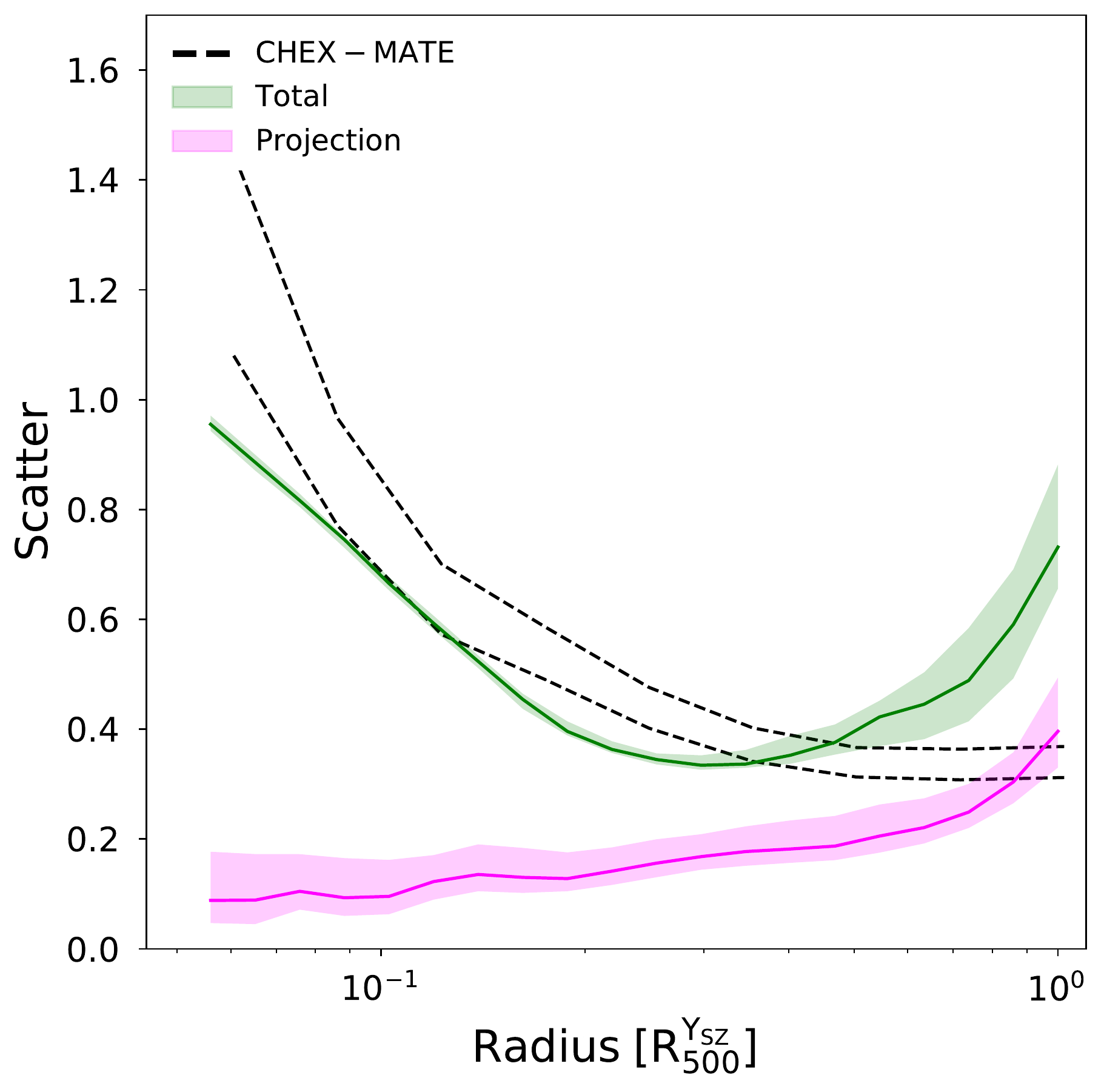}}
\end{center}
\caption{\footnotesize{Comparison between the scatter of the \chxmt\ sample and the \inter\ and \intra\ terms of the simulations. The \chxmt\ scatter is shown using dashed black lines. The medians of the \inter\ and \intra\ terms are represented with green and magenta solid lines, respectively. Their 68\% dispersions are represented using envelopes coloured accordingly. }
}
\label{fig:scatter_final}
\end{figure}

We can break down the contributions to the scatter in the \chxmt\ sample by using the numerical simulation scatter terms as a test bed. 
The Simx \inter\ and \intra\ scatter medians and their dispersion are shown in \figiac{fig:scatter_final}. 
The \chxmt\ scatter dispersion is also shown in the same figure. We recall that it is computed using the EM profiles scaled according to the self-similar model using Equation \ref{eq:monique_em_formula} and is greater than the one shown in  \figiac{fig:scatter_obs} due to the  residual dependency on mass and redshift discussed in Section \ref{sec:emission_profiles} in the $[0.2-0.8]$ radial range. However, the scatters reach the value of approximately $0.4$ at $\Rvysz$, indicating that the differences between the profiles are dominated by clumpy patches in the ICM distribution due to sub-halos and filamentary structures.

The \chxmt\ and Simx \inter\ scatters are in excellent agreement at R$<0.1 \Rvysz$ and marginally consistent within $2 \sigma$ in the [0.1-0.3] radial range. Generally speaking, they exhibit the same behaviour, rapidly declining from the maximum value of the scatter of 1.2 to 0.4. The \intra\ scatter on the other hand is at a minimum value of 0.1 and is almost constant up to $0.3\Rvysz$. This result implies that the observed scatter between the \EMSS\ profiles within $0.5 \Rvysz$ is dominated by genuine differences between objects and not by the projection along one line of sight, as explained in Sect. \ref{sec:inter_intra_explanation}. In other words, we are not limited by the projection on the plane of the sky when studying galaxy cluster population properties at such scales.

The \inter\ and \chxmt\ scatters reach the minimum value, approximately $0.4$, in the [$0.3-0.5] \Rvysz$ radial range and remain almost constant within these radii. This minimum value quantifies the narrow distribution of the profiles shown in \figiac{fig:profili_mediani_chexmate} and \figiac{fig:sim_vs_simx}. Furthermore, the slopes between morphologically relaxed and disturbed objects become consistent at such radii, as shown in the right panel of Fig. \ref{fig:profile_slopes}. This suggests that the differences between EM profiles are minimum at such intermediate scales despite their morphological statuses, mass or redshift. As for the inner regions, the \intra\ term increases mildly in these regions and provides a small contribution.  

The \threehun\ scatters rapidly increase at $R > 0.5 \Rvysz$, with the \inter\ scatter reaching the value of approximately $ 0.7$ at $\Rv$. The \intra\ scatter reaches the \chxmt\ scatter at $\Rvysz$. We argue that this effect is due to a combination of 
not masking the sub-structures when extracting the Simx profiles and the deliberate wrong background subtraction discussed in Sect. \ref{sec:intra_scatter_term}. Indeed, the use of median profiles reduces this effect, and we discuss this effect in detail in Appendix \ref{sec:appendix_subhalo}, where we show that the use of azimuthal median profiles is efficient for removing part of these spatial features. 
The fact that the scatter terms increase in a similar manner despite the use of median profiles reinforces that this behaviour is likely related to analysis techniques rather than genuine differences within the profiles and projection effects.

\section{Discussion and conclusion}\label{sec:conclusions}
 We have studied the properties of the SX and EM radial profiles of the \chxmt\ sample, which comprises 116 SZ selected clusters observed for the first time with deep and homogeneous \xmm\ observations.  Our main findings are as follows:
\begin{itemize}
    \item The choice of making the centre between the peak and the centroid for extraction of the SX profiles yields consistent results in the [0.05-1]$\Rvysz$ radial range. Significant differences can be seen within $\sim 0.05 \Rvysz$.
     \item The use of azimuthal average and median techniques to extract the profiles impacts the overall profile normalisation by a factor of $5\%$ on average. The shape is mostly affected at $R>0.8\Rvysz, $ with azimuthal averaged profiles being flatter at this scale.
     \item The EM profiles exhibit a dependency on the mass and a mild dependency on redshift, which is not accounted for by the  computed scaling according to the self-similar scenario, as found also by \citet{pratt22} and \citet{ettori22}. 
    \item Morphologically disturbed and relaxed cluster \EMS\ profiles have different normalisations and shapes within  $\sim 0.4 \Rvysz$. The differences at larger radii are on average within $10\%$ and are consistent within the dispersion of the full sample. 
    \item The shape and normalisation of the \EMS\ profiles present a continuum distribution within the [0.2-0.4]$\Rvysz$ radial range. 
    The extreme cases of morphologically relaxed and disturbed objects are characterised by power law indexes, $\alpha = 2.51 \pm 0.13$ and $\alpha=1.38 \pm 0.2$, respectively, that are  not consistent at the $3\sigma$ level. The picture changes at $\Rvysz > 0.4$, where the slopes of these extremes becomes marginally consistent at $1\sigma $ in the $[0.4-0.6]\Rvysz$ radial bin. The slopes in the last bin are in excellent agreement.
    \item The scatter of the \chxmt\ sample depends on the scale. The scatter maximum is $\sim 1.1$ within $0.3 \Rvysz$, reflecting  the wide range of profile shapes within the cluster cores that range from the flat emission of disturbed objects to the peaked emission of the relaxed clusters. 
    The scatter decreases towards its minimum value, 0.2, at $0.4 \Rvysz$ and increases rapidly to 0.4 at $\Rvysz$.  This result is coherent with the overall picture of a characteristic scale, R$\sim 0.4 \Rvysz$, at which the differences between profiles in terms of shape and normalisation are minimum. The increase of the scatter at $\Rvysz$ is expected, as this is the scale at which merging related phenomena and patchy distribution of the ICM become important. 
    \item The scatters of the morphologically relaxed cluster and the disturbed cluster are different within $0.4 \Rvysz$, the former being smaller. Above this radius, they are in excellent agreement between themselves and with the entire sample as well, implying that the properties of EM profiles in the outer parts are not affected by the properties in the core. There are no differences in the scatter of the sub-samples formed by high- and low-mass objects, and we found no evolution of the scatter for high-mass objects.
 \end{itemize}
 The overall emerging picture is that there is a characteristic scale, R$\sim 0.4 \Rvysz$, where the differences between profiles in terms of shape and normalisation are minimum. The exceptional data quality has allowed us to provide to the scientific community the scatter of SX and \EMS\ radial profiles of a representative cluster sample with an unprecedented precision of approximately $5\%$. 

The results from observations were compared to a sample drawn from the numerical simulation suite \threehun\ formed by 115 galaxy clusters selected to reflect the \chxmt\ mass and redshift distribution. For each cluster, we computed the EM along 40 randomly distributed lines of sight, which allowed for the investigation of projection effects for the first time. Our main findings can be summarised as follows:

 \begin{itemize}
     \item The properties derived using the Sim or the Simx profiles are similar within $\Rv$, confirming the statistical quality of the mock X-ray images, which were calibrated to match the \chxmt\ average statistical quality.
     \item The simulation \EMS\ profiles appear systematically steeper than those from observations. The hydrostatic bias might play a key role in explaining this difference. The scaling of the \chxmt\ profiles by $\Rvysz/(0.8^{1/3})$,  assuming a $20\%$ bias, alleviates these differences, and the ratio between the \chxmt\ and simulation medians becomes closer to one, with the exception of the centre where simulations typically have a greater gas density.
     \item The \inter\ scatter of the simulation sample follows the same behaviour as that of the observations up to $0.6 \Rvysz$ and then increases more rapidly to an average value of approximately $0.7,$ whereas the observation reaches the value of $0.4$ at $\Rvysz$. The comparison with the \intra\ scatter at such scales hints at a contribution from projection effects on the order of $0.3$.
     \item The \intra\ scatter allowed us to study the spherical symmetry of clusters. This term slightly increases from approximately $0.1$ at $0.1\Rvysz$ up to approximately $0.3$ at $\sim \Rvysz$, exhibiting a rapid gradient at $\Rvysz$. This term is smaller than the \inter\ in the entire $[0.1-0.9]\Rvysz$ radial range considered, and its dispersion is on the order of $10\%$. This implies that the difference we observe between objects is due to a genuine difference in the gas spatial distribution. 
     \item The background subtraction process  becomes crucial at $\Rv$ for determining of the profile shape at $\Rvysz$. The deliberate over- or underestimation significantly contributes to increasing both the \inter\ and \intra\ scatter at such large scales. Furthermore, the rapid increase of both scatters can be also explained by the fact that sub-structures are not masked in simulated images.
\end{itemize}
 The large statistics offered by the simulation dataset allowed us for the first time to investigate the origin of the scatter and break down the components, namely the \intra\ and \inter\ terms, and study them as a function of $\Rvysz$. The overall picture emerging is that there are three regimes amongst the scatter:
 \begin{itemize}
\item $[0.1-0.4]$: The differences between profiles are genuinely due to a different distribution of the gas and also influenced by feedback processes and their implementation (see, e.g., \citealt{gaspari2014}), which   translates into a plethora of profile shapes and normalisations. 
\item $[0.4-0.6]$:  In this range, the scatter is sensitive to the scaling applied, suggesting that this is the scale where clusters are closer to being within the self-similar scenario.
\item $[0.6-1]$: The \chxmt\ scatter and the \inter\ scatter increase at such scales and are greater by a factor of approximately two than the \intra, showing that profile differences are genuine and not due to projection effects. The emission of sub-structures and filamentary structures and the correct determination of the background play a crucial role in determining the shape of the profiles at such scales. 
 \end{itemize}

We were able to investigate the origin of the scatter by combining the statistical power of the \chxmt\ sample not only because of the great number of objects observed with sufficient exposure time to measure surface brightness profiles above $\Rvysz$ but also because of the sample's homogeneity and the uniqueness of the simulation sample. The latter allowed us to discriminate the scatter due to genuine differences between profiles and those related to projection. The \chxmt\ sample allowed us to measure  the scatter up to $\Rvysz$ with the sufficient precision to clearly discriminate the contribution from the \intra\ term at all scales.
\begin{acknowledgements} 
The authors thank the referee for his/her comments. We acknowledge financial contribution from the contracts ASI-INAF Athena 2019-27-HH.0,
``Attivit\`a di Studio per la comunit\`a scientifica di Astrofisica delle Alte Energie e Fisica Astroparticellare''
(Accordo Attuativo ASI-INAF n. 2017-14-H.0), and
from the European Union’s Horizon 2020 Programme under the AHEAD2020 project (grant agreement n. 871158). 
This research was supported by the International Space Science Institute (ISSI) in Bern, through ISSI International Team project \#565 ({\it Multi-Wavelength Studies of the Culmination of Structure Formation in the Universe}).
The results reported in this article are based on data obtained with \xmm, an ESA science mission with instruments and contributions directly funded by ESA Member States and NASA. GWP acknowledges financial support from CNES, the French space agency. 

 \end{acknowledgements}
\bibliographystyle{aa}
\bibliography{lib_articoli}
\FloatBarrier

\clearpage
\onecolumn

\begin{appendix}

\section{The impact of sub-structures in simulations}\label{sec:appendix_subhalo}
The presence of sub-structures within the region of extraction of the radial profiles modifies the shape of the surface brightness and emission measure profiles. This translates into an increase  of the scatter between them. 
In this work, we are interested in the distribution of the gas within the cluster halo filtering the contribution of sub-structures whose emission is detectable within or near $\Rvysz$. This filtering is achieved by masking the sub-structures in observations. The same procedure is difficult to apply to simulations.

\begin{figure}[!ht]
\begin{center}
\resizebox{0.5\textwidth}{!}{\includegraphics[]{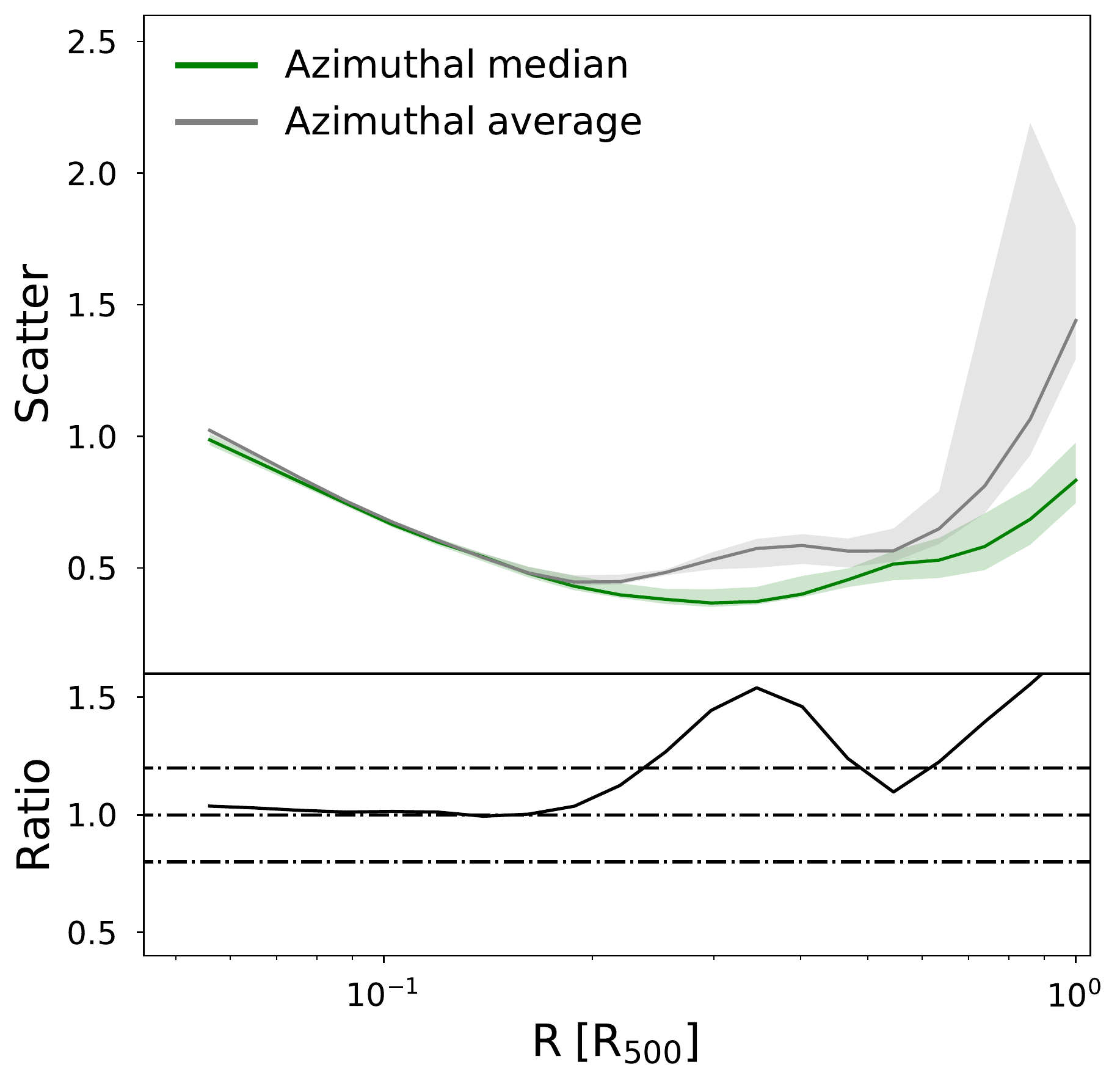}}
\end{center}
\caption{\footnotesize{Comparison between the \inter\ scatters computed using the azimuthal mean and median profiles. Top: Medians of the total from  \threehun\ scatters computed using the azimuthal mean and median EM profiles. These are shown with green and grey lines, respectively. The $68\%$ dispersion is shown with the coloured envelopes. \textit{Bottom:} Ratio between the median of the total scatters computed using the azimuthal averaged profiles over the median computed using the azimuthal median profiles. The dashed-dotted lines indicate the identity line and the $\pm 20\%$ levels.}}
\label{fig:substructure}
\end{figure}

Generally speaking, automatic detection algorithms in X-ray analyses are calibrated to detect point source emission only, as the detection of extended sources would cause the algorithm to also detect the cluster emission itself. For this reason, the identification of extended emission associated to sub-structures is done via eye inspection, but this approach cannot be taken with large datasets comprised of thousands of maps, such as the one we used in this work. The fact that we do not mask sub-structures in the simulated maps constitutes one of the main differences between the X-ray analysis and \threehun\ analysis. However, we could qualitatively investigate the impact of sub-structures on the scatter by comparing the results obtained following the procedures of Section \ref{sec:scatter_origin} that used the azimuthal average and median profiles shown in \figiac{fig:substructure}. 

The bottom panel shows that the scatters are nearly identical, within $0.2 \Rv$, and are on average around the $20\%$ level in the [0.2-0.6]$\Rv$ radial range. The scatter of the mean profiles increases rapidly above that radius. The same behaviour is observed for the scatter of the median even if the increase is less rapid, as shown by the ratio in the bottom panel at R$>0.6\Rv$. 

The azimuthal median in a given annulus does not completely remove the emission from the extended sub-structure, which can only be achieved by masking it. However, we argue that the scales at which the sub-structures become important, R$>0.6\Rv$, correspond to annuli whose size is typically larger than the size of a sub-halo. For this reason, the azimuthal median is marginally affected. 

Indeed, sub-structure masking is a key difference between observations and simulations, and it does affect the computation of the total scatter. However, we suggest that using the median profiles is an effective way to reduce the impact of sub-structures at the scales at which they are important. For this reason, the rapid increase of the \inter\ scatter at $\sim \Rv$ is more likely to be due to a genuine difference between the profiles and to the background subtraction effect discussed in Section \ref{sec:intra_scatter_term}.
\section{ROSAT-\xmm\ background relation}\label{appendix:rosat}
\begin{figure*}[!ht]
\begin{center}
\resizebox{0.7\textwidth}{!}{
\includegraphics[]{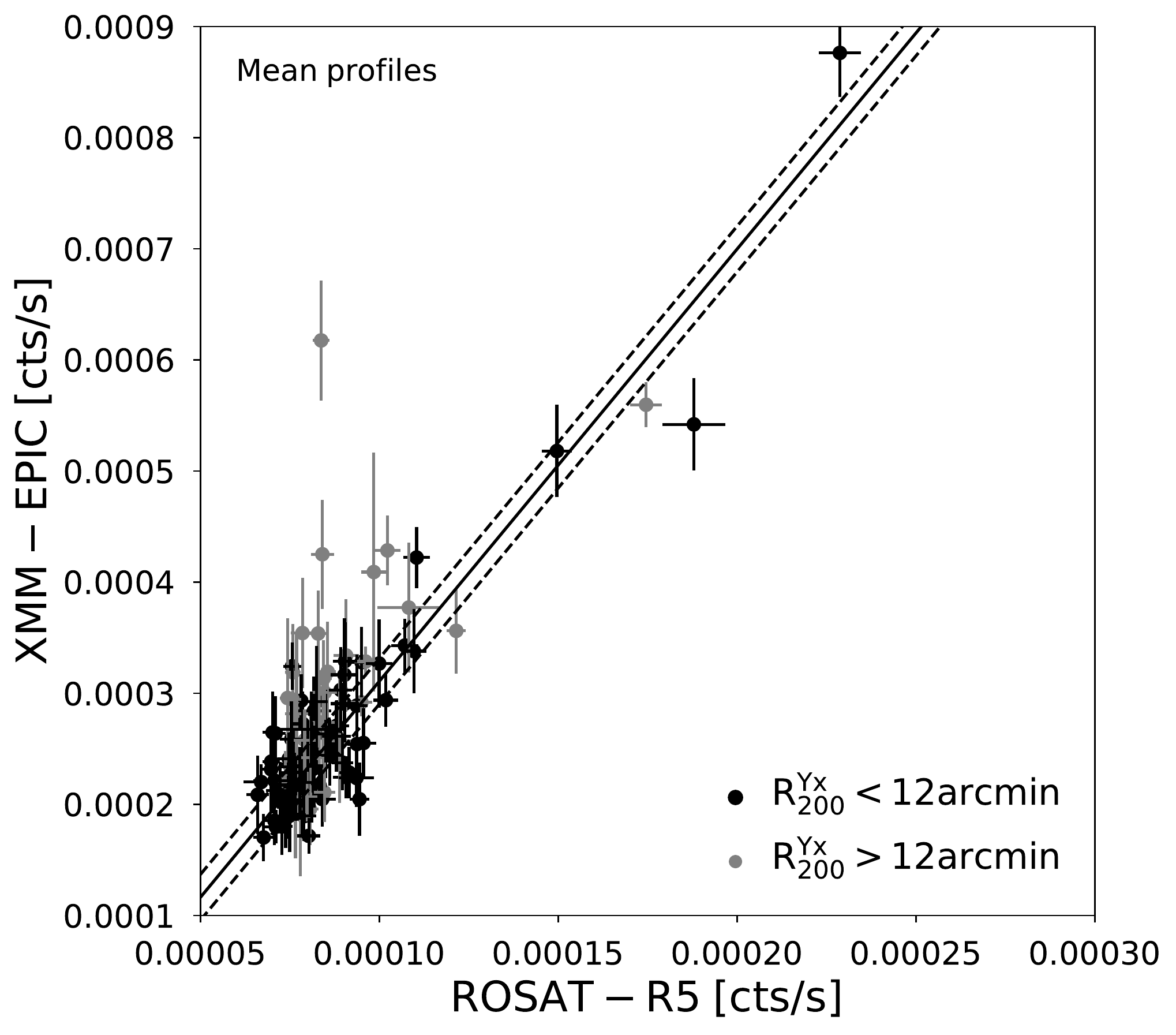}
\includegraphics[]{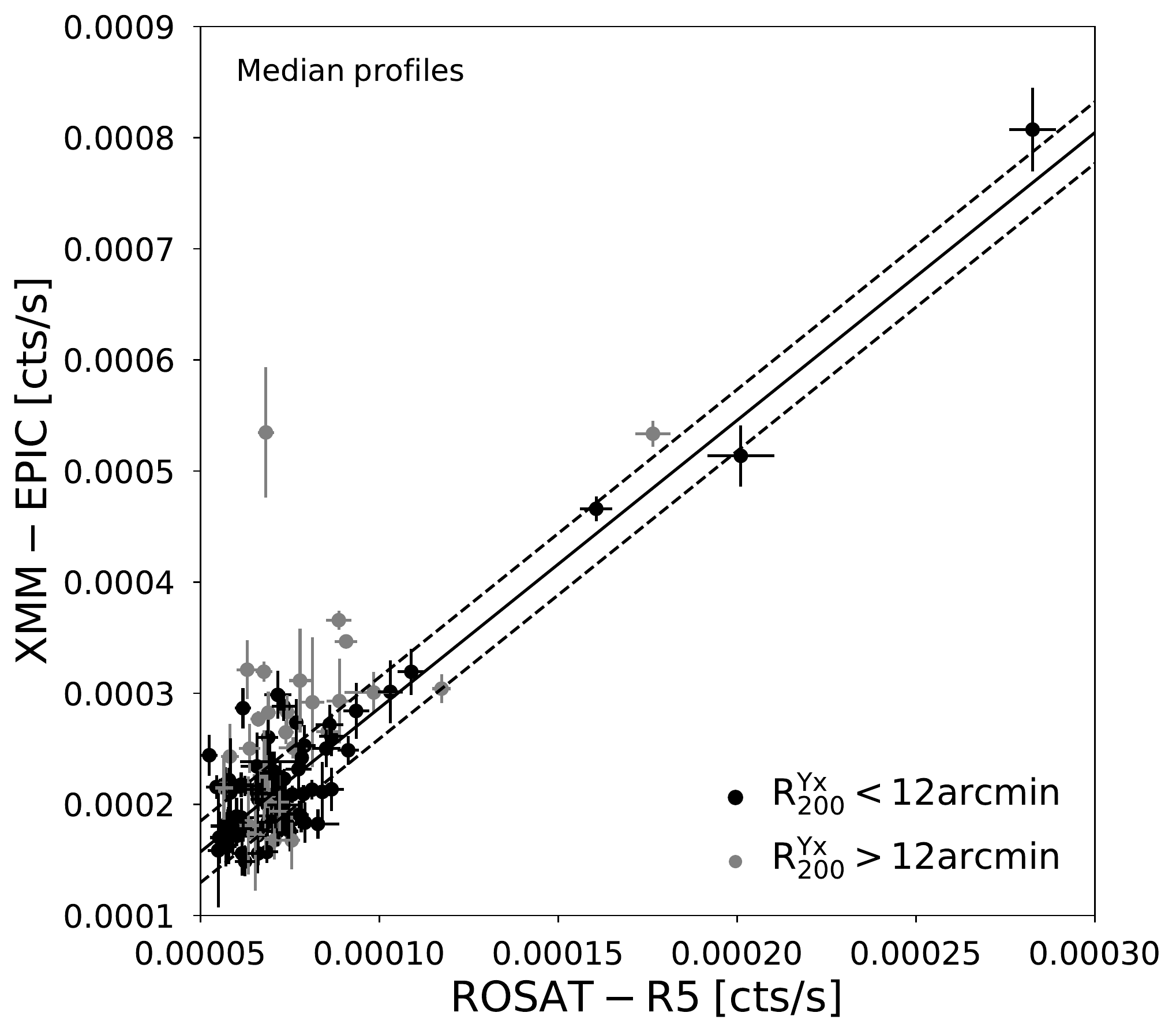}
}
\end{center}
\caption{\footnotesize{Calibration of the sky background count rate between \xmm\ and ROSAT-PSPC. \textit{Left panel}: Relation between the sky background count rate as measured using the \xmm\ mean SX profiles and ROSAT-PSPC in the R5, [0.56-1.21] keV, energy band. The black points represent the clusters for which $\Rvvysz$ is less than 12 arcmin and that have been used to fit the relations. The grey points are the clusters filling the field of view, their $\Rvvysz$ being greater than 12 arcmin. The solid line represents the cross correlation obtained via the linear regression. The dashed lines represent the intrinsic scatter of the relation. \textit{Right panel:} Same as the left panel except for the fact that the \xmm\ count rate is measured using the median SX profiles.}}
\label{fig:rosat_bck}
\end{figure*}
The determination of the sky background level was performed in a region free from cluster emission. In this work, we used the annular region between $\Rvvyx$ and 13.5 arcminutes to measure the photon count rate associated with the sky background. We considered that we had sufficient statistics for the background estimation if the width of this region is at least 1.5 arcmin (i.e. $\Rvvysz<12$ arcmin).  
The $\Rvvysz$ of nearby clusters at z $\lesssim 0.2$ are generally larger than $12$ arcmin, and it was not possible to define a sky background region unless offset observations were available.
For this reason, we predicted the sky background for these objects using the ROSAT All-Sky Survey diffuse background maps obtained with the Position Sensitive Proportional Counters (PSPC). 

We determined the ROSAT photon count rate, ROSAT$_\textrm{cr}$, for each \chxmt\ object in the R5 band, [$0.73-1.56$] keV,  within an annular region centred on the X-ray peak and with the minimum and maximum radius being $\Rvvysz$ and 1.5 degrees, respectively, using the sxrbg tool \citep{rassbkg}.
We then calibrated the relation between the ROSAT$_\textrm{cr}$ and the \xmm\ background sky count rate, XMM$_\textrm{cr}$, for clusters whose $\Rvvyx$   was less than 12  by performing a linear regression using the linmix package \citep{kelly07}:
\begin{equation}\label{eq:back_fit}
    \mathrm{XMM_{cr}} = \alpha + \beta \times \mathrm{ROSAT_{cr}}.
\end{equation}
The results of the linear regression for the mean and median profiles are shown in the left and right panels of \figiac{fig:rosat_bck}, respectively. 
The values of the linear minimisation $y = \alpha + \beta*x$ and the intrinsic scatter $\epsilon$ are reported in Table \ref{tab:rosat}. We used these relations to estimate the \xmm\ sky background for the objects where  $\Rvvyx$ is greater than 12 arcmin in the \chxmt\ sample.

\begin{table}[!ht]
\caption{ {\footnotesize Results of the linear minimisation of the ROSAT-R5 vs \xmm\ sky background count rate shown in Equation \ref{eq:back_fit}.}\label{tab:rosat}}
\begin{center}
\resizebox{0.4\columnwidth}{!} {
\begin{tabular}{lccc}
\hline        
Parameter   &  Val &  & Val  \\
\hline
\hline
                             &   Mean  & & Median \\
\hline
$\alpha$ [$10^{-5}$ ct/s]    & 3.974  & &   2.346        \\
$\beta$                      & 2.730   & &   2.630        \\
$\epsilon$ [$10^{-5}$ ct/s]  & 2.375   & &   2.902      \\
\end{tabular}
}
\end{center}
\footnotesize{\textbf{Notes:} The term $\epsilon$ represents the intrinsic scatter.}
\end{table}

\section{Power law fit}
We report in Table \ref{tab:pow_fit_results} the results of the fit of the median \EMS\ profiles centred on the X-ray peak profiles using the power law shown in Eq. \ref{eq:pow} and described in Sect. \ref{sec:profile_shape}. 
The fit was performed using the mean value of each bin as the pivot for the radius, that is, 0.3, 0.5, 0.7, and 0.9 for the [0.2-0.4], [0.4-0.6], [0.6-0.8], and [0.8-1] $\Rvysz$ radial bins, respectively.
\begin{table*}[!ht]
\caption{ {\footnotesize Results of the power law fit shown in Equation \ref{eq:pow}. }}\label{tab:pow_fit_results}
\begin{center}
\resizebox{1\textwidth}{!} {
\begin{tabular}{ccccc|cccc}
\hline        
Radial bin   &  $\alpha_{\mathrm{CHX}}$ & $\alpha_{\mathrm{Simx}}$ &  $\alpha_{\mathrm{CHX}}$  MR & $\alpha_{\mathrm{CHX}}$ MD & A$_{\mathrm{CHX}}$   & A$_{\mathrm{Simx}}$ &  A$_{\mathrm{CHX}}$ MR &  A$_{\mathrm{CHX}}$ MD   \\

[$\Rv$]     &                          &                          &                        &                      & \multicolumn{4}{c}{[ $10^{-6} \mathrm{cm}^{-6}$ Mpc]}                \\
\hline
\hline
0.2-0.4     &  $2.01 \pm 0.36$   &   $2.37 \pm 0.36$        &  $2.51 \pm 0.13$  & $1.38 \pm 0.20$ & $1.88 \pm 0.41$ & $2.04 \pm 0.27$ & $2.13 \pm 0.38$ & $1.75 \pm 0.31$ \\
0.4-0.6     &  $2.58 \pm 0.33$   &   $2.98 \pm 0.38$        &  $2.93 \pm 0.22$  & $2.25 \pm 0.18$ & $0.59 \pm 0.11$ & $0.50 \pm 0.11$ & $0.55 \pm 0.05$ & $0.66 \pm 0.14$ \\
0.6-0.8     &  $3.03 \pm 0.27$   &   $3.54 \pm 0.56$        &  $3.17 \pm 0.32$  & $3.00 \pm 0.20$ & $0.22 \pm 0.04$ & $0.18 \pm 0.05$ & $0.19 \pm 0.04$ & $0.27 \pm 0.04$ \\ 
0.8-1.0     &  $3.27 \pm 0.36$   &   $3.55 \pm 0.99$        &  $3.44 \pm 0.42$  & $3.49 \pm 0.27$ & $0.10 \pm 0.02$ & $0.07 \pm 0.02$ & $0.07 \pm 0.02$ & $0.12 \pm 0.02$  \\
\hline

\end{tabular}
}
\end{center}
\footnotesize{\textbf{Notes:} The letters MR and MD in columns 4, 5, 8, and 9 stand for morphologically relaxed and disturbed, respectively. }
\end{table*}

\section{Surface brightness profiles}
\begin{figure*}[!ht]
\begin{center}
\resizebox{0.6\textwidth}{!}{
\includegraphics[]{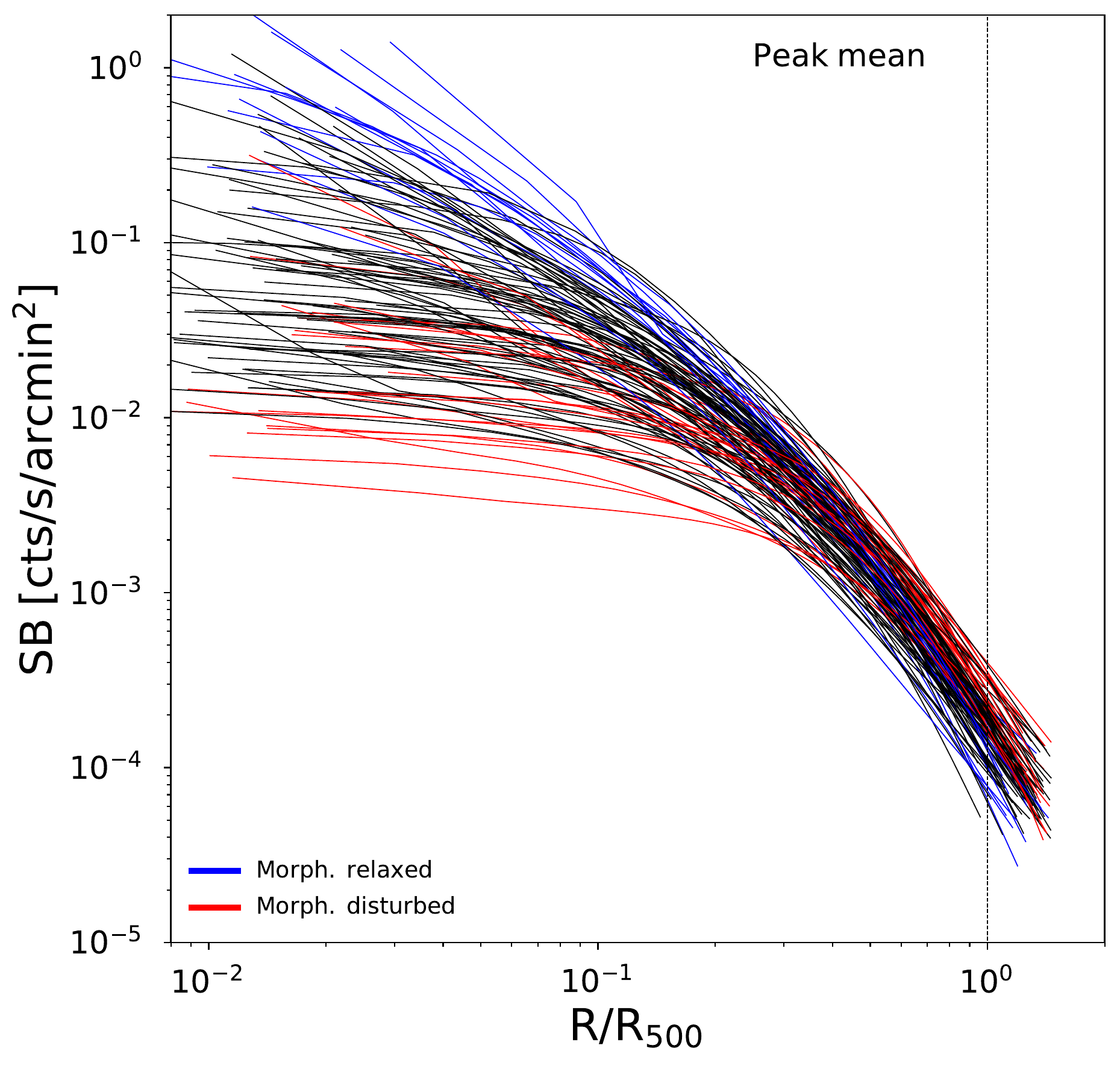}
\includegraphics[]{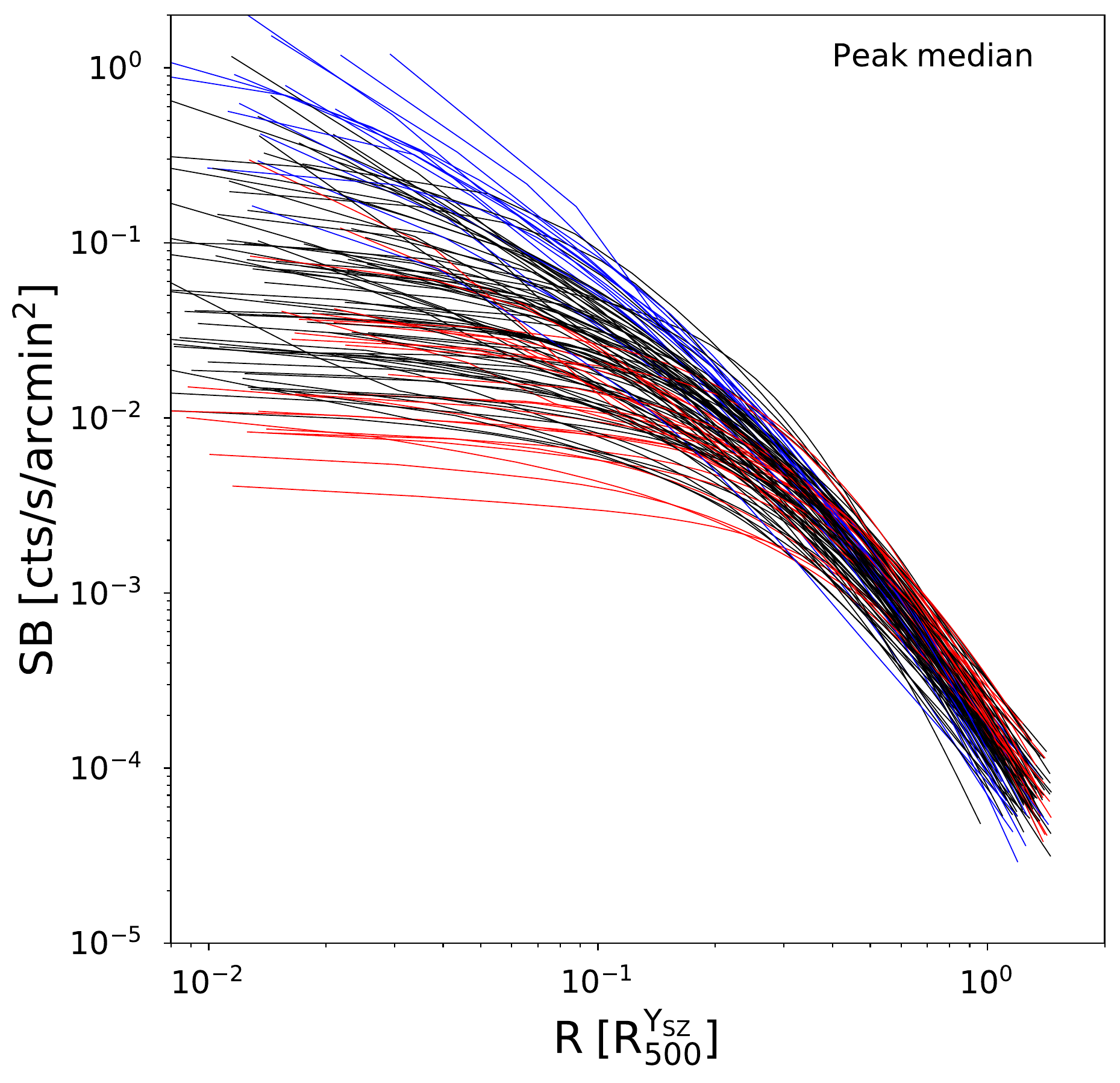}
}
\resizebox{0.6\textwidth}{!}{
\includegraphics[]{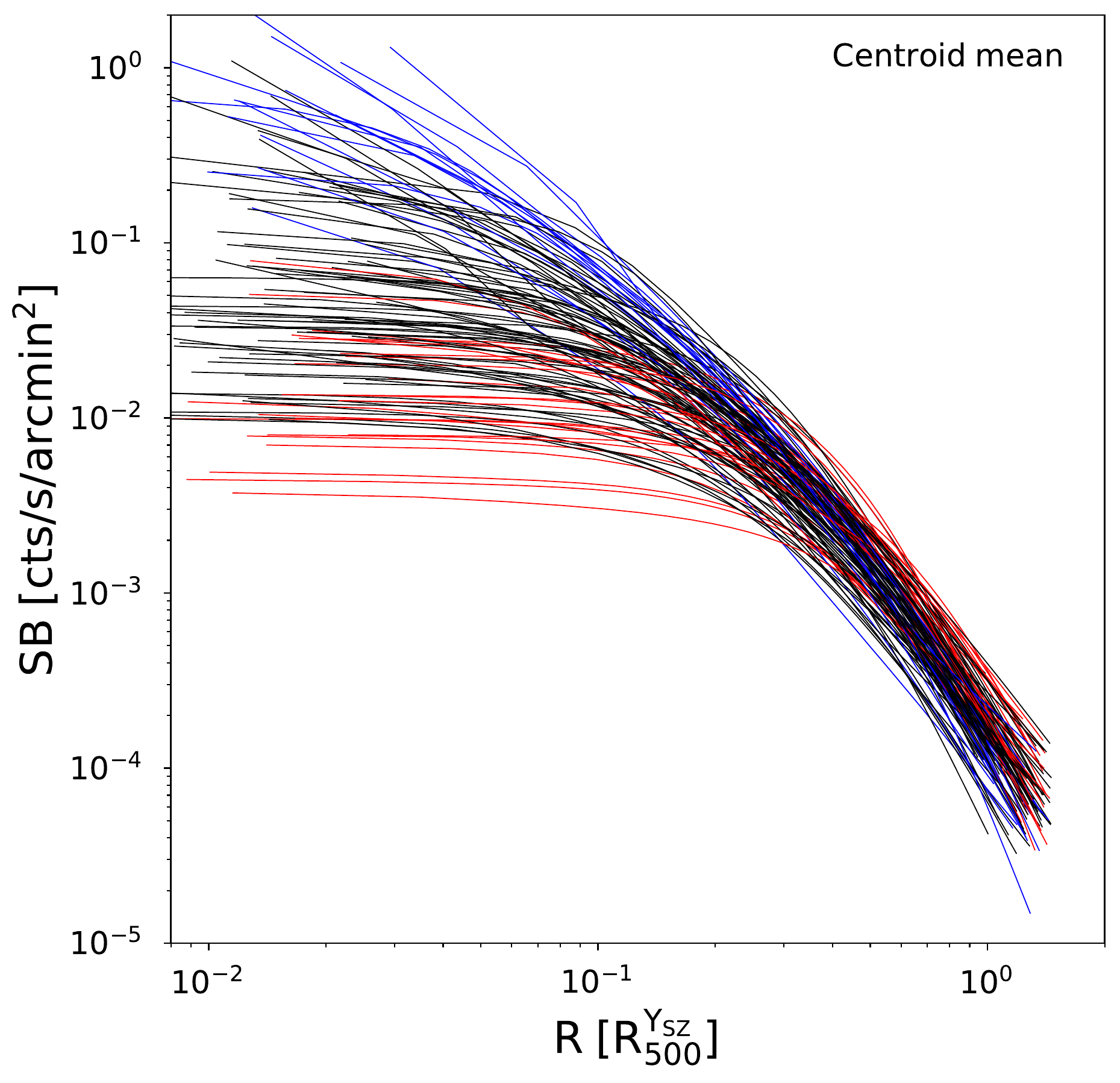}
\includegraphics[]{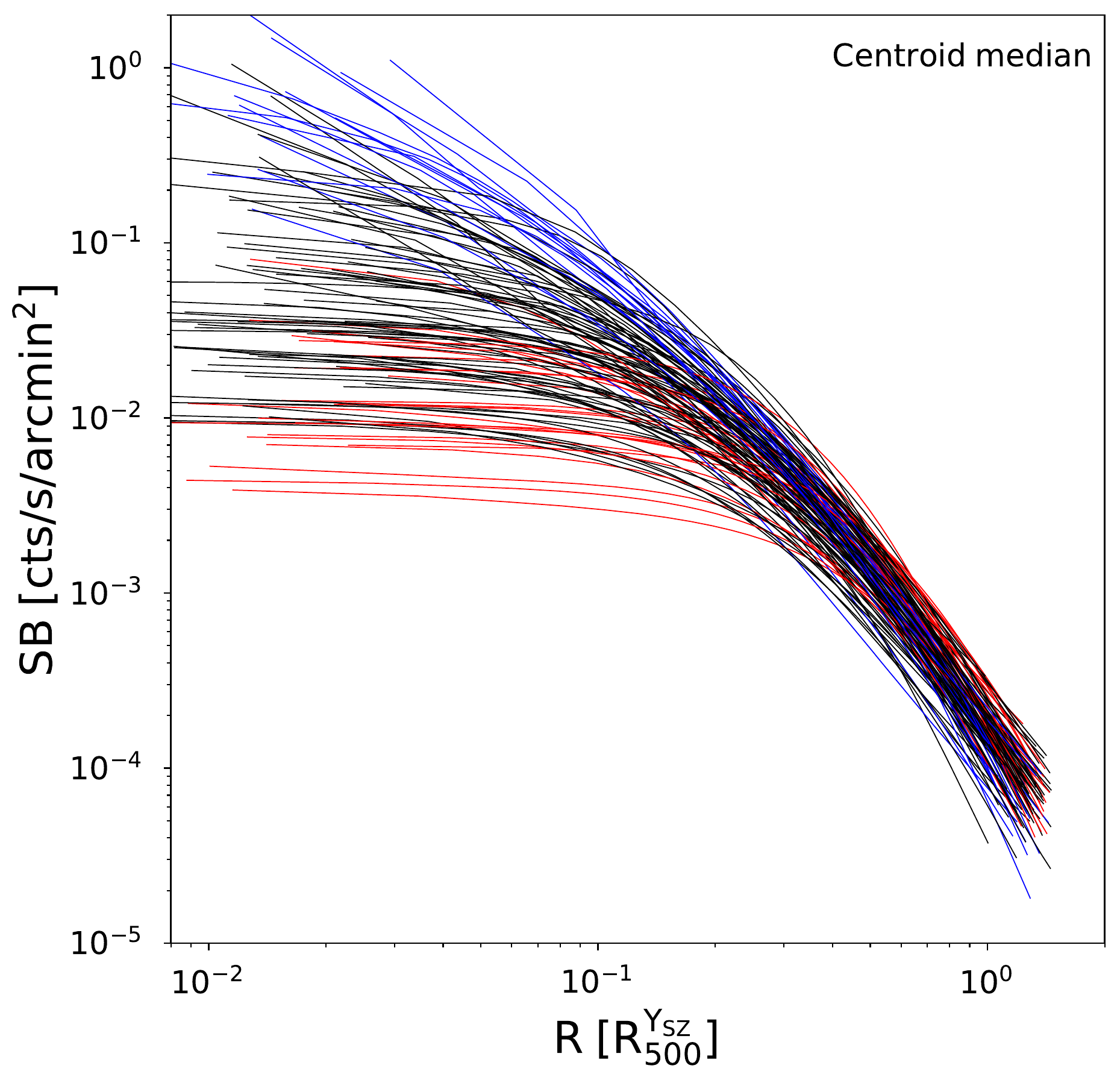}
}
\end{center}
\caption{\footnotesize{Surface brightness radial profiles of the \chxmt\ sample. Left column:  Azimuthal averaged surface brightness profiles of the \chxmt\ sample centred on the X-peak and the centroid in the top and bottom panels, respectively. Blue and red solid lines represent morphologically relaxed and disturbed clusters, respectively. The black solid vertical line identifies $\Rvysz$. Right column: Same as left column but for profiles extracted computing the azimuthal median.}}
\label{fig:sx_all_profiles}
\end{figure*}

We show in \ref{fig:sx_all_profiles} the surface brightness profiles of the \chxmt\ sample that we extracted as described in Section \ref{sec:sx_profiles}. Thedotted line shown in the top-left panel indicates $\Rvysz$ and highlights the data quality of the sample as most of the profiles extend beyond that radius.

\begin{center}
\begin{landscape}
\begin{longtable}{lccccccc}
\caption{ {\footnotesize Observational, morphological, and global properties of the \chxmt\ sample.}}\label{tab:500_prop}\\
\hline        
\hline
\planck\ name & z & X-peak  & Centroid &  N$_\mathrm{H}$    &   $\Rvysz$ & $\Mvysz$               & Morphology  \\
              &   &  RA \phantom{pippopippo} DEC       &  RA \phantom{pippopippo} DEC        &                    &             &                         &           \\
             &   & [J2000] & [J2000]  & [$10^{20}$cm$^{-2}$]  & [arcmin]    & [$10^{14}$ M$_{\odot}$] &            \\
\hline\\
\endfirsthead
\caption{ {\footnotesize continuation}}\\
\hline        
\hline
\planck\ name & z & X-peak  & Centroid &  N$_\mathrm{H}$    &   $\Rvysz$ & $\Mvysz$               & Morphology  \\
              &   &  RA \phantom{pippopippo} DEC       &  RA \phantom{pippopippo} DEC        &                    &             &                         &           \\
             &   & [J2000] & [J2000]  & [$10^{20}$cm$^{-2}$]  & [arcmin]    & [$10^{14}$ M$_{\odot}$] &            \\
\hline\\
\endhead
PSZ2 G075.71+13.51    & $ 0.056 $ & $ 19:21: 12.415$ \, \,  $ 43:56: 50.623$ &  $ 19:21:  9.426$ \, \, $ 43:57: 57.118 $  & $ 1.91 $ & $ 22.020 $ & $  8.740 $ & M\\
PSZ2 G068.22+15.18    & $ 0.057 $ & $ 18:57: 37.625$ \, \,  $ 38: \phantom{0}0: 31.119$ &  $ 18:57: 42.458$ \, \, $ 38: \phantom{0}0: 12.868 $  & $ 3.31 $ & $ 13.548 $ & $  2.142 $ & M\\
PSZ2 G040.03+74.95    & $ 0.061 $ & $ 13:59: 15.106$ \, \,  $ 27:58: 34.080$ &  $ 13:59: 14.381$ \, \, $ 27:59:  6.593 $  & $ 1.48 $ & $ 12.980 $ & $  2.342 $ & M\\
PSZ2 G033.81+77.18    & $ 0.062 $ & $ 13:48: 52.939$ \, \,  $ 26:35: 26.919$ &  $ 13:48: 52.646$ \, \, $ 26:35: 44.943 $  & $ 1.69 $ & $ 15.848 $ & $  4.463 $ & R\\
PSZ2 G057.78+52.32    & $ 0.065 $ & $ 15:44: 59.000$ \, \,  $ 36: \phantom{0}6: 40.699$ &  $ 15:44: 57.836$ \, \, $ 36: \phantom{0}7: 34.614 $  & $ 5.67 $ & $ 12.147 $ & $  2.316 $ & M\\
PSZ2 G105.55+77.21    & $ 0.072 $ & $ 13:11:  \phantom{0}4.980$ \, \,  $ 39:13: 26.314$ &  $ 13:11:  \phantom{0}8.588$ \, \, $ 39:13: 55.277 $  & $ 3.43 $ & $ 10.901 $ & $  2.196 $ & M\\
PSZ2 G042.81+56.61    & $ 0.072 $ & $ 15:22: 29.473$ \, \,  $ 27:42: 27.922$ &  $ 15:22: 25.737$ \, \, $ 27:43: 22.702 $  & $ 5.36 $ & $ 13.499 $ & $  4.219 $ & M\\
PSZ2 G031.93+78.71    & $ 0.072 $ & $ 13:41: 48.706$ \, \,  $ 26:22: 21.527$ &  $ 13:41: 50.464$ \, \, $ 26:22: 46.747 $  & $ 5.14 $ & $ 11.643 $ & $  2.718 $ & M\\
PSZ2 G287.46+81.12    & $ 0.073 $ & $ 12:41: 17.552$ \, \,  $ 18:34: 26.322$ &  $ 12:41: 17.618$ \, \, $ 18:33: 48.907 $  & $ 2.87 $ & $ 11.329 $ & $  2.563 $ & M\\
PSZ2 G040.58+77.12    & $ 0.075 $ & $ 13:49: 23.507$ \, \,  $ 28: \phantom{0}6: 23.519$ &  $ 13:49: 23.701$ \, \, $ 28: \phantom{0}5: 58.161 $  & $ 1.06 $ & $ 11.082 $ & $  2.569 $ & M\\
PSZ2 G057.92+27.64    & $ 0.076 $ & $ 17:44: 14.333$ \, \,  $ 32:59: 29.115$ &  $ 17:44: 14.092$ \, \, $ 32:59: 27.068 $  & $ 5.60 $ & $ 11.084 $ & $  2.659 $ & M\\
PSZ2 G006.49+50.56    & $ 0.077 $ & $ 15:10: 56.232$ \, \,  $  \phantom{0}5:44: 40.654$ &  $ 15:10: 56.030$ \, \, $  \phantom{0}5:44: 40.937 $  & $ 1.55 $ & $ 15.168 $ & $  7.045 $ & R\\
PSZ2 G048.10+57.16    & $ 0.078 $ & $ 15:21:  \phantom{0}8.350$ \, \,  $ 30:38:  \phantom{0}7.594$ &  $ 15:21: 15.377$ \, \, $ 30:38: 16.582 $  & $ 4.34 $ & $ 11.898 $ & $  3.540 $ & D\\
PSZ2 G172.74+65.30    & $ 0.079 $ & $ 11:11: 39.825$ \, \,  $ 40:50: 23.218$ &  $ 11:11: 39.716$ \, \, $ 40:50: 21.227 $  & $ 2.10 $ & $ 10.226 $ & $  2.388 $ & M\\
PSZ2 G057.61+34.93    & $ 0.080 $ & $ 17: \phantom{0}9: 47.051$ \, \,  $ 34:27: 14.010$ &  $ 17: 9: 47.446$ \, \, $ 34:27: 23.362 $  & $ 1.43 $ & $ 11.727 $ & $  3.705 $ & M\\
PSZ2 G243.64+67.74    & $ 0.083 $ & $ 11:32: 51.251$ \, \,  $ 14:27: 15.521$ &  $ 11:32: 50.980$ \, \, $ 14:28: 24.900 $  & $ 1.69 $ & $ 11.226 $ & $  3.624 $ & M\\
PSZ2 G080.16+57.65    & $ 0.088 $ & $ 15: \phantom{0}1:  \phantom{0}7.943$ \, \,  $ 47:16: 38.232$ &  $ 15: \phantom{0}0: 59.165$ \, \, $ 47:16: 51.937 $  & $ 2.91 $ & $  9.474 $ & $  2.513 $ & D\\
PSZ2 G044.20+48.66    & $ 0.089 $ & $ 15:58: 20.468$ \, \,  $ 27:13: 48.074$ &  $ 15:58: 21.094$ \, \, $ 27:13: 43.075 $  & $ 3.79 $ & $ 14.133 $ & $  8.773 $ & M\\
PSZ2 G114.79-33.71    & $ 0.094 $ & $  \phantom{0}0:20: 36.995$ \, \,  $ 28:39: 36.116$ &  $  \phantom{0}0:20: 36.917$ \, \, $ 28:39: 42.461 $  & $ 2.47 $ & $ 10.198 $ & $  3.788 $ & M\\
PSZ2 G056.77+36.32    & $ 0.095 $ & $ 17: \phantom{0}2: 42.568$ \, \,  $ 34: \phantom{0}3: 36.225$ &  $ 17: \phantom{0}2: 41.971$ \, \, $ 34: \phantom{0}3: 32.091 $  & $ 2.04 $ & $ 10.535 $ & $  4.338 $ & M\\
PSZ2 G049.32+44.37    & $ 0.097 $ & $ 16:20: 30.894$ \, \,  $ 29:53: 30.425$ &  $ 16:20: 32.043$ \, \, $ 29:53: 35.492 $  & $ 1.38 $ & $  9.866 $ & $  3.763 $ & M\\
PSZ2 G080.37+14.64    & $ 0.098 $ & $ 19:26:  \phantom{0}9.543$ \, \,  $ 48:33:  \phantom{0}0.121$ &  $ 19:26: 10.342$ \, \, $ 48:32: 54.284 $  & $ 1.29 $ & $  9.205 $ & $  3.127 $ & M\\
PSZ2 G099.48+55.60    & $ 0.105 $ & $ 14:28: 38.408$ \, \,  $ 56:51: 36.620$ &  $ 14:28: 34.413$ \, \, $ 56:53:  9.043 $  & $ 1.76 $ & $  8.271 $ & $  2.749 $ & D\\
PSZ2 G080.41-33.24    & $ 0.107 $ & $ 22:26:  \phantom{0}5.889$ \, \,  $ 17:21: 51.623$ &  $ 22:26:  \phantom{0}4.592$ \, \, $ 17:22: 24.699 $  & $ 5.57 $ & $  9.028 $ & $  3.774 $ & M\\
PSZ2 G113.29-29.69    & $ 0.107 $ & $  \phantom{0}0:11: 45.583$ \, \,  $ 32:24: 53.523$ &  $  \phantom{0}0:11: 45.554$ \, \, $ 32:24: 52.973 $  & $ 1.59 $ & $  8.857 $ & $  3.573 $ & M\\
PSZ2 G053.53+59.52    & $ 0.113 $ & $ 15:10: 12.594$ \, \,  $ 33:30: 29.924$ &  $ 15:10: 11.521$ \, \, $ 33:30: 12.360 $  & $ 2.06 $ & $  9.580 $ & $  5.209 $ & M\\
PSZ2 G046.88+56.48    & $ 0.115 $ & $ 15:24:  \phantom{0}8.181$ \, \,  $ 29:53:  \phantom{0}6.523$ &  $ 15:24:  \phantom{0}9.401$ \, \, $ 29:53: 31.977 $  & $ 4.02 $ & $  9.402 $ & $  5.104 $ & D\\
PSZ2 G204.10+16.51    & $ 0.122 $ & $  \phantom{0}7:35: 47.144$ \, \,  $ 15: \phantom{0}6: 48.825$ &  $  7:35: 47.232$ \, \, $ 15: \phantom{0}6: 58.589 $  & $ 1.16 $ & $  7.978 $ & $  3.706 $ & M\\
PSZ2 G192.18+56.12    & $ 0.124 $ & $ 10:16: 22.500$ \, \,  $ 33:38: 19.329$ &  $ 10:16: 24.056$ \, \, $ 33:38: 18.491 $  & $ 3.28 $ & $  7.801 $ & $  3.620 $ & M\\
PSZ2 G098.44+56.59    & $ 0.132 $ & $ 14:27: 24.855$ \, \,  $ 55:44: 55.015$ &  $ 14:27: 22.753$ \, \, $ 55:44: 59.931 $  & $ 3.92 $ & $  6.800 $ & $  2.826 $ & M\\
PSZ2 G273.59+63.27    & $ 0.134 $ & $ 12: \phantom{0}0: 24.093$ \, \,  $  \phantom{0}3:20: 40.725$ &  $ 12: \phantom{0}0: 23.716$ \, \, $  \phantom{0}3:20: 31.335 $  & $ 0.82 $ & $  8.353 $ & $  5.465 $ & M\\
PSZ2 G217.09+40.15    & $ 0.136 $ & $  \phantom{0}9:24:  \phantom{0}5.764$ \, \,  $ 14:10: 25.122$ &  $  \phantom{0}9:24:  \phantom{0}6.079$ \, \, $ 14:10: 19.625 $  & $ 1.34 $ & $  7.369 $ & $  3.890 $ & M\\
PSZ2 G218.59+71.31    & $ 0.137 $ & $ 11:29: 52.386$ \, \,  $ 23:48: 39.122$ &  $ 11:29: 51.064$ \, \, $ 23:48: 49.086 $  & $ 7.13 $ & $  7.219 $ & $  3.759 $ & D\\
PSZ2 G179.09+60.12    & $ 0.137 $ & $ 10:40: 44.656$ \, \,  $ 39:57: 10.577$ &  $ 10:40: 44.381$ \, \, $ 39:57: 12.733 $  & $ 1.03 $ & $  7.265 $ & $  3.839 $ & M\\
PSZ2 G226.18+76.79    & $ 0.143 $ & $ 11:55: 17.827$ \, \,  $ 23:24: 16.924$ &  $ 11:55: 18.153$ \, \, $ 23:24: 15.915 $  & $ 7.24 $ & $  8.129 $ & $  5.974 $ & M\\
PSZ2 G077.90-26.63    & $ 0.147 $ & $ 22: \phantom{0}0: 53.218$ \, \,  $ 20:58: 19.022$ &  $ 22: \phantom{0}0: 53.101$ \, \, $ 20:58: 23.993 $  & $ 2.24 $ & $  7.456 $ & $  4.989 $ & M\\
PSZ2 G021.10+33.24    & $ 0.151 $ & $ 16:32: 46.985$ \, \,  $  \phantom{0}5:34: 30.355$ &  $ 16:32: 46.967$ \, \, $  \phantom{0}5:34: 32.936 $  & $ 1.19 $ & $  8.427 $ & $  7.788 $ & R\\
PSZ2 G028.89+60.13    & $ 0.153 $ & $ 15: \phantom{0}0: 19.607$ \, \,  $ 21:22:  \phantom{0}9.023$ &  $ 15: \phantom{0}0: 19.783$ \, \, $ 21:22: 11.193 $  & $ 1.57 $ & $  6.940 $ & $  4.473 $ & R\\
PSZ2 G313.88-17.11    & $ 0.153 $ & $ 16: \phantom{0}1: 48.622$ \, \, $-75:45: 16.177$ &  $ 16: \phantom{0}1: 48.193$ \, \, $-75:45:  \phantom{0}8.926 $  & $ 5.79 $ & $  8.374 $ & $  7.858 $ & R\\
PSZ2 G071.63+29.78    & $ 0.156 $ & $ 17:47: 14.949$ \, \,  $ 45:13: 16.921$ &  $ 17:47: 13.660$ \, \, $ 45:12: 37.672 $  & $ 2.11 $ & $  6.624 $ & $  4.131 $ & D\\
PSZ2 G062.46-21.35    & $ 0.162 $ & $ 21: \phantom{0}4: 53.306$ \, \,  $ 14: \phantom{0}1: 29.123$ &  $ 21: \phantom{0}4: 52.969$ \, \, $ 14: \phantom{0}1: 29.734 $  & $ 5.86 $ & $  6.431 $ & $  4.106 $ & R\\
PSZ2 G094.69+26.36    & $ 0.162 $ & $ 18:32: 32.424$ \, \,  $ 64:50:  \phantom{0}0.613$ &  $ 18:32: 30.227$ \, \, $ 64:49: 38.915 $  & $ 1.72 $ & $  5.818 $ & $  3.081 $ & M\\
PSZ2 G066.68+68.44    & $ 0.163 $ & $ 14:21: 40.389$ \, \,  $ 37:17: 30.114$ &  $ 14:21: 39.902$ \, \, $ 37:17: 31.735 $  & $ 1.52 $ & $  6.218 $ & $  3.802 $ & R\\
PSZ2 G050.40+31.17    & $ 0.164 $ & $ 17:20:  \phantom{0}8.584$ \, \,  $ 27:40: 13.923$ &  $ 17:20:  \phantom{0}9.163$ \, \, $ 27:40: 11.458 $  & $ 4.91 $ & $  6.403 $ & $  4.219 $ & M\\
PSZ2 G049.22+30.87    & $ 0.164 $ & $ 17:20:  \phantom{0}9.866$ \, \,  $ 26:37: 31.037$ &  $ 17:20:  \phantom{0}9.690$ \, \, $ 26:37: 21.341 $  & $ 1.24 $ & $  7.146 $ & $  5.904 $ & R\\
PSZ2 G263.68-22.55    & $ 0.164 $ & $  \phantom{0}6:45: 29.064$ \, \, $-54:13: 39.415$ &  $  \phantom{0}6:45: 30.011$ \, \, $-54:13: 28.896 $  & $ 4.91 $ & $  7.893 $ & $  7.955 $ & M\\
PSZ2 G285.63+72.75    & $ 0.165 $ & $ 12:30: 47.325$ \, \,  $ 10:33:  \phantom{0}7.622$ &  $ 12:30: 46.802$ \, \, $ 10:33: 28.002 $  & $ 5.00 $ & $  7.002 $ & $  5.606 $ & M\\
PSZ2 G238.69+63.26    & $ 0.169 $ & $ 11:12: 54.419$ \, \,  $ 13:26:  \phantom{0}4.625$ &  $ 11:12: 54.181$ \, \, $ 13:26: 33.234 $  & $ 7.26 $ & $  6.211 $ & $  4.167 $ & M\\
PSZ2 G149.39-36.84    & $ 0.170 $ & $  \phantom{0}2:21: 34.396$ \, \,  $ 21:21: 56.224$ &  $  \phantom{0}2:21: 34.675$ \, \, $ 21:22:  \phantom{0}9.957 $  & $ 8.46 $ & $  6.714 $ & $  5.346 $ & M\\
PSZ2 G187.53+21.92    & $ 0.171 $ & $  \phantom{0}7:32: 20.308$ \, \,  $ 31:37: 57.919$ &  $  \phantom{0}7:32: 20.464$ \, \, $ 31:37: 55.784 $  & $ 1.35 $ & $  6.604 $ & $  5.165 $ & M\\
PSZ2 G000.13+78.04    & $ 0.171 $ & $ 13:34:  \phantom{0}8.218$ \, \,  $ 20:14: 27.123$ &  $ 13:34:  \phantom{0}9.672$ \, \, $ 20:14: 28.105 $  & $ 3.37 $ & $  6.586 $ & $  5.122 $ & M\\
PSZ2 G067.17+67.46    & $ 0.171 $ & $ 14:26:  \phantom{0}2.032$ \, \,  $ 37:49: 33.518$ &  $ 14:26:  \phantom{0}0.963$ \, \, $ 37:49: 39.424 $  & $ 1.89 $ & $  7.350 $ & $  7.143 $ & M\\
PSZ2 G218.81+35.51    & $ 0.175 $ & $  \phantom{0}9: \phantom{0}9: 12.697$ \, \,  $ 10:58: 30.523$ &  $  \phantom{0}9: \phantom{0}9: 12.554$ \, \, $ 10:58: 33.929 $  & $ 2.62 $ & $  6.501 $ & $  5.241 $ & D\\
PSZ2 G067.52+34.75    & $ 0.175 $ & $ 17:17: 18.999$ \, \,  $ 42:26: 59.275$ &  $ 17:17: 18.625$ \, \, $ 42:26: 54.180 $  & $ 3.01 $ & $  6.167 $ & $  4.494 $ & M\\
PSZ2 G041.45+29.10    & $ 0.178 $ & $ 17:17: 44.795$ \, \,  $ 19:40: 35.724$ &  $ 17:17: 47.051$ \, \, $ 19:40: 37.372 $  & $ 3.24 $ & $  6.477 $ & $  5.411 $ & M\\
PSZ2 G085.98+26.69    & $ 0.179 $ & $ 18:19: 57.539$ \, \,  $ 57: \phantom{0}9: 39.886$ &  $ 18:19: 54.155$ \, \, $ 57:10:  \phantom{0}9.398 $  & $ 1.51 $ & $  5.910 $ & $  4.172 $ & D\\
PSZ2 G111.75+70.37    & $ 0.183 $ & $ 13:13:  \phantom{0}6.819$ \, \,  $ 46:17: 25.912$ &  $ 13:13:  \phantom{0}5.636$ \, \, $ 46:16: 31.997 $  & $ 2.91 $ & $  5.876 $ & $  4.342 $ & D\\
PSZ2 G313.33+61.13    & $ 0.183 $ & $ 13:11: 29.410$ \, \, $- \phantom{0}1:20: 29.175$ &  $ 13:11: 29.520$ \, \, $- \phantom{0}1:20: 25.528 $  & $ 1.02 $ & $  7.421 $ & $  8.771 $ & R\\
PSZ2 G083.86+85.09    & $ 0.183 $ & $ 13: \phantom{0}5: 50.907$ \, \,  $ 30:53: 43.423$ &  $ 13: \phantom{0}5: 51.368$ \, \, $ 30:53: 56.263 $  & $ 7.36 $ & $  6.043 $ & $  4.735 $ & M\\
PSZ2 G217.40+10.88    & $ 0.189 $ & $  \phantom{0}7:38: 18.558$ \, \,  $  \phantom{0}1: \phantom{0}2: 15.325$ &  $  \phantom{0}7:38: 18.362$ \, \, $  \phantom{0}1: \phantom{0}2: 16.088 $  & $ 1.15 $ & $  6.123 $ & $  5.340 $ & R\\
PSZ2 G224.00+69.33    & $ 0.190 $ & $ 11:23: 58.030$ \, \,  $ 21:28: 57.124$ &  $ 11:23: 58.286$ \, \, $ 21:29:  \phantom{0}0.392 $  & $ 3.08 $ & $  5.994 $ & $  5.106 $ & M\\
PSZ2 G124.20-36.48    & $ 0.197 $ & $  \phantom{0}0:55: 50.350$ \, \,  $ 26:24: 35.217$ &  $  \phantom{0}0:55: 53.140$ \, \, $ 26:24: 31.763 $  & $ 1.49 $ & $  6.541 $ & $  7.253 $ & D\\
PSZ2 G195.75-24.32    & $ 0.203 $ & $  \phantom{0}4:54:  \phantom{0}9.827$ \, \,  $  \phantom{0}2:55: 29.225$ &  $  \phantom{0}4:54: 10.087$ \, \, $  \phantom{0}2:55: 53.424 $  & $ 1.33 $ & $  6.535 $ & $  7.800 $ & M\\
PSZ2 G159.91-73.50    & $ 0.206 $ & $  \phantom{0}1:31: 53.353$ \, \, $-13:36: 42.371$ &  $  \phantom{0}1:31: 53.063$ \, \, $-13:36: 47.658 $  & $ 1.69 $ & $  6.632 $ & $  8.464 $ & M\\
PSZ2 G346.61+35.06    & $ 0.223 $ & $ 15:15:  \phantom{0}2.849$ \, \, $-15:23: 10.574$ &  $ 15:15:  \phantom{0}1.989$ \, \, $-15:22: 39.963 $  & $ 5.53 $ & $  6.197 $ & $  8.409 $ & D\\
PSZ2 G055.59+31.85    & $ 0.224 $ & $ 17:22: 27.224$ \, \,  $ 32: \phantom{0}7: 56.518$ &  $ 17:22: 26.675$ \, \, $ 32: \phantom{0}7: 51.890 $  & $ 1.03 $ & $  5.992 $ & $  7.724 $ & M\\
PSZ2 G092.71+73.46    & $ 0.228 $ & $ 13:35: 17.944$ \, \,  $ 41: \phantom{0}0:  1.126$ &  $ 13:35: 18.896$ \, \, $ 41: \phantom{0}0:  \phantom{0}9.805 $  & $ 2.21 $ & $  5.977 $ & $  8.003 $ & M\\
PSZ2 G072.62+41.46    & $ 0.228 $ & $ 16:40: 20.160$ \, \,  $ 46:42: 31.218$ &  $ 16:40: 20.457$ \, \, $ 46:42: 26.700 $  & $ 3.32 $ & $  6.727 $ & $ 11.426 $ & M\\
PSZ2 G073.97-27.82    & $ 0.233 $ & $ 21:53: 36.775$ \, \,  $ 17:41: 42.122$ &  $ 21:53: 36.848$ \, \, $ 17:41: 48.852 $  & $ 2.20 $ & $  6.218 $ & $  9.516 $ & M\\
PSZ2 G340.94+35.07    & $ 0.236 $ & $ 14:59: 29.073$ \, \, $-18:10: 44.774$ &  $ 14:59: 29.213$ \, \, $-18:10: 44.609 $  & $ 3.04 $ & $  5.760 $ & $  7.795 $ & M\\
PSZ2 G208.80-30.67    & $ 0.248 $ & $  \phantom{0}4:54:  \phantom{0}6.672$ \, \, $-10:13: 12.866$ &  $  \phantom{0}4:54:  \phantom{0}8.829$ \, \, $-10:14: 19.927 $  & $ 1.71 $ & $  5.400 $ & $  7.255 $ & D\\
PSZ2 G340.36+60.58    & $ 0.253 $ & $ 14: \phantom{0}1:  \phantom{0}2.040$ \, \,  $  \phantom{0}2:52: 41.925$ &  $ 14: \phantom{0}1:  \phantom{0}1.901$ \, \, $  \phantom{0}2:52: 39.166 $  & $ 3.91 $ & $  5.743 $ & $  9.199 $ & R\\
PSZ2 G266.83+25.08    & $ 0.254 $ & $ 10:23: 50.072$ \, \, $-27:15: 20.572$ &  $ 10:23: 49.951$ \, \, $-27:15: 23.133 $  & $ 2.99 $ & $  5.282 $ & $  7.258 $ & R\\
PSZ2 G229.74+77.96    & $ 0.269 $ & $ 12: \phantom{0}1: 14.612$ \, \,  $ 23: \phantom{0}6: 28.724$ &  $ 12: \phantom{0}1: 16.318$ \, \, $ 23: \phantom{0}6: 30.125 $  & $ 1.92 $ & $  5.084 $ & $  7.441 $ & D\\
PSZ2 G087.03-57.37    & $ 0.278 $ & $ 23:37: 37.705$ \, \,  $  \phantom{0}0:16:  \phantom{0}2.925$ &  $ 23:37: 39.382$ \, \, $  \phantom{0}0:16: 13.669 $  & $ 6.63 $ & $  4.926 $ & $  7.329 $ & M\\
PSZ2 G107.10+65.32    & $ 0.280 $ & $ 13:32: 38.954$ \, \,  $ 50:33: 30.306$ &  $ 13:32: 44.154$ \, \, $ 50:32: 56.523 $  & $ 1.13 $ & $  4.999 $ & $  7.800 $ & D\\
PSZ2 G186.37+37.26    & $ 0.282 $ & $  \phantom{0}8:42: 57.144$ \, \,  $ 36:21: 57.014$ &  $  \phantom{0}8:42: 57.517$ \, \, $ 36:21: 51.864 $  & $ 1.35 $ & $  5.573 $ & $ 10.998 $ & M\\
PSZ2 G259.98-63.43    & $ 0.284 $ & $  \phantom{0}2:32: 18.459$ \, \, $-44:20: 47.910$ &  $  \phantom{0}2:32: 17.649$ \, \, $-44:20: 57.756 $  & $ 5.90 $ & $  4.872 $ & $  7.451 $ & M\\
PSZ2 G106.87-83.23    & $ 0.292 $ & $  \phantom{0}0:43: 24.811$ \, \, $-20:37: 24.747$ &  $  \phantom{0}0:43: 24.574$ \, \, $-20:37: 22.358 $  & $ 6.83 $ & $  4.812 $ & $  7.732 $ & M\\
PSZ2 G262.27-35.38    & $ 0.295 $ & $  \phantom{0}5:16: 37.024$ \, \, $-54:30: 56.483$ &  $  \phantom{0}5:16: 39.379$ \, \, $-54:31:  1.180 $  & $ 3.54 $ & $  4.978 $ & $  8.759 $ & D\\
PSZ2 G266.04-21.25    & $ 0.296 $ & $  \phantom{0}6:58: 30.077$ \, \, $-55:56: 37.797$ &  $  \phantom{0}6:58: 29.777$ \, \, $-55:56: 43.551 $  & $ 4.49 $ & $  5.580 $ & $ 12.470 $ & M\\
PSZ2 G008.94-81.22    & $ 0.307 $ & $  \phantom{0}0:14: 19.107$ \, \, $-30:23: 28.159$ &  $  \phantom{0}0:14: 17.229$ \, \, $-30:23:  \phantom{0}7.230 $  & $ 3.04 $ & $  4.870 $ & $  8.989 $ & D\\
PSZ2 G278.58+39.16    & $ 0.308 $ & $ 11:31: 54.653$ \, \, $-19:55: 44.575$ &  $ 11:31: 56.147$ \, \, $-19:55: 44.025 $  & $ 4.02 $ & $  4.730 $ & $  8.290 $ & M\\
PSZ2 G008.31-64.74    & $ 0.312 $ & $ 22:58: 48.237$ \, \, $-34:48:  \phantom{0}1.673$ &  $ 22:58: 48.340$ \, \, $-34:48: 16.957 $  & $ 1.88 $ & $  4.506 $ & $  7.421 $ & D\\
PSZ2 G325.70+17.34    & $ 0.315 $ & $ 14:47: 33.340$ \, \, $-40:20: 36.772$ &  $ 14:47: 32.611$ \, \, $-40:20: 33.188 $  & $ 3.53 $ & $  4.495 $ & $  7.570 $ & M\\
PSZ2 G349.46-59.95    & $ 0.347 $ & $ 22:48: 44.575$ \, \, $-44:31: 47.748$ &  $ 22:48: 44.883$ \, \, $-44:31: 42.214 $  & $ 3.99 $ & $  4.768 $ & $ 11.359 $ & M\\
PSZ2 G207.88+81.31    & $ 0.353 $ & $ 12:12: 18.311$ \, \,  $ 27:32: 54.422$ &  $ 12:12: 18.904$ \, \, $ 27:33:  5.442 $  & $ 3.30 $ & $  4.090 $ & $  7.440 $ & M\\
PSZ2 G143.26+65.24    & $ 0.363 $ & $ 11:59: 14.275$ \, \,  $ 49:47: 41.128$ &  $ 11:59: 15.703$ \, \, $ 49:47: 51.346 $  & $ 1.33 $ & $  3.966 $ & $  7.257 $ & D\\
PSZ2 G271.18-30.95    & $ 0.370 $ & $  \phantom{0}5:49: 19.297$ \, \, $-62: \phantom{0}5: 14.978$ &  $  \phantom{0}5:49: 18.796$ \, \, $-62: \phantom{0}5: 11.902 $  & $ 6.20 $ & $  3.932 $ & $  7.373 $ & R\\
PSZ2 G113.91-37.01    & $ 0.371 $ & $  \phantom{0}0:19: 41.833$ \, \,  $ 25:18:  \phantom{0}4.618$ &  $  \phantom{0}0:19: 39.458$ \, \, $ 25:17: 28.226 $  & $ 4.69 $ & $  3.958 $ & $  7.582 $ & D\\
PSZ2 G172.98-53.55    & $ 0.373 $ & $  \phantom{0}2:39: 53.403$ \, \, $- \phantom{0}1:34: 43.974$ &  $  \phantom{0}2:39: 54.152$ \, \, $- \phantom{0}1:34: 48.158 $  & $ 7.51 $ & $  3.906 $ & $  7.367 $ & M\\
PSZ2 G216.62+47.00    & $ 0.383 $ & $  \phantom{0}9:49: 51.753$ \, \,  $ 17: \phantom{0}7: 10.527$ &  $  \phantom{0}9:49: 51.757$ \, \, $ 17: \phantom{0}7: 18.162 $  & $ 5.01 $ & $  4.012 $ & $  8.469 $ & M\\
PSZ2 G046.10+27.18    & $ 0.389 $ & $ 17:31: 38.965$ \, \,  $ 22:51: 47.874$ &  $ 17:31: 40.056$ \, \, $ 22:51: 52.331 $  & $ 4.60 $ & $  3.861 $ & $  7.840 $ & D\\
PSZ2 G286.98+32.90    & $ 0.390 $ & $ 11:50: 49.017$ \, \, $-28: \phantom{0}4: 36.574$ &  $ 11:50: 49.731$ \, \, $-28: \phantom{0}4: 51.557 $  & $ 2.71 $ & $  4.646 $ & $ 13.742 $ & M\\
PSZ2 G057.25-45.34    & $ 0.397 $ & $ 22:11: 45.747$ \, \, $- \phantom{0}3:49: 45.974$ &  $ 22:11: 45.754$ \, \, $- \phantom{0}3:49: 35.485 $  & $ 3.19 $ & $  4.070 $ & $  9.624 $ & M\\
PSZ2 G206.45+13.89    & $ 0.410 $ & $  \phantom{0}7:29: 50.887$ \, \,  $ 11:56: 28.523$ &  $  \phantom{0}7:29: 51.061$ \, \, $ 11:56: 26.337 $  & $ 6.87 $ & $  3.648 $ & $  7.459 $ & M\\
PSZ2 G243.15-73.84    & $ 0.410 $ & $  \phantom{0}1:59:  \phantom{0}1.932$ \, \, $-34:12: 54.069$ &  $  \phantom{0}1:59:  \phantom{0}1.523$ \, \, $-34:13: 18.610 $  & $ 3.37 $ & $  3.747 $ & $  8.086 $ & M\\
PSZ2 G083.29-31.03    & $ 0.412 $ & $ 22:28: 33.296$ \, \,  $ 20:37: 12.319$ &  $ 22:28: 33.252$ \, \, $ 20:37: 16.322 $  & $ 2.10 $ & $  3.664 $ & $  7.642 $ & M\\
PSZ2 G241.11-28.68    & $ 0.420 $ & $  \phantom{0}5:42: 56.882$ \, \, $-36: \phantom{0}0:  0.975$ &  $  \phantom{0}5:42: 56.210$ \, \, $-35:59: 54.191 $  & $ 3.89 $ & $  3.566 $ & $  7.361 $ & M\\
PSZ2 G262.73-40.92    & $ 0.421 $ & $  \phantom{0}4:38: 17.598$ \, \, $-54:19: 24.413$ &  $  \phantom{0}4:38: 17.538$ \, \, $-54:19: 18.041 $  & $ 2.74 $ & $  3.576 $ & $  7.461 $ & M\\
PSZ2 G239.27-26.01    & $ 0.430 $ & $  \phantom{0}5:53: 26.252$ \, \, $-33:42: 36.972$ &  $  \phantom{0}5:53: 24.362$ \, \, $-33:42: 35.104 $  & $ 1.84 $ & $  3.714 $ & $  8.772 $ & M\\
PSZ2 G225.93-19.99    & $ 0.435 $ & $  \phantom{0}6: \phantom{0}0:  \phantom{0}8.022$ \, \, $-20: \phantom{0}8:  \phantom{0}5.678$ &  $  \phantom{0}6: \phantom{0}0: 10.214$ \, \, $-20: \phantom{0}7: 38.748 $  & $ 2.93 $ & $  3.819 $ & $  9.789 $ & D\\
PSZ2 G277.76-51.74    & $ 0.438 $ & $  \phantom{0}2:54: 16.472$ \, \, $-58:56: 59.907$ &  $  \phantom{0}2:54: 23.102$ \, \, $-58:57: 28.650 $  & $ 4.90 $ & $  3.646 $ & $  8.650 $ & D\\
PSZ2 G284.41+52.45    & $ 0.441 $ & $ 12: \phantom{0}6: 12.169$ \, \, $- \phantom{0}8:48:  \phantom{0}3.836$ &  $ 12: \phantom{0}6: 12.264$ \, \, $- \phantom{0}8:48:  \phantom{0}8.038 $  & $ 2.57 $ & $  3.854 $ & $ 10.400 $ & M\\
PSZ2 G205.93-39.46    & $ 0.443 $ & $  \phantom{0}4:17: 34.752$ \, \, $-11:54: 33.373$ &  $  \phantom{0}4:17: 34.246$ \, \, $-11:54: 21.899 $  & $ 0.99 $ & $  3.980 $ & $ 11.542 $ & M\\
PSZ2 G056.93-55.08    & $ 0.447 $ & $ 22:43: 21.951$ \, \, $- \phantom{0}9:35: 43.371$ &  $ 22:43: 22.815$ \, \, $- \phantom{0}9:35: 52.671 $  & $ 1.76 $ & $  3.704 $ & $  9.491 $ & D\\
PSZ2 G324.04+48.79    & $ 0.452 $ & $ 13:47: 30.663$ \, \, $-11:45:  \phantom{0}7.275$ &  $ 13:47: 30.809$ \, \, $-11:45: 13.630 $  & $ 4.41 $ & $  3.811 $ & $ 10.578 $ & R\\
PSZ2 G210.64+17.09    & $ 0.480 $ & $  \phantom{0}7:48: 46.441$ \, \,  $  \phantom{0}9:40:  \phantom{0}5.820$ &  $  \phantom{0}7:48: 47.331$ \, \, $  \phantom{0}9:40: 13.545 $  & $ 1.71 $ & $  3.289 $ & $  7.790 $ & M\\
PSZ2 G044.77-51.30    & $ 0.503 $ & $ 22:14: 57.349$ \, \, $-14: \phantom{0}0: 11.673$ &  $ 22:14: 57.004$ \, \, $-14: \phantom{0}0: 10.935 $  & $ 8.12 $ & $  3.255 $ & $  8.359 $ & M\\
PSZ2 G201.50-27.31    & $ 0.538 $ & $  \phantom{0}4:54: 10.952$ \, \, $- \phantom{0}3: \phantom{0}0: 53.478$ &  $  \phantom{0}4:54: 11.056$ \, \, $- \phantom{0}3: \phantom{0}0: 49.066 $  & $ 1.21 $ & $  3.094 $ & $  8.304 $ & M\\
PSZ2 G004.45-19.55    & $ 0.540 $ & $ 19:17:  \phantom{0}5.068$ \, \, $-33:31: 20.804$ &  $ 19:17:  \phantom{0}5.347$ \, \, $-33:31: 22.095 $  & $ 1.83 5$ & $  3.291 $ & $ 10.090 $ & M\\
PSZ2 G228.16+75.20    & $ 0.545 $ & $ 11:49: 35.358$ \, \,  $ 22:24:  \phantom{0}9.825$ &  $ 11:49: 35.731$ \, \, $ 22:24:  \phantom{0}0.321 $  & $ 7.24 $ & $  3.237 $ & $  9.790 $ & M\\
PSZ2 G111.61-45.71    & $ 0.546 $ & $  \phantom{0}0:18: 33.528$ \, \,  $ 16:26: 11.320$ &  $  \phantom{0}0:18: 33.396$ \, \, $ 16:26:  \phantom{0}8.511 $  & $ 1.56 $ & $  3.085 $ & $  8.499 $ & M\\
PSZ2 G155.27-68.42    & $ 0.567 $ & $  \phantom{0}1:37: 24.847$ \, \, $- \phantom{0}8:27: 20.166$ &  $  \phantom{0}1:37: 24.693$ \, \, $- \phantom{0}8:27: 33.384 $  & $ 3.83 $ & $  2.986 $ & $  8.365 $ & M\\
PSZ2 G066.41+27.03    & $ 0.575 $ & $ 17:56: 51.042$ \, \,  $ 40: \phantom{0}8:  \phantom{0}4.525$ &  $ 17:56: 49.973$ \, \, $ 40: \phantom{0}8:  \phantom{0}8.150 $  & $ 4.77 $ & $  2.875 $ & $  7.695 $ & D\\
PSZ2 G339.63-69.34    & $ 0.596 $ & $ 23:44: 43.777$ \, \, $-42:43: 11.551$ &  $ 23:44: 44.150$ \, \, $-42:43: 14.971 $  & $ 4.26 $ & $  2.846 $ & $  8.051 $ & R\\
\hline
\end{longtable}
\footnotesize{\textbf{Notes:} \textit{Column 8:} R, D, and M stand for object morphologically relaxed, disturbed, and mixed, respectively, following the morphological classification of \citet{campitiello2022}.}
\end{landscape}
\end{center}
\end{appendix}
\end{document}